\documentclass{aa}  

\usepackage{graphicx}
\usepackage{txfonts}

\usepackage[colorlinks=true,allcolors=blue]{hyperref}
\usepackage{hyperref}
\hypersetup{colorlinks = true, citecolor = {blue}}

\usepackage{subfig}

\newcommand{\ha}{H$\alpha\ $}
\newcommand{\kms}{km s$^{-1}$}
\newcommand{\Isophote}{{\it Isophote}}

\begin{document}

   \title{The PUMA project. III. Incidence and properties of ionised gas disks in ULIRGs, associated velocity dispersion and its dependence on starburstiness}
   
   \titlerunning{ PUMA-III. ionised gas disks}
   
   \author{M. Perna          \inst{\ref{i1},\ref{i2}}\thanks{E-mail: mperna@cab.inta-csic.es}
                \and 
           S. Arribas
                \inst{\ref{i1}} 
                \and
           L. Colina
                \inst{\ref{i1}}
                \and
           M. Pereira Santaella
                \inst{\ref{i1}}
                \and
           I. Lamperti
                \inst{\ref{i1}}
                \and
           E. Di Teodoro
                \inst{\ref{i3},\ref{i4}}
                \and
           H. {\"U}bler
                \inst{\ref{i5},\ref{i6}}
                \and
           L. Costantin
                \inst{\ref{i1},\ref{i7}}
                \and
           R. Maiolino
                \inst{\ref{i5}, \ref{i6}, \ref{i8}}
                \and
           G. Cresci
                \inst{\ref{i2}}
                \and
           E. Bellocchi
                \inst{\ref{i9}}
                \and
           C. Catal\'an-Torrecilla
                \inst{\ref{i1}} 
                \and    
           S. Cazzoli
                \inst{\ref{i10}} 
                \and    
           J. Piqueras L\'opez
                \inst{\ref{i1}} 
                }

   \institute{Centro de Astrobiolog\'ia, (CAB, CSIC--INTA), Departamento de Astrof\'\i sica, Cra. de Ajalvir Km.~4, 28850 -- Torrej\'on de Ardoz, Madrid, Spain\label{i1}
       \and
       INAF - Osservatorio Astrofisico di Arcetri, Largo Enrico Fermi 5, I-50125 Firenze, Italy\label{i2}
       \and
       Department of Physics \& Astronomy, Johns Hopkins University, Baltimore, MD 21218, USA\label{i3}
       \and
       Space Telescope Science Institute, 3700 San Martin Drive, Baltimore, MD 21218, USA\label{i4}
       \and
       University of Cambridge, Cavendish Laboratory, Cambridge CB3 0HE, UK\label{i5}
       \and
       University of Cambridge, Kavli Institute for Cosmology, Cambridge CB3 0HE, UK\label{i6}
       \and
       INAF - Astronomical Observatory of Brera, via Brera 28, I-20121 Milano, Italy\label{i7}
       \and
       Department of Physics and Astronomy, University College London, Gower Street, London WC1E 6BT, UK\label{i8}
       \and
       Centro de Astrobiolog\'ia (CSIC-INTA), ESAC Campus, 28692 Villanueva de la Ca\~nada, Madrid, Spain\label{i9}
       \and 
       IAA - Instituto de Astrof{\'i}sica de Andaluc{\'i}a (CSIC), Apdo. 3004, 18008, Granada, Spain\label{i10}
      }

   \date{Received September 15, 1996; accepted March 16, 1997}

 
  \abstract
   {   }
   {A classical scenario suggests that ULIRGs transform colliding spiral galaxies into a spheroid-dominated early-type galaxy. Recent high-resolution simulations have instead shown that, under some circumstances, rotation disks can be preserved during the merging process or rapidly regrown after coalescence. Our goal is to analyze in detail the ionised gas kinematics in a sample of ULIRGs to infer the incidence of gas rotational dynamics in late-stage interacting galaxies and merger remnants.

   }
   {We analysed integral field spectrograph MUSE data of a sample of 20 nearby ($z < 0.165$) ULIRGs (with 29 individual nuclei), as part of the ``Physics of ULIRGs with MUSE and ALMA'' (PUMA) project. We used multi-Gaussian fitting techniques to identify gaseous disk motions, and the 3D-Barolo tool to model them. 
    }
   {We found that 27\% (8/29) individual nuclei are associated with kpc-scale disk-like gas motions. The rest of the sample displays a plethora of gas kinematics, dominated by winds and merger-induced flows, which make the detection of rotation signatures difficult. On the other hand, the incidence of stellar disk-like motions is $\lesssim 2$ times larger than gaseous disks, as the former are probably less affected by winds and streams. The eight galaxies with a gaseous disk present  relatively high  intrinsic gas velocity dispersion ($\sigma_0 \in [30-85]$ \kms), and rotationally-supported  motions (with gas rotation velocity over  velocity dispersion $v_{rot}/\sigma_0 \gtrsim 1-8$), and dynamical masses in the range $(2-7)\times 10^{10}$ M$_\odot$. By combining our results with those of local and high-z (up to $z \sim 2$)  disk galaxies from the literature, we found a significant correlation between  $\sigma_0$ and the offset from the main sequence ($
 \delta MS$), after correcting for their evolutionary trends.}
   {Our results confirm the presence of kpc-scale rotating disks in interacting galaxies and merger remnants in the PUMA sample, with an incidence going from $27\%$ (gas) to $\lesssim 50\%$ (stars). Their gas $\sigma_0$ is up to a factor of $\sim 4$ higher than in local normal MS galaxies,  similar to high-$z$ starbursts as presented in the literature; this suggests that interactions and mergers enhance the star formation rate while simultaneously increasing the velocity dispersion in the interstellar medium.}


   \maketitle
%
\section{Introduction}

Local ultraluminous infrared galaxies (ULIRGs, with rest-frame [$8-1000\ \mu$m] luminosity  $L_{IR}$ in excess of $10^{12} L_\odot $) are  an important class of objects for understanding the formation and evolution of massive galaxies. A classic evolutionary scenario (\citealt{Sanders1988,Springel2005}) suggests that ULIRGs evolve into elliptical galaxies through a merger-induced dissipative collapse. In this scenario, the gas of colliding galaxies loses angular momentum and energy, falling into the coalescing centre of the system. Here it serves as fuel for the starburst (SB) and the growth of a supermassive black hole (BH), in a dust enshrouded environment. Then, the system evolves into an optically bright quasar once it either consumes or removes shells of gas and dust through powerful winds. Finally, the merger remnant becomes an elliptical galaxy. 

Recent theoretical works have pointed out that dissipative mergers can also lead to the formation of new disk galaxies. Gas that is not efficiently forced to collapse and form new stars, nor expelled by SB and active galactic nuclei (AGN) winds, can be preserved in a disk, or re-form a new disk plane and start regrowing a stellar disk (\citealt{Robertson2006,Robertson2008, Bullock2009,Governato2009,Hopkins2009,Hopkins2013}).  Hydrodynamical simulations show that cold flows from filamentary structures  also play a major role in the buildup of disks in galaxies (\citealt{Keres2005,Dekel2009, Governato2009}). The interaction between inflows and outflows, the amount of gas, as well as the mass ratio of the merging galaxies and their orbital parameters (e.g. \citealt{Hopkins2009}), all affect the probability of preserving (or reforming) a disk.

From an observational point of view, ordered disk-like kinematics are generally observed in merger systems, both at low-$z$ (\citealt{Bellocchi2013,Medling2014, Ueda2014, Barrera2015, Perna2019}) and at high-$z$ (up to $z \gtrsim 4$; e.g. \citealt{Hammer2009,Alaghband2012,Harrison2012,Perna2018,Leung2019,Tadaki2020,Cochrane2021}). 

Recently, we started a project aimed at studying, at sub-kpc scales, the 2D multi-phase outflow structure in a representative sample of 25 local ULIRGs, by comparing the capabilities offered by the ALMA interferometer and the VLT/MUSE integral field spectrograph. The project, labelled PUMA - Physics of ULIRGs with MUSE and ALMA - is described in the first paper of a series, Perna et al. (2021; \citealt{Perna2021} hereinafter). First MUSE data results are also presented in \citet{Perna2021}: we derived stellar kinematics for all the PUMA systems, and found that post-coalescence systems are more likely associated with disk-like motions, while interacting (binary) systems are dominated by non-ordered and streaming motions. We also investigated the presence of nuclear outflows associated with the individual nuclei, and found ionised and neutral outflows in almost all individual nuclei of our ULIRGs sample. A more comprehensive study of physical and kinematic properties of the interstellar medium (ISM) of the archetypical ULIRG Arp 220 was instead presented in \citet[][as part of the PUMA project]{Perna2020}.  In Pereira-Santaella (2021; \citealt{Pereira2021} hereinafter)  we instead analysed the $\sim 220$ GHz and CO(2-1) ALMA observations to constrain the hidden energy source of ULIRGs, providing evidence for the ubiquitous presence of obscured AGN that could dominate their infrared emission.

The PUMA sample also allows us to investigate the presence of rotational motions in connection with inflows and outflows in dissipative mergers, and therefore to test the predictions of hydrodynamical simulations.  Hence, the present paper is aimed at investigating the prevalence of gas rotational motions in the inner regions of PUMA systems, as well as their (mis)alignments with the stellar component. In this work, we also characterise the kinematic properties of the associated disk structures in terms of inclination, rotational velocity, velocity dispersion, and dynamical mass. Finally, we compare PUMA properties with those of other local (U)LIRGs and high-$z$ populations of normal main sequence (MS) and SB galaxies,  studying 
the variation of the gas velocity dispersion as a function of the star formation rate SFR and the starburstiness of the system, defined as the ratio between the specific SFR (sSFR = SFR/M$_*$) of a galaxy and the sSFR of a MS galaxy with the same $z$ and M$_*$ ($\delta MS = sSFR/sSFR|_{MS}$; e.g. \citealt{Elbaz2011}).

This paper is organised as follows. In Sect. \ref{Smethods} we briefly summarise the PUMA sample selection, and present the data analysis of spectroscopic (MUSE) and photometric (HST) data. In Sects. \ref{Sdecomposition}-\ref{Smisalignments} we report the main results obtained from the spectroscopic analysis, in terms of incidence of disk-like motions in the gas and stellar components, and also compare the gas and stellar motions along their kinematic major axes. For those systems with disk-like gas motions, in Sect. \ref{S3DB} we present 3D-Barolo modelling and infer a kinematic classification in terms of the ratio between rotational velocity and intrinsic velocity dispersion. Sect.  \ref{Scorrelations} presents the study of correlations between intrinsic velocity dispersion and SFR and starburstiness in an extended sample of MS and SB galaxies in the redshift range $z \sim 0.03-2.6$. Finally, Sect. \ref{Sconclusions} summarises our conclusions.
Throughout this paper, we adopt the cosmological parameters $H_0 =$ 70 km/s/Mpc, $\Omega_m$ = 0.3, and $\Omega_\Lambda =$ 0.7. 

\section{Methods}\label{Smethods}

\subsection{Sample}

The PUMA sample is a volume-limited ($z<0.165$; d $<800$ Mpc) representative sample of 25 local ULIRGs. 
The sample selection is described in detail in \citet{Perna2021}. In brief, the targets were selected among the IRAS 1 Jy Survey (\citealt{Kim1998}), the IRAS Revised Bright Galaxy sample (\citealt{Sanders2003}), and the \citet{Duc1997} catalogue, isolating the sources visible by ALMA and MUSE and uniformly covering the ULIRG luminosity range. 
The sample was also selected to include an equal number of systems with AGN and SB nuclear activity (based on mid-IR spectroscopy) in the pre- and post-coalescence phases of major mergers (with projected nuclear distances lower than 10 kpc). 

So far, we have obtained MUSE observations for 85\% of the systems in the sample (21 systems with 31 individual nuclei; see Tables 1 and 2 in \citealt{Perna2021}), and  ALMA CO(2–1) and $\sim 220$ GHz continuum observations for the entire sample (22 systems, with 32 individual nuclei, have been already presented in \citealt{Pereira2021}). In this work, we focus on the kinematic properties of the ionised gas of the 20 ULIRGs (with 29 individual nuclei) observed with MUSE, therefore excluding the Arp 220 system whose properties have been extensively described in \citet[][but see also Appendix \ref{APV} for a brief description of its kinematics]{Perna2020}. Information about the MUSE data used in this work are collected in \citet{Perna2021}, Table 2. At the mean distance ($\sim$ 400 Mpc), the MUSE spaxel scale, resolution, and FoV correspond to $\sim 0.34$ kpc (0.2$^{\prime\prime}$), $\sim 1$ kpc ($0.6^{\prime\prime}$), and $\sim$ 100 $\times$ 100 kpc$^2$ ($60^{\prime\prime}\times60^{\prime\prime}$).

\subsection{Data analysis}\label{Sanalysis}
In this section we describe the MUSE spectroscopic and HST imaging data analysis we followed to characterise the kinematics and dynamics in our PUMA targets. 

\subsubsection{MUSE spectroscopic analysis}

\begin{figure*}[h]
%
\centering
\includegraphics[width=17.cm,trim= 0 0 0 0,clip]{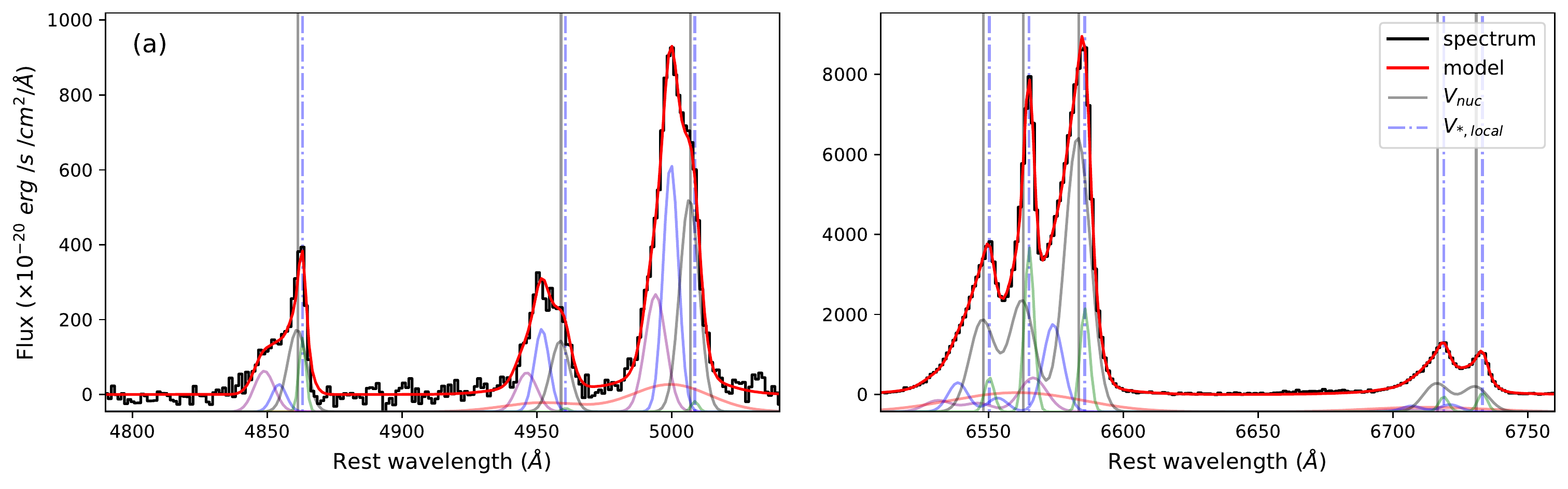}
\includegraphics[width=17.cm,trim= 0 0 0 0,clip]{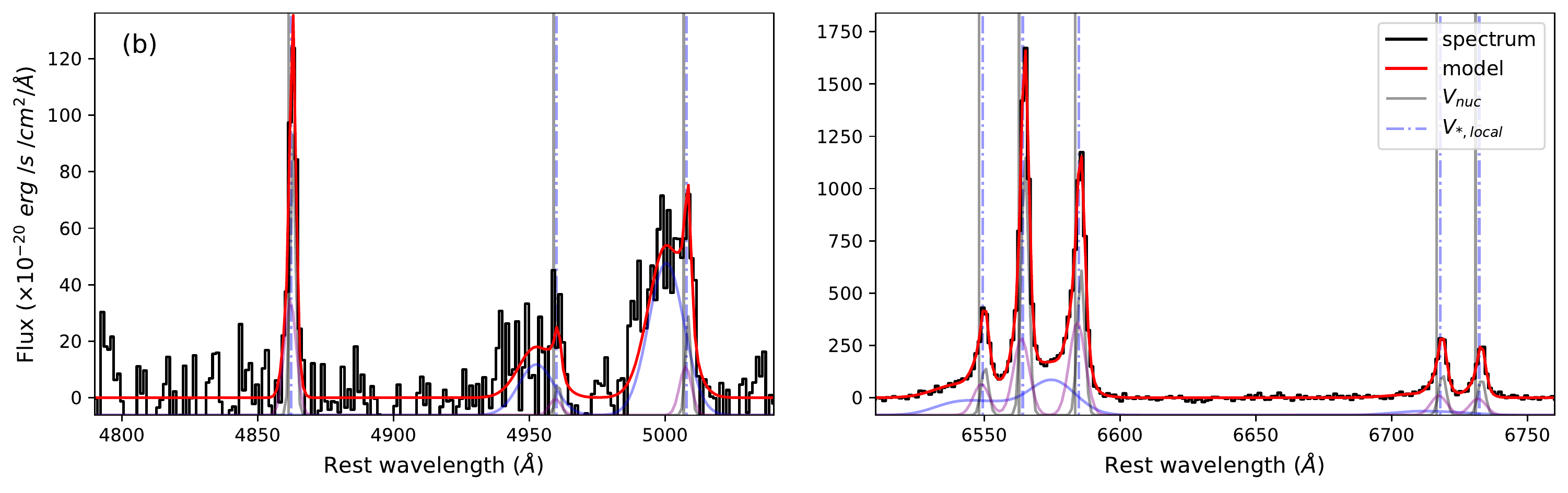}

\caption{\small Examples of our  multi-component Gaussian fit decomposition for two continuum-subtracted spectra of I13120, extracted from single spaxels at $\sim 1^{\prime\prime}$ (top panels) and $2^{\prime\prime}$ (bottom) south-east from the nucleus. The red curves show the best-fit solutions in the H$\beta$-[OIII] (left) and H$\alpha$-[NII] (right) regions; all emission lines are fitted simultaneously. The solid vertical lines mark the rest frame wavelength of the emission lines, derived from the stellar velocities of the nuclear spectrum; the dot-dashed blue vertical lines mark instead the local stellar systemic (i.e. at the position of the spaxel from which the spectrum is extracted). 
These examples demonstrate the diversity
of emission-line profiles observed in the FOV of a single target: the spectrum in panels $(a)$ is dominated by broad and blueshifted Gaussian components, while the one in panel $(b)$ is dominated by bright narrow Gaussians close to the systemic velocity (for all but [OIII] lines). 
}
\label{I13120spectra}
\end{figure*}

The MUSE data reduction and analysis was executed by following an approach similar to that described in \citet{Perna2021}. We briefly summarise it in the following. The data reduction and exposure combination were carried out by using the ESO pipeline (muse - 2.6.2). The  astrometric registration was performed using the Gaia DR2 catalogue (\citealt{Gaia2018}) for all but 5 systems, for which we used registered HST optical images as reference (because of the absence of Gaia stars in the MUSE field of view (FOV); see \citet{Perna2021}, section 3.1 for more details). 

We first fitted and subtracted the stellar continuum from each spaxel. To do so, we initially performed a Voronoi binning (\citealt{Cappellari2003}) on the cube to achieve a minimum signal-to-noise ratio (S/N $>16$) per bin on the continuum. We then fitted the stellar continuum in each bin through the pPXF code (Penalized PiXel-Fitting; \citealt{Cappellari2004,Cappellari2017}), using the Indo-U.S. Coud\'e Feed Spectral Library (\citealt{Valdes2004}) as stellar spectral templates to model the stellar continuum emission and absorption line systems. We then subtracted the stellar
continuum from the total spectra in each spaxel, scaling the fit from
bin to each spaxel according to the observed continuum flux (see \citet{Perna2021}, sect. 5.1).

\begin{table*}
\footnotesize
\begin{minipage}[!h]{1\linewidth}
\setlength{\tabcolsep}{6pt}
\centering
\caption{PUMA geometric properties and stellar and narrow \ha velocities along the kinematic major axis}
\begin{tabular}{lcccccccccc}
IRAS ID (other)   &  $2R_e$ & $i^{morph}$ & PA$^{morph}$ & PA$_{*}^{kin}$ & PA$_{gas}^{kin}$ &   $\delta v_{*}$ & $\delta v_{gas}$ & $\bar \sigma_{*}$ & $\bar \sigma_{gas}$  & disk-like \\
 & (kpc) & (deg) &(deg) &(deg) &(deg) & (\kms) & (\kms)& (\kms) & (\kms) & kin\\ 
\scriptsize{(1)} & \scriptsize{(2)}   &\scriptsize{(3)}   & \scriptsize{(4)} & \scriptsize{(5)} & \scriptsize{(6)} & \scriptsize{(7)} & \scriptsize{(8)}& \scriptsize{(9)} & \scriptsize{(10)}& \scriptsize{(11)}\\
\hline

 F00188$-$0856  &$1.5\pm 0.1$ & $26\pm 12^s$ & $5\pm 1^u$ & $93\pm 9$ & $81\pm 16$  & $145\pm 7$ & $77\pm 2$ & $23_{-15}^{+8}$& $107_{-5}^{+2}$ & s \\
\hline

F00509+1225 (IZw1) & $3.68\pm 0.05$ &  $36\pm 14^s$ & $23\pm  2^u$ & $130\pm 3$ & $130\pm 3$ &  $270\pm 2$ & $370\pm 10$ & $3\pm 2$ & $57_{-18}^{+11}$&  s, g  \\
\hline
F01572+0009  & $3.81\pm 0.05$ & $37\pm 14^s$ & $33\pm 2^u$ & $68\pm 3$ & $-$ &  $178\pm 17$ & $300 \pm 10$ & $100_{-12}^{+2}$&$140_{-17}^{+11}$ & s\\
\hline

F05189$-$2524 & $0.5\pm 0.1$ & $34\pm 14^s$ & $39\pm 4^u$ & $68\pm 3$ & $-$ & $125\pm 3$ & $125\pm 6$ & $10\pm 2$& $100\pm 2$& s\\
\hline

F07251$-$0248 E &$-$&$-$ &$-$ & $-$ & $-19\pm 3$  &  $95\pm 9$ & $170\pm 10$ & $95_{-7}^{+4}$ & $114\pm 1$ &  $-$  \\

F07251$-$0248 W &$-$&$-$ &$-$ & $112\pm 3$ & $106\pm 3$ & $197\pm 7$ & $325\pm 10$ &$69_{-4}^{+2}$& $103_{-6}^{+3}$&  s?, g?\\
\hline

F09022$-$3615& $-$&$-$ &$-$ & $25\pm 3$ & $-$ &  $290\pm 4$ & $190\pm 6$ &$120_{-5}^{+2}$&$125_{-10}^{+6}$ &  $-$ \\
\hline

F10190+1322 E & $-$ & $-$ &$-$ & $68\pm 3$ & $93\pm 3$ &  $278\pm 28$ & $265\pm 8$ &$85\pm 2$&$118_{-8}^{+2}$ &  s, g \\

F10190+1322 W & $5.2\pm 0.3^{**}$ & $46\pm 15^s$ & $131\pm 5^s$ & $118\pm 3$ & $106\pm 3$ &  $244\pm 3$ & $310\pm 2$ &$7_{-5}^{+2}$ &$63\pm 2$ & s, g\\
\hline 

F11095$-$0238 NE &$-$ &$-$ & $-$ & $-$ & $-$ & $-$ & $-$ &$-$ &$-$ & $-$   \\

F11095$-$0238 SW &$-$ &$-$ & $-$ & $-$ & $-$ & $-$ & $-$ & $-$& $-$& $-$   \\
\hline 

F12072$-$0444 N &$-$ &$-$ &$-$ & $99\pm 12$ & $74\pm 6$ & $188\pm 2$ & $130\pm 2$ & $17\pm 14$& $107_{-14}^{+4}$& s, g \\

F12072$-$0444 S &$-$ &$-$ & $-$ & $-$ & $-$ & $-$ & $-$ &$-$ & $-$& $-$  \\
\hline 

13120$-$5453 & $1.1\pm 0.1$ & $25\pm 11^s$ & $102\pm 2^u$ & $99\pm 3$ & $93\pm 3$ &  $214\pm 3$ & $328\pm 2$ & $98_{-18}^{+6}$ & $78_{-5}^{+1}$ & s, g \\
\hline 

F13451+1232 E&  $-$ & $-$ & $-$ & $-$  & $-$ & $-$ & $-$ & $-$& $-$ & $-$ \\

F13451+1232 W &  $-$& $-$ & $-$ & $-$ & $-$ & $-$ & $-$ & $-$& $-$& $-$  \\
\hline 

F14348$-$1447 NE&  $-$ & $-$ & $-$ & $19\pm 3$ & $174\pm 3$ &  $154\pm 8$ & $180\pm 7$ & $90\pm 2$& $90\pm 2$ & s?, g? \\

F14348$-$1447 SW &  $-$& $-$ & $-$ & $174\pm 3$ & $-$ & $181\pm 4$ & $47\pm 5$ & $84\pm 2$ & $101\pm 2$ & s  \\
\hline 

F14378$-$3651& $1.4\pm 0.1$& $25\pm 11^s$ & $29\pm 3^u$ & $12\pm 3$ & $25\pm 9$ &  $129\pm 5$ & $84\pm 3$ &$49\pm 14$ & $85_{-6}^{+1}$& s\\
\hline 

F16090$-$0139& $9.4\pm 0.5^{**}$ &$38\pm 15^s$ &$11\pm 2^s$ & $145\pm 3$ & $168\pm 3$ &  $145\pm 16$ & $182\pm 3$ &$150_{-15}^{+7}$ &$120_{-7}^{+3}$ & s? \\
\hline

F17208$-$0014&$2.4\pm 0.1^{**}$ &  $40\pm 15^s$ & $173\pm 3^s$ & $155\pm 4$ & $155\pm 3$ & $206\pm 10$ & $210\pm 4$ & $177\pm 3$&$126\pm 2$ & s?, g?\\
\hline

F19297$-$0406 S& $-$ &$-$ &$-$ & $59\pm 3$ & $-$ & $217\pm 12$ & $270\pm 5$ &$134_{-15}^{+4}$ & $126\pm 7$& $-$ \\

F19297$-$0406 N&$-$&$-$ &$-$ & $62\pm 3$ & $-$ &  $160\pm 3$ & $160\pm 13$ &$141\pm 4$&$130_{-9}^{+5}$& $-$ \\
\hline

19542+1110 & $1.1\pm 0.1$ & $16\pm 9^s$ & $58\pm 3^s$ & $56\pm 3$ & $-$ &  $182 \pm 2$ & $97\pm 7$ &  $5\pm 2$&$122_{-9}^{+2}$ & s\\
\hline

20087$-$0308  & $2.8\pm 0.1^{**}$  &  $36\pm 14^s$ & $82\pm 2^s$ & $87\pm 6$ & $50\pm 3$ & $137\pm 8$ & $353\pm 6$ &$146_{-22}^{+7}$ &$116\pm 4$ & s?, g?  \\
\hline 

20100$-$4156  NW  &$-$ &$-$ &$-$ &  $-$  &$-$ &$-$  &$-$  &$-$ & $-$& $-$ \\

20100$-$4156  SE &$3.0\pm 0.1^{**}$  & $56\pm 16^s$ & $74\pm 1^s$ & $74\pm 6$ & $19\pm 3$ &  $154\pm 7$ & $100\pm 6$ & $12\pm 1$ & $104\pm 2$ & s, g\\
\hline

F22491$-$1808 E &$-$&$-$ &$-$ & $-$ & $155\pm 6$ & $56\pm 2$ & $77\pm 5$ & $81\pm 5$& $105\pm 2$& $-$ \\

F22491$-$1808 W&$-$ &$-$ &$-$ & $-$  &$-$ &$-$  &$-$  & $-$ & $-$ & $-$  \\

\hline
\hline
\end{tabular}
\label{Tproperties}
\vspace{0.2cm}
\end{minipage}
{\small
{\bf Notes.}\\ Column (1): target name. Column (2): $2R_e$ from \Isophote \ fits of available HST/F160W images for all but the Seyfert 1 systems IZw1 and I01572 (for which we used the \citealt{Veilleux2006} measurements) and the two systems without HST data: I16090 and I10190. The geometric parameters of the latter two sources are derived from MUSE narrow-band images (at $\lambda \sim 7500\AA$).  Columns (3) and (4): inclination and PA of the galaxy. The PA is taken anticlockwise from the North direction on the sky. These parameters are derived with \Isophote, at the distance reported in Column (2). Columns (5) and (6): stellar and gas PA$^{kin}$, taken anticlockwise from the North direction on the sky. Column (7) and (8): maximum stellar and gas velocity variations along PA$^{kin}$, non-corrected for galaxy inclination. The velocity amplitudes are computed along an intermediate PA$^{kin}$ when PA$_{*}^{kin}$ and PA$_{gas}^{kin}$ are  consistent within the errors ($3\sigma$); alternatively, stellar (gas) velocity amplitudes are computed along PA$_{*}^{kin}$ (PA$_{gas}^{kin}$); for those sources with a missing PA$^{kin}$ measurement, the gas and stellar $\delta v$ are computed along the only available PA$^{kin}$ measurement. Velocity uncertainties are derived with a bootstrap.
Columns (9) and (10): median velocity dispersion along PA$^{kin}$. All velocity and velocity dispersion values for the gas component are derived from the narrow \ha velocity maps. Column (11): kinematic classification according to the two criteria defined in Sect. \ref{Skinaxis}: `s' for stellar disk-like kinematics; `g' for gas disk-like kinematics.\\
($^{**}$): The effective radius measurement is highly uncertain, due to prominent tidal tails (and, for I10190 W, the nearby E nucleus). \\
($^s$): in Columns 3 and 4, the label identifies those sources which geometric parameters do not vary significantly at $r > R_e$, i.e. which inclination and PA are relatively stable.\\
($^u$): in Columns 3 and 4, the label identifies those sources which inclination and PA vary significantly at $r > R_e$. These sources generally have prominent tails which strongly affect the {\it isophote} fit.  

}
\end{table*}

A slightly different approach was instead used for the two Seyfert 1 in our sample, IZw1 and I01572: in addition to the stellar spectral templates, we made use of an AGN template constructed on the basis of the observed nuclear spectrum, modelled with a combination of a power-law continuum, forbidden, and permitted emission lines as described in \citet{Perna2021}, Appendix B. Because of the point-spread-function blending, this AGN component accounts for a significant fraction of the total emission in the innermost nuclear regions, and rapidly reduces going to radii $r \gtrsim 1''$. This step allows us to better reconstruct the stellar velocity field in the nuclear regions with respect to our previous analysis results (see Fig. 5 in \citet{Perna2021}, and Fig. \ref{BBaroloizw1}, top-right in this work).

Before proceeding with the fit of the emission lines, we derived a second Voronoi tessellation to achieve a minimum S/N = 8 of the \ha line for each bin. This feature has been preferred to the [OIII]$\lambda 5007$ line, generally used to trace ionised outflows, as the latter is highly absorbed in ULIRG systems due to their large dust content. The use of \ha as a reference for the tessellation allows us to better preserve the important spatial information (both for kinematics and emission line structures).

At this point we fitted the most prominent gas emission lines from the continuum-subtracted cube, by using the Levenberg–Markwardt least-squares fitting code CAP-MPFIT (\citealt{Cappellari2017}). In particular, we  modelled the H$\beta$ and H$\alpha$ lines, the [O {\small III}]$\lambda\lambda$4959,5007, [N {\small II}]$\lambda\lambda$6548,83, [S {\small II}]$\lambda\lambda$6716,31, and [O {\small I}]$\lambda\lambda$6300,64 doublets with a simultaneous fitting procedure. 
To account for broad and asymmetric line profiles, already observed in the nuclear regions of almost all PUMA targets (\citealt{Perna2021}), we performed each spectral fit five times at maximum, with one to five kinematic components (i.e. Gaussian sets, each centred at a given velocity and with a given FWHM). The final number of kinematic components used to model the spectra was derived on the basis of the Bayesian information criterion (BIC, \citealt{Schwarz1978}). A detailed description of the Gaussian fit routine can be found in \citet{Perna2021} and \citet{Perna2020}.

In Fig. \ref{I13120spectra} we show two examples of our continuum-subtracted, high S/N spectra, extracted from two different regions of the target 13120$-$5453 (I13120 hereinafter). The best-fit models show the presence of broad and asymmetric line profiles in the two spectra, and a significant diversity in the relative contribution of narrow and broad components: the emission lines in the spectrum in the $(a)$ panels are dominated by the contribution of extremely broad Gaussian components, while the ones in the $(b)$ panels have a well defined narrow core, especially in the Balmer transitions, associated with less perturbed gas. 
These two examples show that a multi-component simultaneous fit of all prominent optical emission lines is required to properly separate ordered and perturbed motions in our PUMA systems.

\subsubsection{Complementary photometric analysis}\label{Sphot}

We made use of the Photutils (\citealt{Bradley2016}) \Isophote \ package of Astropy (\citealt{Astropy2018}) to perform a basic photometric analysis of ancillary HST near-infrared images, available for most of our targets. This analysis provides important morphological parameters to be compared with those inferred from the kinematic analysis described in the next sections.

We fitted a series of isophotal ellipses to each galaxy: \Isophote \ was instructed to hold the centre position constant, whereas the ellipticity ($\epsilon$) and position angle (PA) of the ellipses interpolating the galaxy isophotes were allowed to vary (e.g. \citealt{Costantin2017,Costantin2018}). \Isophote \ provides the azimuthally averaged surface brightness profile as well as the variation of $\epsilon$ and PA as a function of the semi-major axis length. In Table \ref{Tproperties} we report the inferred inclinations, derived from $\epsilon$ following \citet{Willick1997}, and PA at the galactocentric distance of two times the effective radius $R_e$, defined as the radius which contains half of the galaxy light (and computed adopting the curve-of-growth method; see e.g. \citealt{Crespo2021}). 
Morphological parameters are reported for a small fraction of sources (12/29 individual galaxies), as the intrinsic values of PUMA galaxies are usually distorted by merger interactions, and the presence of companion systems. These distortions are also sometimes responsible of significant variations in the morphological parameters at large radii ($r> R_e$); we therefore report in the table whether the morphological inclination and PA do show significant variations at large distances.

\begin{figure*}[!h]
%
\centering
\includegraphics[width=13.cm,trim= 30 60 30 50,clip]{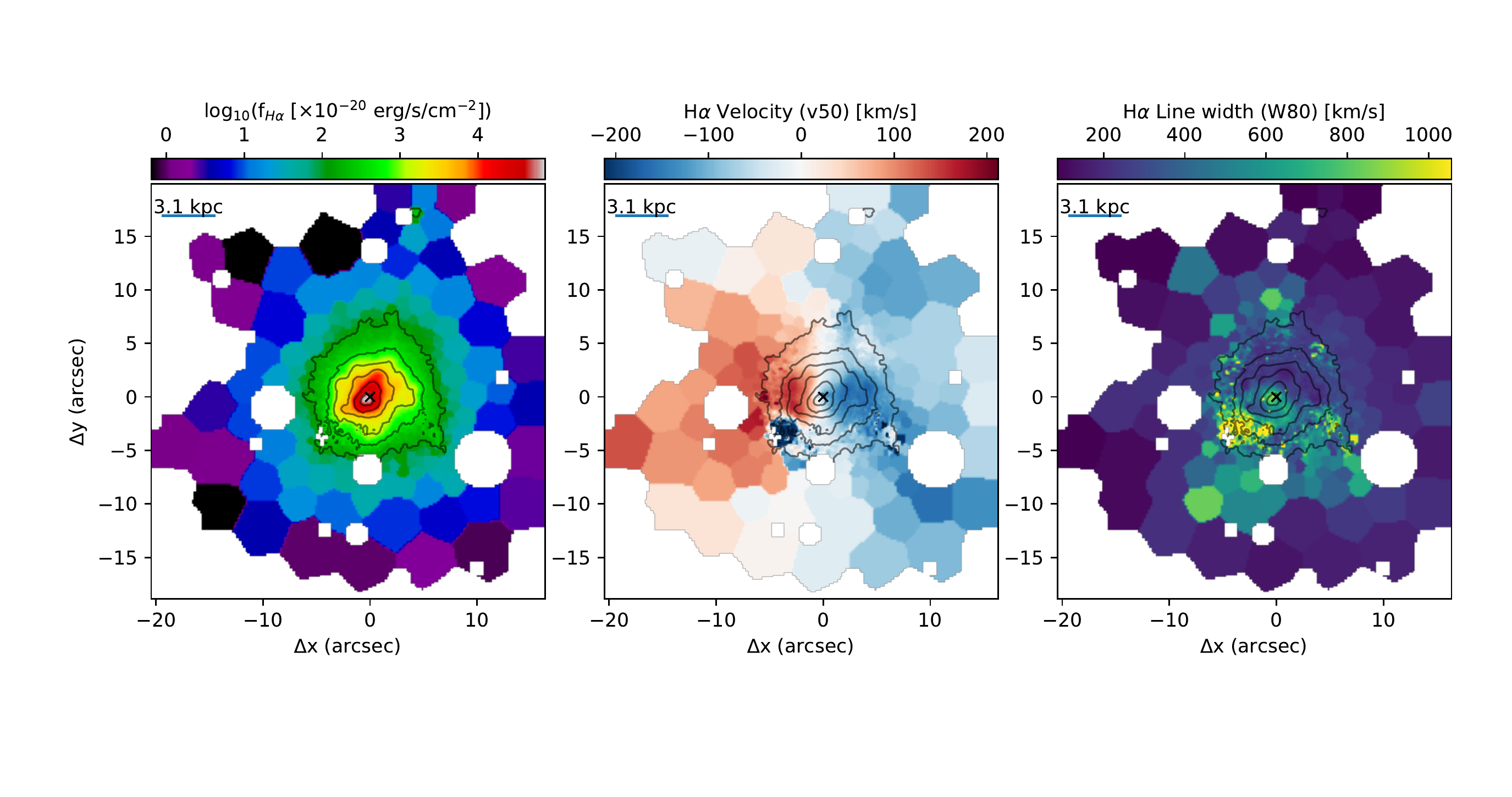}
\includegraphics[width=13.cm,trim= 30 40 30 50,clip]{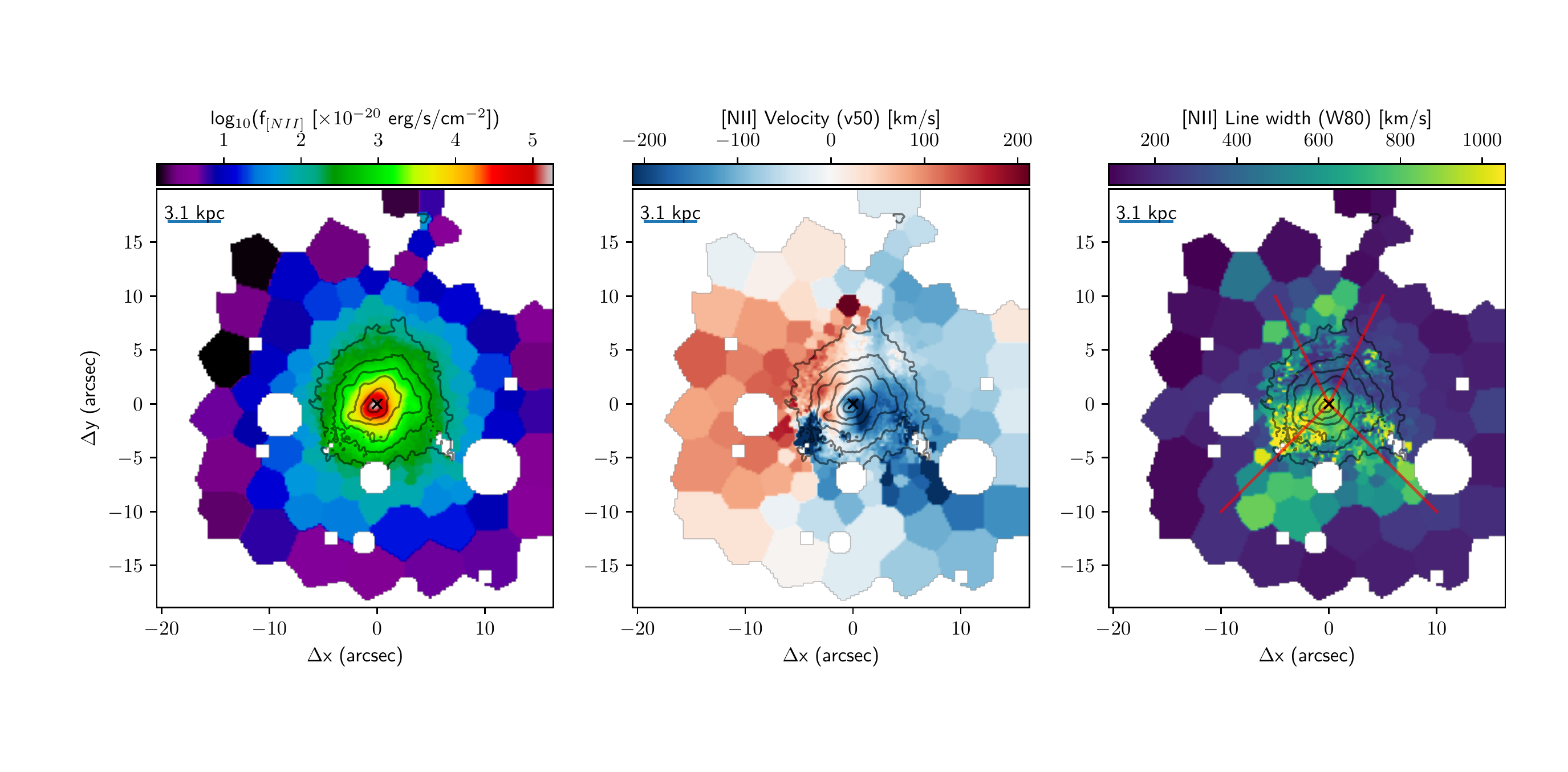}

\caption{\small I13120 maps. 
Top: total \ha integrated flux  (left), \ha centroid ($v50$, centre) and line-width ($W80$, right) obtained from the multi-component Gaussian fit. Bottom: Similar panels for [NII];  in the $W80$ map, the red lines indicate the wide biconical outflow along the north-south direction.
Masked regions mark the spaxels contaminated by the presence of background and foreground sources, and excluded as disturbing the data analysis. The first solid contour is 3$\sigma$ and the jump is 0.5 dex. The cross marks the nucleus. North is up and West is right.
}
\label{I13120linemaps}
\end{figure*}

\begin{figure*}[!h]
%
\centering

\includegraphics[width=5.cm,trim= 0 10 0 15,clip]{{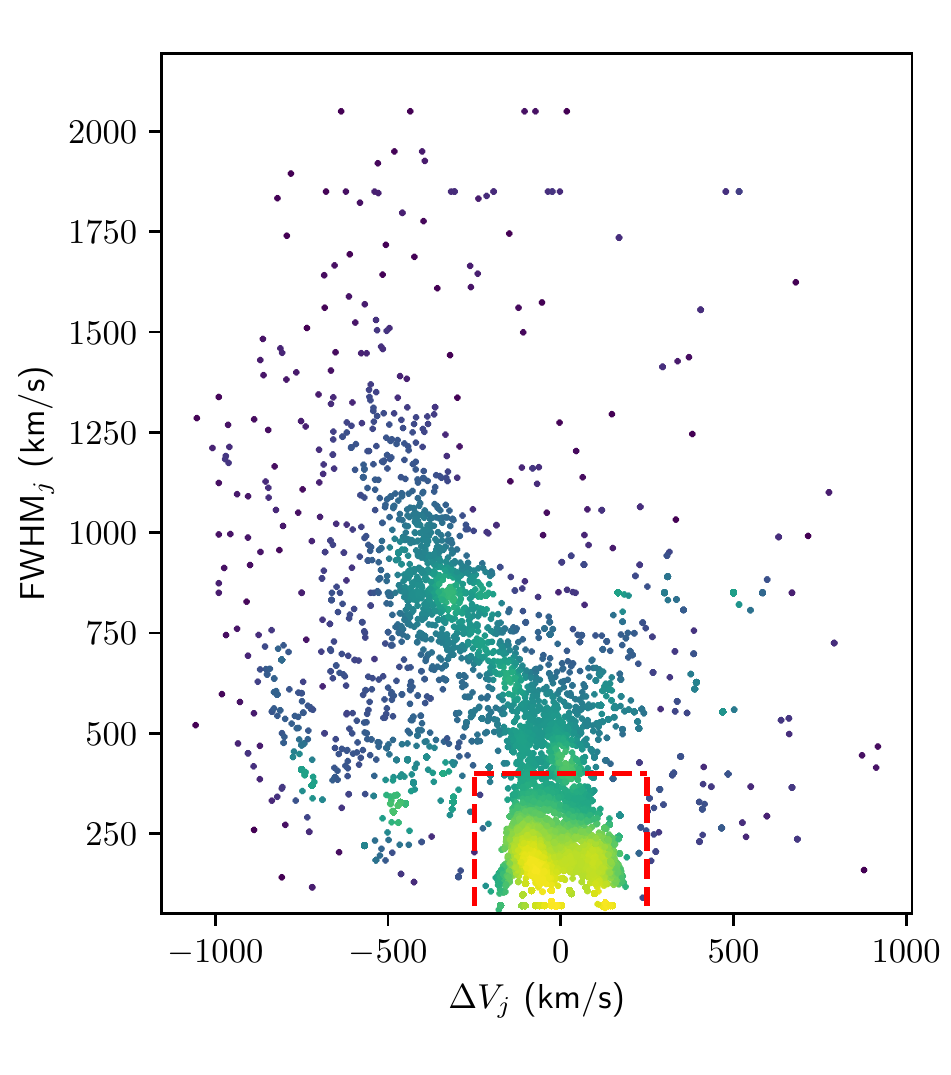}}
\includegraphics[width=13.cm,trim= 30 50 30 35,clip]{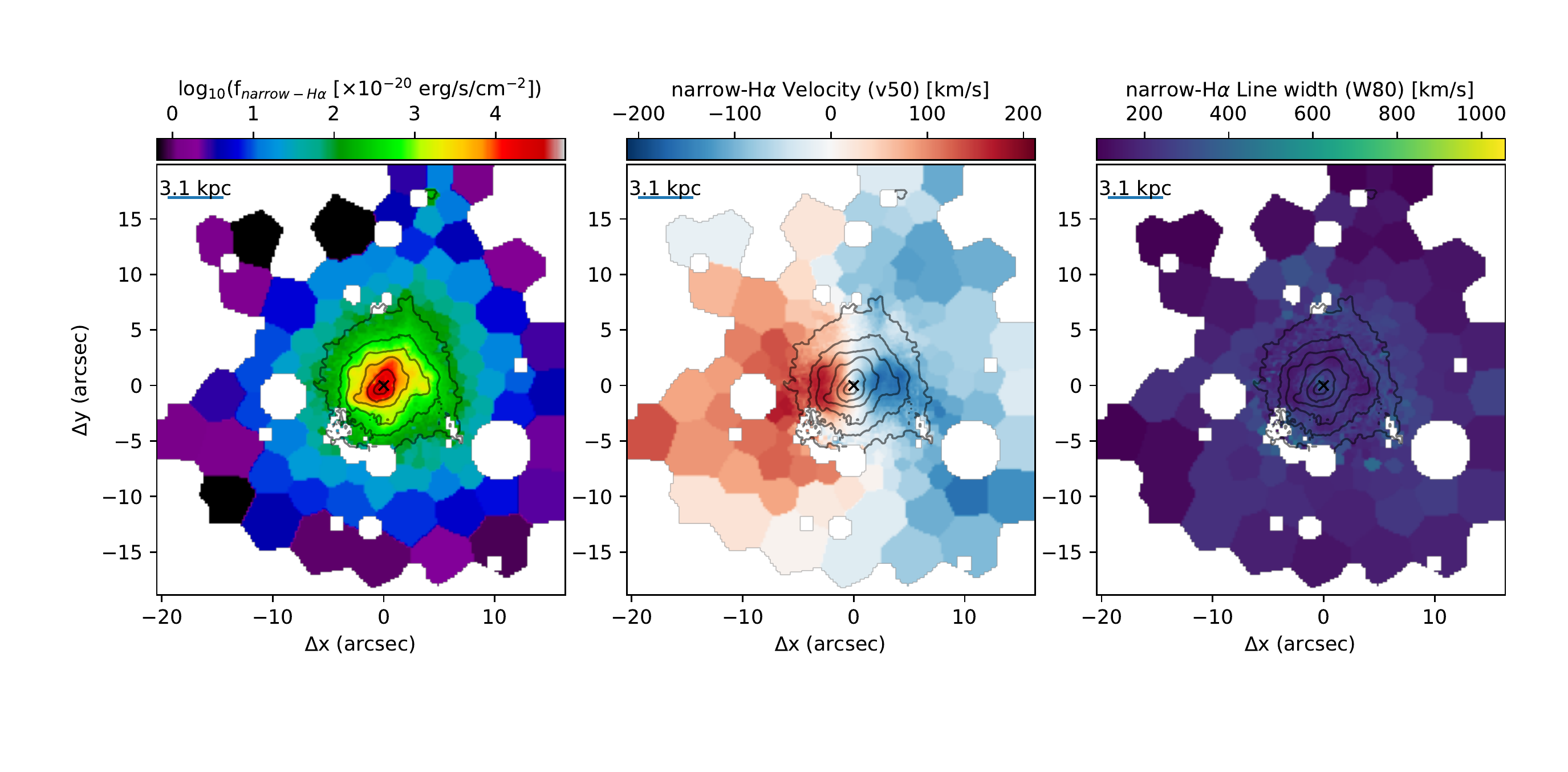}

\caption{\small {\it Left panel}: I13120 velocity shift $\Delta V_j$-FWHM$_j$ diagram (colored by density in log-scale) for the individual Gaussian components used to model the emission line profiles in the data cube. The red dashed lines isolate the Gaussian components used to reconstruct the narrow \ha data cube, with $|\Delta V_j| < 250$ \kms and FWHM$_j< 400$ \kms. {\it Right panels}: the narrow \ha flux distribution,  velocity ($v50$) and line width ($W80$) maps are reported in the 2nd, 3rd and 4th panels respectively (see Fig. \ref{I13120linemaps} for details).
}
\label{I13120narrowlinemaps}
\end{figure*}

Our rough estimates for $R_e$ are in general agreement with those obtained in previous works, by a factor of 2 (\citealt{Veilleux2006, Haan2011, Kim2013}), for all but the two Seyfert 1 in our sample, IZw1 and I01572. For these two sources we therefore considered the distance obtained by \citet{Veilleux2006}, who performed a more rigorous multi-component two-dimensional image decomposition to separate the host galaxy from its bright active nucleus. 

In Sect. \ref{Smisalignments} we will compare the  morphological PAs we derived with {\it Isophote} with the kinematic ones derived from MUSE spectroscopic analysis. The inclination measurements will be instead used to model the gas kinematics with 3D-Barolo (\citealt{Diteodoro2015}) in Sect. \ref{S3DB}.

\subsection{Emission line tracers and velocity parameters}

Throughout this paper, we differentiate between the individual kinematic component parameters, FWHM$_j$ and $\Delta V_j$, with $j$ from 1 to 5 at maximum, and the non-parametric velocities $v10$, $v50$, $v90$ and $W80$ (e.g. \citealt{Liu2013}). The former identify the width of a specific kinematic component, $j$, and its velocity shift with respect to the systemic, defined as the stellar velocity in the nuclear position (see Sect. 5.2 in \citealt{Perna2021});  FWHM$_j$ and $\Delta V_j$ are common to all emission lines fitted simultaneously. Instead, the non-parametric velocities are defined as follows: 
$v10$, $v50$ and $v90$ are the 10th, 50th, and 90th-percentile velocities, respectively, calculated on the (multi-component) fitted line profile with respect to the systemic. Therefore, they correspond to the velocities at which 10, 50 and 90\% of the line flux is accumulated. $W80$ is defined as $v90 - v10$.

Observations of relatively large samples of AGN and star-forming galaxies (SFGs) have indicated that the \ha emission is not necessarily dominated by outflows as is the case for [OIII] emission lines (e.g. \citealt{Bae2014, Cicone2016}), as the Balmer line has a significant contribution from star-forming regions. Indeed, the spectral analysis of the line emission coming from the nuclear regions of our PUMA systems already revealed that  $v10$ and $W80$ of \ha are, on average, 20\% smaller than those of the [OIII] (\citealt{Perna2021}). Similarly, [NII] lines have a larger velocity-width than H$\alpha$, consistent with those of the [OIII] ( $\langle W80$ [O {\small III}] / $W80$ [NII] $\rangle \sim  \langle v10$ [O {\small III}] / $v10$ [NII] $\rangle= 1.05$). This empirical evidence can be explained taking into account the fact that [NII] is brighter than \ha both in AGN ionisation cones, often affected by gas flows (e.g. \citealt{Fischer2013}), and in shocks (e.g. \citealt{Allen2008}). Therefore, the nitrogen line may be better analogous to the [O III] emission, as preferentially traces outflows and more unsettled material compared to H$\alpha$ (see also e.g. \citealt{Harrison2016}).  Because of the high extinction in PUMA systems, [NII] has to be preferred to the (fainter) [OIII] as outflow tracer (see also \citealt{Perna2020}).

For each PUMA target, we produced emission line maps for the flux, $v50$, and $W80$ non-parametric velocities, obtained considering the total modelled line profiles made up of the sum of the fitted Gaussians. The \ha and [NII] maps of I13120 are presented in Fig. \ref{I13120linemaps}.  These maps mark the dissimilarities mentioned above: the \ha velocity field appears more regular than the [NII] one; indeed the \ha map shows slightly smaller velocity-widths.  
The \ha and, in particular, the [NII] velocity maps show a biconical outflow structure, the approaching part to the south and the receding part to the north. Precisely, the $v50$ map shows high-$v$ [NII] gas blueshifted to the south, within a conical region with a large opening angle, and high-$v$ redshifted [NII] extending to the north. The outflow biconical structure is also associated with high $W80$ (up to $\sim 1000$ \kms). 
We therefore conclude that, with respect to [NII] lines, \ha is less affected by outflows and highly perturbed kinematics. 

In this paper we focus on the disentangling of ionised gas rotation dynamics in the PUMA sample; we therefore present all \ha maps in Appendix \ref{AHa}, and leave the [NII] maps to a following investigation.

\section{Results and discussion}\label{Sresults}

\subsection{Ionised gas kinematic decomposition}\label{Sdecomposition}

The velocity field of I13120 (Fig. \ref{I13120linemaps}) shows a regular velocity gradient along the east-west direction, in addition to the typical features observed in biconical outflows, like blue- and red-shifted emitting gas with increased line widths in regions preferentially located along the perpendicular direction. In order to understand if this system is rotationally supported, we take advantage of the kinematic decomposition described in Sect. \ref{Sanalysis}. 
Figure \ref{I13120narrowlinemaps}, left, shows the distribution of $\Delta V_j$ and FWHM$_j$ for each Gaussian component $j$ used to model the emission line profiles in I13120. 
The figure shows a clear trend: highest FWHM$_j$ ($\gtrsim 700$ \kms) are associated with  significant blueshifts ($\Delta V_j \lesssim -400$ \kms), while Gaussian components with smaller FWHM$_j$ have $|\Delta V_j| \lesssim 200$ \kms. The highest velocity shifts and line widths are associated with the outflow (see also e.g. \citealt{Woo2016} for similar diagrams obtained from SDSS integrated spectra of nearby AGN); instead, the smallest velocities are associated with less perturbed kinematics. To better investigate the presence of rotationally supported motions, we select in the $\Delta V_j - $ FWHM$_j$ plane the components with $|\Delta V_j| < 250$ \kms and FWHM$_j< 400$ \kms (see e.g. \citealt{Mingozzi2019} for a similar approach), and construct a new data cube for the H$\alpha$ emission, labelled narrow \ha data cube. The line width threshold FWHM$_j = 400$ \kms was chosen taking into account the fact that in our targets the stellar component, more sensitive to gravitational motions, has a velocity dispersion significantly smaller (with $\sigma_*$ up to 200 \kms only in the innermost nuclear regions, due to beam smearing effects). The flux distribution as well as the velocity field and velocity dispersion of narrow \ha are reported in Fig. \ref{I13120narrowlinemaps}.
As expected, the narrow \ha maps display more regular velocity patterns (and significantly lower line widths) with respect to those obtained from the total \ha (and [NII]) lines in Fig. \ref{I13120linemaps}. 

In the next section, we investigate the presence of rotation in our PUMA systems, taking advantage of this kinematic decomposition between more and less perturbed Gaussian components in the $\Delta V_j - $FWHM$_j$ plane, and extracting for each target the position-velocity (PV) diagrams along the kinematic major axis of narrow \ha data cubes.

\begin{figure*}[ht]
\centering
\includegraphics[width=18.cm,trim= 30 520 0 390,clip]{{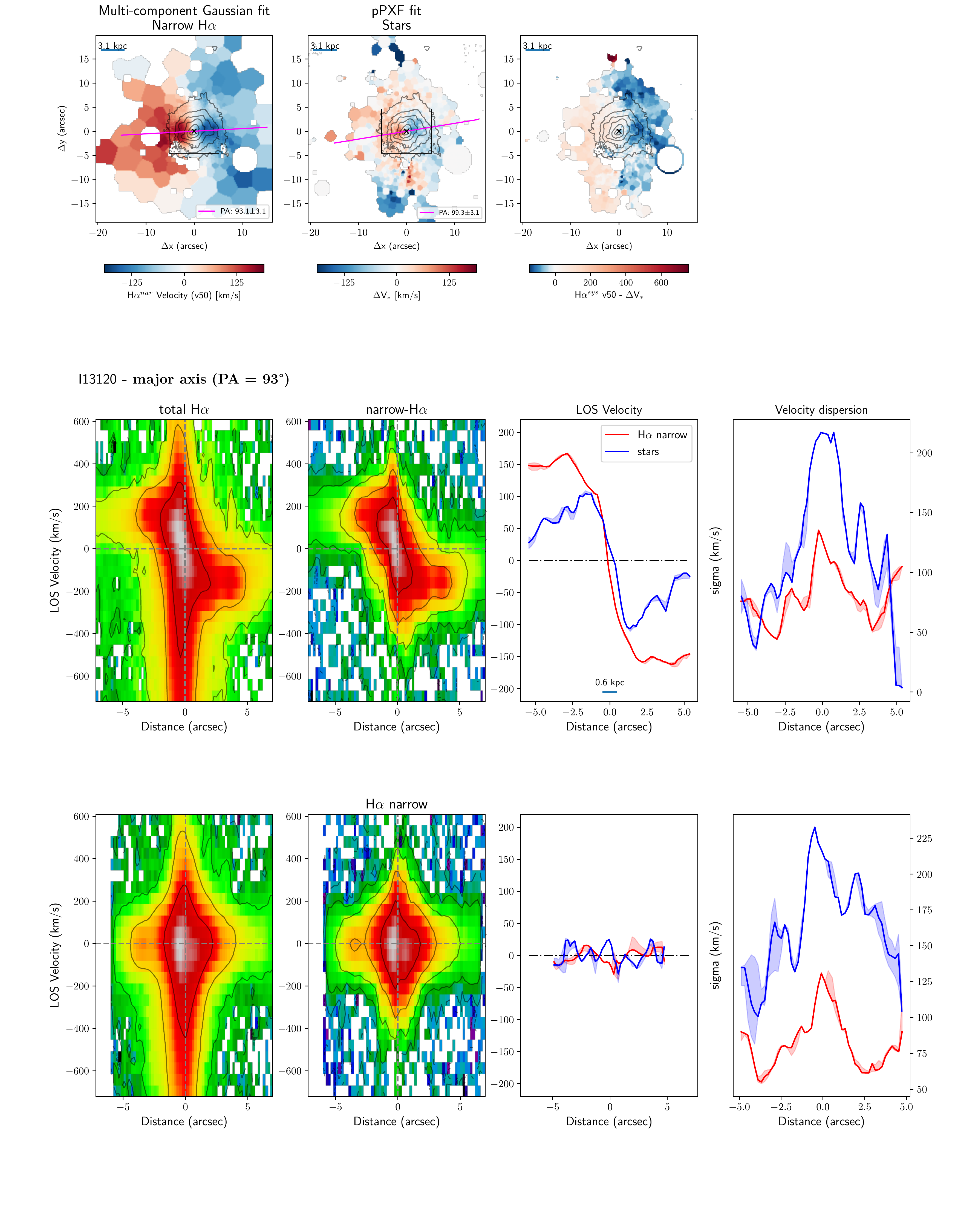}}

\caption{\small 
I13120 position-velocity diagrams along the galaxy major axis. {\it From left to right:} PV maps of the total \ha and the narrow \ha  emission; the last two panels show the extracted line velocity centroids and line width of the narrow H$\alpha$, as well as the stellar velocities along the same axis. $1''$ corresponds to $\sim 0.6$ kpc at the distance of I13120, as labeled in the third panel. 
}
\label{PV_I13120}
\end{figure*}

\subsection{Kinematics along the major axis}\label{Skinaxis}

Figure \ref{PV_I13120} shows the PV plots along the kinematic major axis position angle (PA$^{kin}$) of I13120, for both the total \ha emission (first panel) and the narrow H$\alpha$ (second panel). A clear velocity gradient from $\sim 200$ \kms\ to $\sim - 200$ \kms\ is observed in both panels; therefore, the exclusion of very-high velocity components does not introduce or alter significantly this gradient. In the same figure, we report the extracted line velocity centroids and velocity dispersion of the narrow H$\alpha$, as well as the stellar velocities (obtained from pPXF analysis, see \citealt{Perna2021}) along the stellar component major axis. At this stage, no correction for the beam smearing is performed in the reported velocity dispersion. Both the narrow \ha and stellar components exhibit  i) a well defined velocity gradient along their major axes, and ii) a peak in the velocity dispersion diagram at the position of the nucleus. These two conditions provide initial evidence for a rotation-dominated system (e.g. \citealt{Schreiber2018}).

In Fig. \ref{PV_all} we show the comparison between the narrow \ha and stellar velocities along their kinematic major axis, for the 18 targets for which we can observe a clear velocity gradient (and measure a PA$^{kin}$)  together with a peak in the velocity dispersion diagram at the position of the nucleus for at least one component (i.e. gas or stars). The PA$^{kin}$ measurements, obtained with the python PaFit package (\citealt{Krajnovic2006}), the velocity amplitudes and median velocity dispersion measured along PA$^{kin}$ are reported in Table \ref{Tproperties} (columns 5 to 10), for both stellar and gas components.

The simple visual comparison between gas and stellar kinematics along PA$^{kin}$ allows us to isolate 5 nuclei reasonably associated with more regular, disk-like kinematics for both gas and stellar components, according to the two criteria highlighted before: IZw1, I10190 W\footnote{We exclude I10190 E, as its gas kinematics in the receding part are dominated by those of the W nucleus.}, I12072 N, I13120, I20100 S. 
The relatively small number of systems with such characteristics is due to the fact that PUMA consists of advanced interacting ULIRGs systems\footnote{With the exception of IZw1, a minor merger system with log $L_{IR}/ L_\odot = 11.3$ but log $L_{bol}/ L_\odot > 12 $.} with nuclear projected separations smaller than 10 kpc (i.e. systems classified as IIIb, IV, and V in the \citealt{Veilleux2002} scheme).   
The small number of sources (5) with stellar and gas disk-like kinematics does not allow us to infer any specific conclusion about the conditions possibly related to more regular motions, also because of the different intrinsic properties of these  systems: IZw1 has a small companion at $\sim 18$ kpc; I10190 and I20100 have two nuclei separated by $\gtrsim 5$ kpc, and prominent tidal features; I12072 has two nuclei at a projected distance of $\sim 2.3$ kpc; I13120 has a single nucleus and extended tails and loops surrounding the main
body of the galaxy (up to $\gtrsim 20$ kpc from the nucleus; see e.g. Fig. 1 in \citealt{Privon2016}). 

The rest of the sample displays a variety of kinematics. In the last column of Table \ref{Tproperties} we distinguish among systems with evidence for disk-like kinematics in gas and stellar components.   
For instance, I07251 W, I14348 NE and I17208 do not have well defined kinematic properties: 
they might present disk-like motions, but with kinematic centres possibly not coincident with the nuclear position (see Fig. \ref{PV_all}).  We note however that these small offsets might also be due to different amount of dust or the presence of tidal streams along the major axis PA (see e.g. \ha flux distribution in Fig. \ref{I17208linemaps}). A better investigation of these offsets is reported in Appendix \ref{A3DBarolo}. 
I20087 has peculiar outflow features, reasonably responsible of the observed velocity gradient (with a maximum variation $\delta v_{gas} \sim 353$ km/s, a factor 2.6 higher than $\delta v_*$). 
Finally, there are 7 systems with regular stellar kinematics but highly perturbed gas motions, with $\sigma_{gas}$ generally higher than 100 km/s  across the major axis, and without clear trends in the LOS velocity: I00188, I01572, I05189, I14348 SW, I14378 and I19542 (with the possible addition of I16090, affected by poor data quality). A more detailed description of individual targets is reported in Appendix \ref{APV}. 

Summarising, we found five systems with regular, disk-like kinematics traced by the narrow H$\alpha$ and stars on scales of $\gtrsim 6$ kpc (in diameter),  IZw1, I10190 W, I12072 N,  I13120 and I20100 S, with the possible inclusion of three additional targets (I14348 NE, I17208 and I07251). The remaining targets (21/29 individual nuclei) show more complex gas kinematics, dominated by tidal streams (e.g. I09022 in Fig. \ref{I090221linemaps}), loops (e.g. I05189 in Fig. \ref{I0589linemaps}) and outflows (e.g. I13451 in Fig. \ref{I13451linemaps}) that prevent a clear identification of possible features due to more regular, disk-like motions. On the other hand, the incidence of stellar disk-like  motions is slightly higher than rotating gas, with 13 (17, including the more uncertain systems in Table \ref{Tproperties}) systems out of 29. This higher incidence is probably due to the fact that the stellar component is less affected by non-gravitational perturbations (shocks, outflows). 
Therefore, the incidence of gas disk-like  kinematics in our PUMA sample, of $27\%$ (8/29), has to be considered as lower limit. In support of this perspective, we note that in a merger, the gas has shorter dissipative timescales than stars, thus it should settle back on a rotating disk earlier than stars (e.g. \citealt{Springel2005b}). 

\citet{Rodriguez2017} analysed the morphology of $\sim 18000$ central galaxies at $z \sim 0$ from the Illustris cosmological hydrodynamic simulation (\citealt{Sijacki2015}). They found that, for objects with $M_* \lesssim 10^{11}$ M$_\odot$, mergers do not seem to play any significant role in determining the galaxy morphology: remnants are associated with both spheroidal and disk-dominated galaxies (see also \citealt{Sparre2017} for similar results). An incidence of $\sim 27-50$\% for rotating disks in our PUMA sample is therefore consistent with these theoretical predictions.  

We stress however that the PUMA sample, with its relatively small number of targets and the different intrinsic properties of each ULIRG, limits the statistical meaning of our results. Among the 8/29 systems with gaseous disk-like kinematics, I10190 W, I14348 N and I20100 SE are associated with less advanced stages of the merger (wide binaries with nuclear separations $\gtrsim 5$ kpc in projection), and we cannot exclude a disk destruction in subsequent phases;  IZw1 is instead a minor merger. On the other hand, the four remaining targets have a unique kinematic centre and kpc-scale rotation signatures, regardless the presence of double nuclei (in the binary systems I07251 and I12072) or strong streams (in the remnants I13120 and I17208). Among the other systems, we identified 5 merger remnants with a stellar disk (I00188, I01572, I05189, I14378 and I19542) but highly perturbed gas kinematics which might prevent the detection of a gaseous disk. These 9 targets (4 with a gaseous disk and 5 with a stellar disk) represent the strongest evidence for the preserving (or reforming) of a gaseous disk in major merger processes within the PUMA sample.

\subsection{Differences between gas and stellar kinematics along major axis PA$^{kin}$}

The narrow \ha and stellar PV diagrams in Fig. \ref{PV_I13120} (3rd and 4th panel) display significant dissimilarities: the maximum gas and stellar velocity variations along PA$^{kin}$ are $\delta v_{gas} \sim 328$ \kms and $\delta v_* \sim 214$ \kms respectively, while the peak velocity dispersion are $\sigma_{gas} \sim 140$ \kms and $\sigma_* \sim 210$ \kms. Differences between gas and stellar kinematics along PA$^{kin}$ are usually observed in (U)LIRGs (e.g. \citealt{Cazzoli2014, Crespo2021}), and can be interpreted as due to the presence of different dynamical structures or distinct levels of obscuration (see below). 

A precise comparison between stellar and gas rotation along PA$^{kin}$ can be performed only for a couple of systems in our sample. Among the 8 systems isolated in the previous section (IZw1, I07251 W, I10190 W, I12072 N, I13120, I14348 NE, I17208, and I20100 S), only two targets show regular gas PV diagrams, without significant  contributions from perturbed components: I10190 W and I13120\footnote{IZw1 PV diagrams are strongly affected by the AGN in the vicinity of the nucleus; similarly, the gas velocity profiles of remaining systems are affected by residual outflow and stream components; see Appendix \ref{APV}.}. These two systems show similar maximum stellar and gas velocities, with $\delta v_{gas} \sim 200$ \kms and $\delta v_* \sim 300$ \kms, and similar maximum gas velocity dispersion, of $\sim 130$ \kms; vice versa, their maximum stellar velocity dispersion are significantly different, with $\sigma_* \sim 200$ \kms in I13120 and $\sim 80$ \kms in I10190 W. This difference might be due to intrinsic dissimilarities, as for instance the nuclear obscuration. As dust preferentially obscures young stars, which tend to be dynamically cooler than older stellar populations, an higher obscuration in the nuclear regions could translate into a higher velocity dispersion. The continuum color of I13120, defined as log ($f_R/f_B$), with $f_R$ and $f_B$ being the flux at $\sim 9000\AA$ and $\sim 4500\AA$ respectively, is a factor $\sim 3$ higher than I10190 W in the nuclear regions (\citealt{Perna2021}). 
The different $R_e$ of the two systems (see Table \ref{Tproperties}) could play a role as well: at fixed disk mass, a more compact disk has a steeper inner velocity gradient, resulting into a higher velocity dispersion peak.

Instead, the observed differences between stellar and gas velocity amplitudes in I10190 W and I13120 ($\delta v_{gas} \sim  1.5 \delta v_*$) can be related to the star formation activity during the merger stages. For instance, numerical simulations by \citet{Cox2006} showed that old stars that are present prior to the merger, i.e. the oldest stellar populations, are the slowest rotators in a merger remnant; on the contrary, younger stars forming during the first passage of the galaxies and the final merger event are the fastest rotators (see their Fig. 7). We might speculate that
youngest stars do not significantly contribute to the measured stellar velocity dispersion, as they are more embedded in dusty regions. 
In Catal{\'a}n-Torrecilla et al. (in prep.) we will present the stellar population synthesis and its spatial distribution, in order to test this scenario.

\begin{figure}[t!]
%
\centering
\includegraphics[width=9.cm,trim= 0 0 0 0,clip]{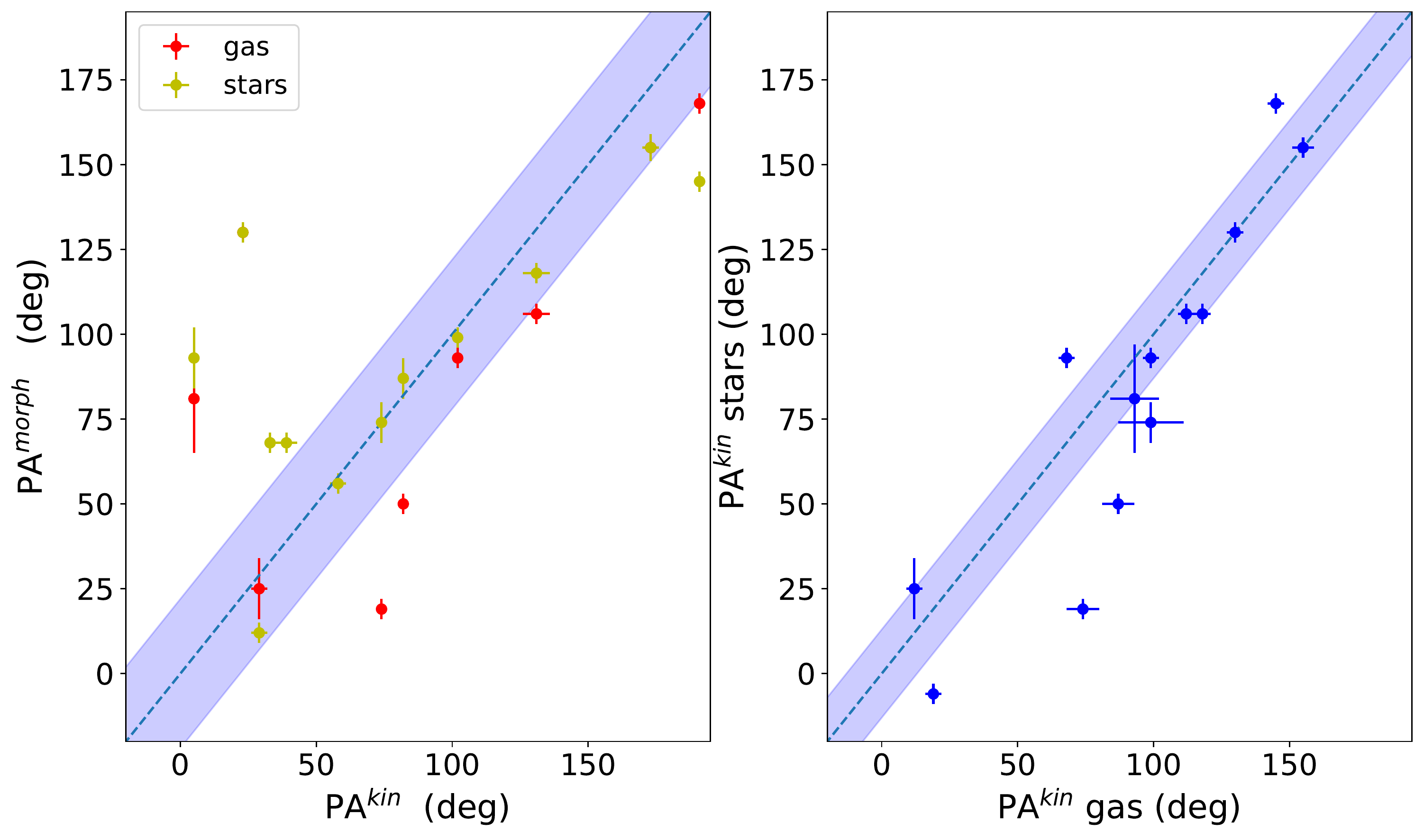}

\caption{\small {\it Left}: comparison between morphological and kinematic major axis PA  for the stellar (yellow) and gas (red) components. The blue dashed line indicates the 1:1 relation; the blue shaded region ($\pm 22^\circ$) includes 90\% of the morpho-kinematic PA misalignments of a sample of 80 non-interacting galaxies from the CALIFA survey (\citealt{Barrera2015}).  {\it Right}: comparison between stellar and gas kinematic major axis PAs. The  dashed line indicates the 1:1 relation; the  shaded region ($\pm 15^\circ$) includes 90\% of the kinematic PA misalignments of CALIFA non-interacting galaxies (\citealt{Barrera2015}).  
}
\label{PAaligmnent}
\end{figure}

\subsection{Morpho-kinematic PA (mis)alignments}\label{Smisalignments}

Figure \ref{PAaligmnent}, left, shows the comparison between morphological and (gas and stellar) kinematic major axis position angles, for all PUMA systems where it was possible to determine PA$^{kin}$. Following \citet{Barrera2015}, we compare our morpho-kinematic (mis)alignments with those of a control sample of 80 non-interacting galaxies from the CALIFA survey,  
whose spatial sampling (from $\sim 0.3$ to 1.5 kpc) and FOV coverage (sizes from 7 to 40 kpc) are comparable with those of our PUMA galaxies. In the figure, we report the 1:1 relation, with a shaded region including 90\% of the CALIFA non-interacting sources, i.e. with a misalignment smaller than $22^\circ$. The relatively small PUMA sample does not allow us to derive strong conclusions about the general behaviour of ULIRGs systems; nevertheless, we note that 57\% (4/7) of PUMA systems have $|PA^{morph} - PA_{gas}^{kin}|$ misalignments larger than $22^\circ$, and 42\% (5/12) have $|PA^{morph} - PA_{*}^{kin}| > 22^\circ$. These results are consistent with those reported by \citet{Barrera2015}, who analysed the morpho-kinematic misalignments in a larger sample of $\sim 80$ interacting CALIFA galaxies, considering both stellar and gas kinematic PAs.

Figure \ref{PAaligmnent}, right, shows instead the comparison between gas and stellar kinematic position angles, for the 13 PUMA systems where it was possible to determine the major axis PAs. Also in this case, we compare our results with those in \citet{Barrera2015}: in the figure, we report the 1:1 relation, with a shaded region including 90\% of the CALIFA non-interacting sources, i.e. with a misalignment smaller than $15^\circ$. About 38\% (5/13) PUMA kinematic misalignments are larger than $15^\circ$, roughly consistent with \citet{Barrera2015}, who find that 20\%  of the CALIFA interacting sample has  kinematic misalignments larger than $15^\circ$. 

The most deviating points in Fig. \ref{PAaligmnent}, right, are associated with I10190 E, I14348 NE, I16090, I20087, I20100 SE (see Table \ref{Tproperties}). The slightly larger number of PUMA systems with more extreme kinematic misalignments might be due to their more advanced merger stage with respect to CALIFA interacting galaxies, which sample contains $\gtrsim 43\%$ pre-merger systems without any visual feature of interaction and projected distances up to 160 kpc. In fact, the presence of more close companions, prominent tidal streams, and strong nuclear winds in our PUMA systems might all contribute to the kinematic misalignments (but see also \citealt{Chen2016}).

These results indicate that interactions and mergers do have an impact on the internal kinematic alignment of galaxies. However, we note that stellar and gas PAs are roughly aligned, while more significant misalignments can be found between the  morphological and kinematic PAs, consistent with the  
\citet{Barrera2015} results.

\subsection{3D-Barolo and gas kinematics classification}\label{S3DB}

To test whether the systems with more regular gas kinematics are compatible with a rotationally supported system, we modelled the narrow \ha data cubes with 3D-Barolo (\citealt{Diteodoro2015}). In particular, we modelled the gas kinematics of the following systems: IZw1, I07251, I10190 W, I12072 N, I13120,  I14348 NE, I17208 and I20100 SE. In this section, we present the general strategy adopted for I13120; more details about the fitting procedure per individual targets are reported in Appendix \ref{A3DBarolo}.

The main assumption of the 3D-Barolo model is that all the emitting material of the galaxy is confined to a geometrically thin disk and its kinematics are dominated by pure rotational motion. The possible presence of residual components associated with the outflow might affect the 3D-Barolo modelling, especially in the innermost nuclear regions, where the outflow is stronger. Nevertheless, this model allows us to asses the presence of such disks, and to infer a simple kinematic classification through the standard $v_{rot}/\sigma_0$ ratio, where  $v_{rot}$ is the intrinsic maximum rotation velocity (corrected for inclination, $v_{rot}= v_{LOS} /sin (i)$) and $\sigma_0$ is  the intrinsic velocity dispersion of the rotating disk, related to its thickness. In this work, we define $\sigma_0$ as the measured line width in the outer parts of the galaxy, corrected for the instrumental spectral resolution (e.g. \citealt{Schreiber2018}).

3D-Barolo best-fit results have been obtained following two different approaches. The first one consists of a two-step strategy. First, we tried different azimuthal models spanning a range of disk inclination angles $i$ with respect to the observer (5 to 85$^\circ$ spaced by 5$^\circ$, with 0$^\circ$ for face-on); during this step, the $i$ parameter is fixed, and the fitting minimization is performed considering the following free parameters: $v_{rot}$, the rotation velocity, $\sigma$, the velocity dispersion, and $\phi$, the major axis PA. The disk center is fixed to the position of the nucleus (inferred from registered HST/F160W images; see \citealt{Perna2021}). We therefore inferred the disk inclination angle considering the best-fit configuration with the minimal residuals, defined using the Eqs. 2 and 3b in \citet{Diteodoro2015}. Then, we run 3D-Barolo with a local normalization, letting it minimize the $v_{rot}$, $\sigma$, $\phi$ and $i$ parameters. In this second step, the inclination is left free to vary in a few degrees around the best-fit $i$ defined in the previous step. 
For the second method, we simply run 3D-Barolo with a local normalization, but initialising the inclination to the value derived from the isophote modelling of HST data (Sect. \ref{Sphot}), hence assuming that continuum and narrow \ha have the same geometry. As for the PA measurements, 3D-Barolo fit analysis is performed on the innermost nuclear position, excluding the regions with poor S/N ($<3$), for which a Voronoi tesselation would be required in the 3D-barolo modelling.

\begin{figure}[!t]
%
\centering
\includegraphics[width=9.cm,trim= 0 0 0 0,clip]{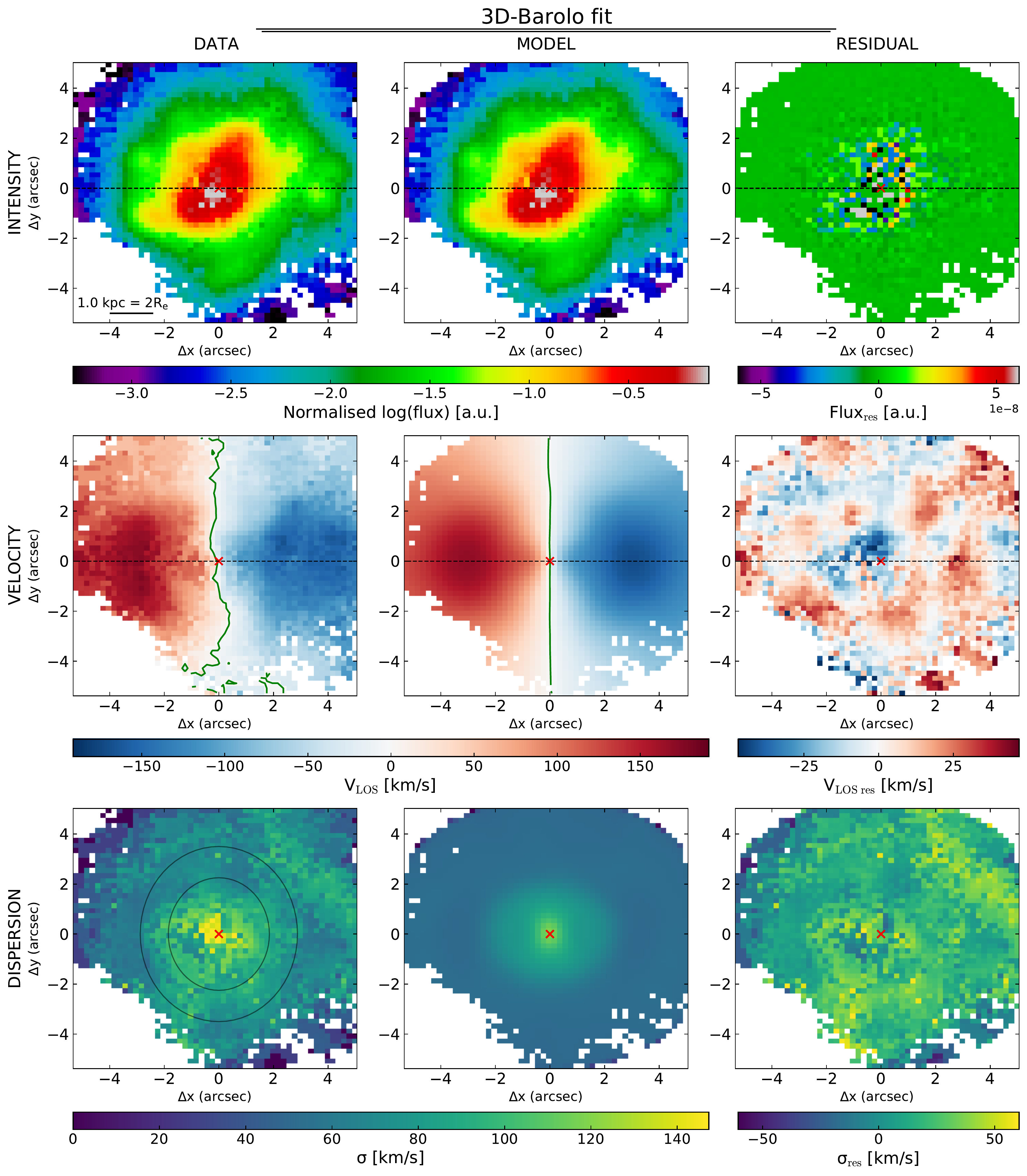}

\includegraphics[width=9cm,trim= 0 0 0 27,clip]{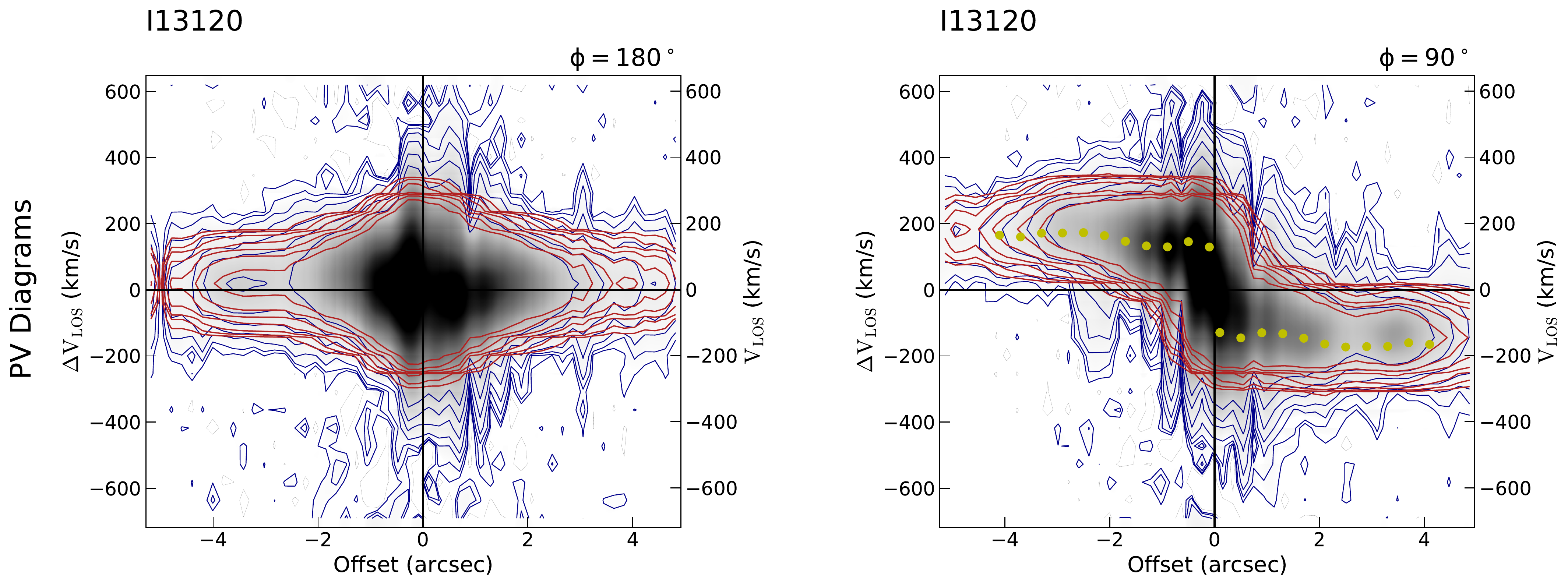}

\caption{\small 
I13120 narrow \ha  3D-Barolo disk kinematic best-fit of the moment 0, 1 and 2 (first to third rows) and PV diagrams along the minor and major disk kinematic axes (bottom). 
The black curves in the velocity dispersion map obtained from the data (third row, left panel) identify the region from which $\sigma_0$ is extracted. In the PV diagram along the major axis (bottom right) and minor axis (left), data are indicated with a grey-scale map and  blue contours, while  best-fit model are shown with red contours. 
}
\label{BBaroloI13120}
\end{figure}

The resulting best fit plots for I13120 are shown in Fig. \ref{BBaroloI13120}, while best-fit parameters are reported in Table \ref{Tdisk}, together with those of the remaining 7 targets with evidence of rotation (see Appendix \ref{A3DBarolo} for their 3D-Barolo fit analysis). 
We note that the 3D-Barolo best-fit inclination of I13120 obtained with the two methods,  $i_{I} = 34^{\circ}\pm 3^{\circ}$ and $i_{II} = 27^{\circ}\pm 3^{\circ}$, are still consistent with the value we derived from the isophote modelling of HST/F160W data, $i_{morph} = 25^{\circ} \pm 11^{\circ}$ (Sect. \ref{Sphot}). However, their slightly different $i$ values translate in different rotational velocities; we therefore decided to report in the table the best-fit results obtained with both methods. 
Similarly, for each target in the table we indicate if both methods provide totally consistent results (I10190 W and I20100 SE) or not (I13120 and IZw1), or alternatively, if the results are obtained from the first (I07251,  I12072 and I14348 NE, with no isophote analysis) or the second method only (I17208, with unconstrained $i$ when fitted with the first approach).

\begin{table*}
\footnotesize
\begin{minipage}[!h]{1\linewidth}
\setlength{\tabcolsep}{6pt}
\centering
\caption{\ha disk parameters}
\begin{tabular}{lccccccccc}
target     & 3DB method(s) &   $i^{3DB}$& $\phi^{3DB}$  & $v_{rot}^{3DB}$ & $\sigma_0$ & $v_{rot}^{3DB} / \sigma_0$ & $R_e$ & $M_{dyn}$& $M_{*}$\\
            &  &  (deg) &  (deg) & (km/s)  & (km/s) & & (kpc) & ($\times 10^{10}\ M_\odot$) & ($\times 10^{10}\ M_\odot$)\\
\scriptsize{(1)} & \scriptsize{(2)}   &\scriptsize{(3)}   & \scriptsize{(4)} & \scriptsize{(5)} & \scriptsize{(6)} & \scriptsize{(7)} & \scriptsize{(8)}& \scriptsize{(9)}& \scriptsize{(10)}\\
\hline

IZw1  & I &  $47\pm 4$  & $138\pm 9$  & $195\pm 20$ & $35\pm 14$ & $8\pm 3$ & $1.84\pm 0.05^\dagger$  ($1.0\pm 0.1$)& $3.6_{-1.0}^{+0.8}$ & $-$ \\
      & II &  $38\pm 3$  & $136\pm 8$  & $220\pm 20$ & $32\pm 12$ & $7\pm 4$ & $1.84\pm 0.05^\dagger$ ($1.0\pm 0.1$)& $4.4_{-1.0}^{+0.8}$ & $-$ \\
\hline

I07251 & I & $45\pm 6$ &   $285\pm6$ & $185\pm 15$ & $80\pm 30$ & $2\pm 1$ & $2.8\pm 0.6$* ($2.3\pm 0.4$)& $7.0_{-3.4}^{+2.8}$ & $-$ \\
\hline

I10190 W  & I, II & $48\pm 4$ & $117\pm 9$ & $210\pm 11$ & $56\pm 10$ & $3\pm 1$ & $2.6\pm 0.1$ ($2.9\pm 0.3$)& $6.5_{-1.1}^{+0.9}$  & 4\\

\hline

I12072 N  & I &  $48 \pm 4$  & $87\pm 6$ & $70\pm 9$ &  $85\pm 25$ & $0.8\pm 0.3$ & $2.8\pm 0.6$* ($1.6\pm 0.3$)& $3.5_{-2.8}^{+2.2}$ & 3.6\\
\hline

I13120  & I & $34\pm 3$   & $88\pm 8$ & $270\pm 12$ & $58 \pm 10$ & $5\pm 1$ & $0.5\pm 0.1$ ($1.0 \pm 0.3$)& $1.9_{-0.5}^{+0.4}$ & $3\pm 0.1$\\
   & II & $27\pm 3$   & $88\pm 9$ & $315\pm 20$ & $58 \pm 10$ & $6\pm 1$ & $0.5\pm 0.1$ ($1.0 \pm 0.3$) & $2.5_{-0.7}^{+0.6}$ & $3\pm 0.1$\\
\hline

I14348  NE & I & $52\pm 3$   & $185\pm 7$& $109\pm 8$& $73 \pm 15$ & $1.5\pm 0.3$ & $2.8\pm 0.6$* ($6.0 \pm 0.7$)& $3.6_{-1.9}^{+1.6}$* & 10.8\\
\hline

I17208 & II & $47\pm 4$   & $142\pm 5$  & $155\pm 15$ & $70\pm 20$ & $2\pm 1$ & $1.3\pm 0.1$ ($1.7\pm 0.4$) & $2.3_{-0.9}^{+0.8}$ & $13.5\pm 4.0$\\
\hline

I20100 SE & I, II & $58\pm 4$   & $287\pm 6$ & $110\pm 10$ & $78\pm 18$ & $1.4\pm 0.3$ & $1.5\pm 0.2$ ($3.7\pm 0.6$) &  $2.1_{-1.1}^{+0.9}$ & $-$ \\

\hline
\end{tabular}
\label{Tdisk}
\vspace{0.2cm}
\end{minipage}
{\small
{\it Notes.} Column (1): Target name. (2): 3D-Barolo fit analysis methodology: 'I' for a two-step strategy, to first constrain the inclination and then all remaining disk parameters, and 'II' for a single-step strategy considering the $i^{morph}$ (Sect. \ref{Sphot}) as initial guess for the inclination. (3): 3D-Barolo disk inclination $i$. (4): 3D-Barolo kinematic PA of the major axis on the receding half of the galaxy, taken anticlockwise from the North direction on the sky. (5): 3D-Barolo rotation velocity. (6) measured velocity dispersion in the outer part of the galaxy, after subtracting the instrumental resolution (in quadrature). (7): maximum rotation velocity over velocity dispersion. (8): effective radius measurements from Table \ref{Tproperties}; these values have been preferred to those obtained from the \ha flux map, reported in parenthesis, as less affected by dust obscuration. They are however totally consistent with $R_e$(H$\alpha$), in the range $1-2.9$ kpc. The only exceptions are represented by I14348 NE and I20100 SE, which $R_e$(H$\alpha$) are strongly affected by the presence of strong off-nuclear \ha emission.  (9): dynamical masses within 2$R_e$. (10): (SED-based) stellar masses from \citet{Rodriguez2010} and \citet{daCunha2010}. For the former, no uncertainties were reported in the original paper. \\
($^\dagger$): $R_e$ from \citet{Veilleux2006}, see Sect. \ref{Sphot}.
($^*$): mean $R_e$ of local (U)LIGs, from \citet{Bellocchi2013}. 
}
\end{table*}

The $\sigma_0$ values reported in Table \ref{Tdisk} are estimated from the narrow \ha velocity dispersion map, as the median value in a radial elliptical annulus which takes care of the disk inclination (as shown in the velocity dispersion panels, see e.g. Fig. \ref{BBaroloI13120}).  This was preferred to the (beam-smearing corrected) value that could be inferred from 3D-Barolo, because of the significant fit residuals in the velocity maps, and for consistency with previous works in the literature (see next sections). The use of an annulus region allows us to mitigate the beam-smearing effects or residual outflow contributions, which are higher in the centre than the outside (see e.g. PV diagrams in Fig. \ref{BBaroloI13120}), and - more in general - remove different contributions which  artificially increase (by a $\sim 20\%$, on average) the velocity dispersion, e.g. due to tidal streams and companion systems. 
We used these $\sigma_0$ values to measure the ratio $v_{rot}/\sigma_0$ for all the galaxies with indication of gas rotation (column 7 in Table \ref{Tdisk}).

It is important to note that our best-fit models have important limitations and systematic uncertainties, and the small  formal errors on a parameter do not necessarily imply a good fit  (see \citealt{Neeleman2021}). Both the very simplified disk models and a (possible imprecise) separation between narrow and perturbed \ha components (Sect. \ref{Sanalysis}) might be responsible of the significant residuals we observe in the velocity and velocity dispersion maps (e.g. Fig. \ref{BBaroloI13120}). Nevertheless, our 3D kinematical analysis shows that, on average, this small sub-sample of PUMA systems have a ratio of rotational velocity to velocity dispersion of $v_{rot}/\sigma_0 \sim 1-8$. Although slightly lower 
than that of spiral galaxies in the local Universe ($v_{rot}/\sigma_0 \sim 10$), our values are still comparable with \ha measurements of other low-z (U)LIRGs in the literature (e.g. \citealt{Bellocchi2013,Crespo2021}) and systems at z $\sim 0.5-1$ (\citealt{Rizzo2021} and references therein). Therefore, 3D-Barolo results provide further indication of rotationally-supported gas motions in these targets.

\subsection{Gas velocity dispersion in (U)LIRGs and high-$z$ populations: Dependence on starburstiness}\label{Scorrelations}

Many theoretical and observational studies suggest that gas in
high-$z$ galaxies has larger random motions compared to nearby galaxies: in particular, the ionised gas velocity dispersion goes from $\sim 20$ km/s in nearby spirals to $\sim 45$ \kms in massive main sequence star-forming disk galaxies at $z \sim 2$, although with a significant scattering of values (e.g. \citealt{Ubler2019}).

\begin{figure}[h!]
%
\centering
\includegraphics[width=8.35cm,trim= 0 0 0 0,clip]{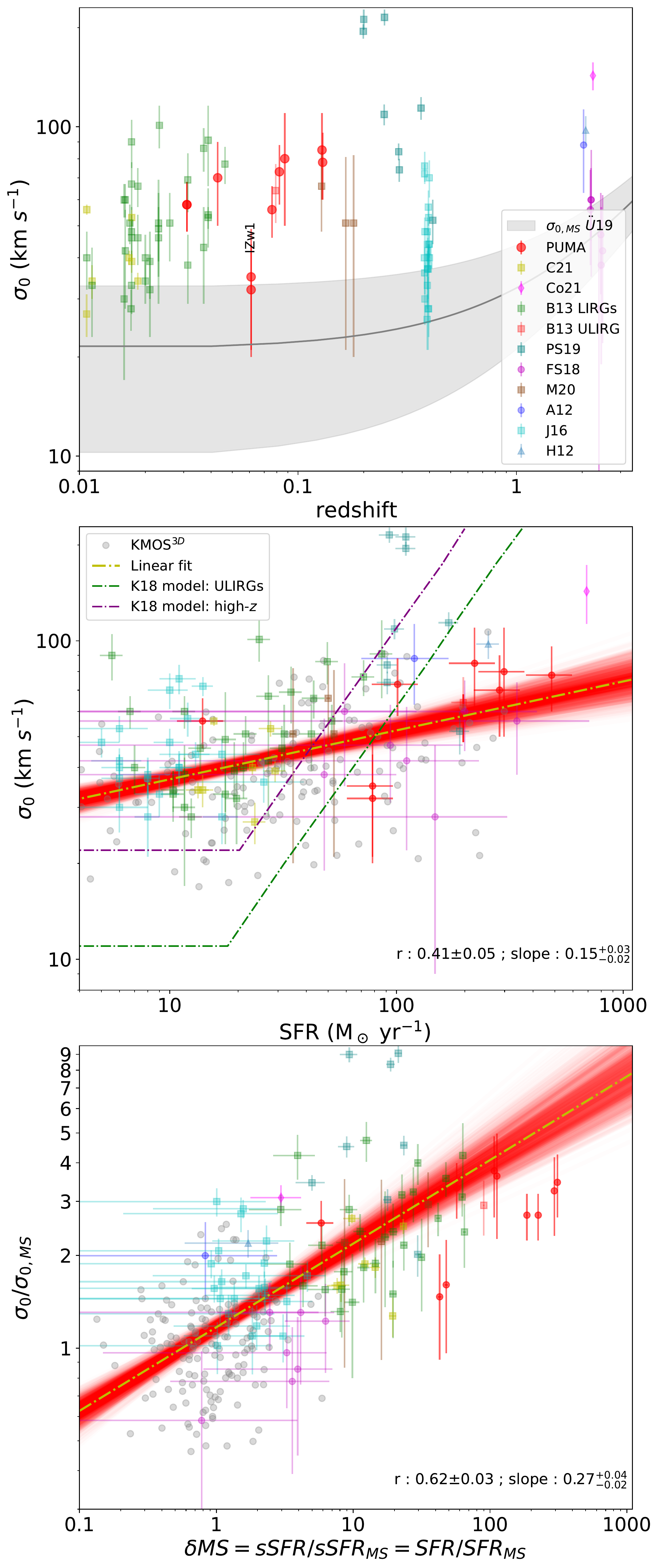}

\caption{\small 
{\it Top}: Velocity dispersion $\sigma_0$ as a function of $z$ for PUMA sub-sample (red dots) and other individual ionised gas measurements from the literature, 
distinguishing between the different samples presented in Appendix \ref{Aextendedsample}, as labeled. 
The \citet{Ubler2019} evolutionary trend of MS galaxies is shown with a solid curve (shaded area: $1\sigma$ scatter around the average trend).  
{\it Middle}: $\sigma_0$ as a function of the SFR, for all targets already reported in the first panel, and the KMOS$^{3D}$ galaxies (grey points; \citealt{Ubler2019}). K18 models and a linear fit (with scatter) are also reported, as labeled.
{\it Bottom}: $\sigma_0$ normalised to the evolutionary trend of MS galaxies as a function of $\delta MS$ for all targets already reported above. A linear fit is also reported.  
}
\label{sigmaz}
\end{figure}

Figure \ref{sigmaz}, top, shows the narrow \ha velocity dispersion of our rotationally supported PUMA systems (red circles) as a function of the redshift, together with the \citet{Ubler2019} evolutionary trend of star-forming galaxies. This trend mostly traces the velocity dispersion evolution of normal MS galaxies (e.g. \citealt{Ubler2019,Schreiber2018}); we therefore labelled it as $\sigma_{0, \ MS}$ hereinafter. The $\sigma_0$ of our PUMA systems, in the range $30-85$ \kms, are more compatible with - or possibly higher than - those of high-$z$ galaxies rather than nearby spirals. This can be explained taking into account the following arguments.  
On the one hand, the velocity dispersion increases as natural consequence of the availability of huge gas reservoirs and intense star formation that is taking place in ULIRGs and high-$z$ galaxies (e.g. \citealt{Lehnert2009, Arribas2014,Johnson2018}). 
On the other hand, both the gravitational instabilities due to the galaxy interactions and the presence of non-circular motions in our PUMA targets can contribute to the $\sigma_0$ enhancement with respect to isolated nearby galaxies. 

To better investigate the origin of the differences between PUMA  systems and normal MS galaxies at different redshifts, in Fig. \ref{sigmaz} we report  additional individual ionised gas measurements of SB disk galaxies from the literature (see Appendix \ref{Aextendedsample} for details). Many of them show a significant deviation from the $\sigma_{0, \ MS}$ evolutionary trend, similar to our PUMA systems.

All these measurements from the literature have been obtained from IFS data; therefore, they are not strongly affected by beam-smearing effects and other systematics that tend to overestimate the intrinsic dispersion (see discussion in \citealt{Ubler2019}). We also stress here that all selected individual sources are disk galaxies. This ensures relatively small contribution of outflows in the velocity dispersion measurements, which incidence increases with SFR and AGN activity (e.g. \citealt{Cicone2016,Villar2020}). A few additional caveats should be kept in mind regarding this compilation of sources: all of them are presented as SB galaxies in the original papers, but we note that i) there is no rigorous definition of a SB galaxy, but
several different criteria are often used, 
and ii) especially at high-$z$, stellar mass and SFR measurements can be highly uncertain,  depending on the availability of multi-wavelength information. This aspect is further discussed below.

\subsubsection{The $\sigma_0 - SFR$ correlation}

Most of the galaxies presented in Fig. \ref{sigmaz}, top, significantly deviate from the $\sigma_{0, \ MS}$ evolutionary trend. In order to understand if this deviation is due to the extreme SFR in these systems, we show in Fig. \ref{sigmaz}, middle, the velocity dispersion $\sigma_0$ as a function of the SFR, for all the targets already mentioned, in addition to the sample of normal MS galaxies used to derive the $\sigma_{0, \ MS}$ trend (\citealt{Ubler2019}). All SFR measurements of SB reported in the figure are obtained from IR luminosities (as reported in the original papers, or using the \citealt{Kennicutt1998} relation, assuming the Chabrier IMF).  For the PUMA targets, we considered the nuclear IR luminosities reported in \citet{Pereira2021}, table 7; for the binary systems, the fraction of the IR luminosity assigned to each nucleus is based on their relative ALMA continuum fluxes. 
We observe a relatively poor correlation (Spearman rank correlation coefficient 0.4), reasonably due to a bias selection: a more clear correlation is in fact observed when combining samples of galaxies covering a larger dynamical range, i.e. also including local MS galaxies (e.g. \citealt{Yu2019, Varidel2020}). By performing a linear regression fit we derive 
\begin{equation}
    log (\sigma_0) = (0.15_{-0.02}^{+0.03}) \times log (SFR) +  (1.46 \pm 0.04),
\end{equation}

compatible (within $1\sigma$) with \citet{Arribas2014} previous results, obtained from a sample of (U)LIRGs observed with the optical spectrograph VIMOS. We compare this fit with the model predictions by \citet{Krumholz2018}, for both high-$z$ galaxies and local ULIRGs (dash-dotted lines in Fig. \ref{sigmaz}, middle). The slight differences between the two theoretical curves in the figure are due to the distinct ISM physical conditions of these two classes of sources (see Table 3 in \citealt{Krumholz2018}). These models explain the observed slow increase in $\sigma_0$ as a function of SFR  in the range [$10^{-3}, 10$] M$_\odot$ yr$^{-1}$, followed by a steeper increase up to 10s \kms considering two different regimes. The $\sigma_0$ floor is due to stellar feedback processes, while at higher SFR the velocity dispersion is regulated by gravitational turbulence.  We note that the comparison between these theoretical curves and our collected data is strongly limited by caveats. To begin, all SB galaxies in our plot are disk galaxies: this could result in the exclusion of all targets with more extreme (i.e. higher) $\sigma_0$. The next problem is that the \citet{Bellocchi2013} and PUMA velocity dispersion measurements in this plot are derived from the narrow H$\alpha$, i.e. after removing the more extreme kinematic components. 
As shown in Figs. \ref{I00188linemaps}-\ref{I22491linemaps} and Fig. \ref{APV}, significantly higher velocity dispersion (up to several 100s \kms) is measured in the total \ha line profiles of our PUMA galaxies, reasonably associated with extended outflows and streaming motions.  
Finally, the ISM of (U)LIRGs and high-$z$ SB galaxies might not be in vertical pressure nor energy balance, as assumed in \citet{Krumholz2018} models: instead, their velocity dispersion might be strongly affected by non-circular motions. Specifically, these $\sigma_0$ measurements might not be dominated by the turbulent component, that is relevant for a comparison with the SFR (e.g. \citealt{Bacchini2020}).

\subsubsection{The $\sigma_0/\sigma_{0, \ MS} - \delta MS$ correlation}

In the $\sigma_0-$SFR plane, local (U)LIRGs and high-$z$ galaxies tend to occupy the same region, regardless of their intrinsic difference in terms of morphology, gas fraction and starburstiness. In order to distinguish between normal MS and SB galaxies,  we show in Fig. \ref{sigmaz}, bottom, the velocity dispersion $\sigma_0$ normalised to $\sigma_{0,\ MS}$ (solid line in the top panel; \citealt{Ubler2019}) as a function of the starburstiness $\delta MS = sSFR / sSFR|_{MS}$ for all targets already reported in the previous panels. 
The $sSFR|_{MS}$ is derived from the \citet{Speagle2014} relation, starting from the available stellar mass  measurements from the literature for \citet{Johnson2016}, \citet{Schreiber2018}, \citet{Molina2020}, \citet{Cochrane2021}, and KMOS$^{3D}$ individual targets; stellar masses of LIRGs and ULIRGs from this study, from \citet[][]{Bellocchi2013}, \citet{Pereira2019} and \citet[][]{Crespo2021}, are instead derived from the dynamical mass estimates, assuming $M_* = (1-f_{gas}) \times (1-f_{DM}) \times M_{dyn}$, where $f_{gas}$ and $f_{DM}$ are the gas and dark matter fractions\footnote{In this work, we define $f_{gas} = M_{gas}/M_{bar}$, with $M_{bar} = (M_* + M_{gas})$, and $f_{DM} = M_{DM}/M_{dyn}$.}, respectively, and $M_{dyn}$ is the dynamical mass within 2$R_e$ (see next section). For the gas fraction, we considered a conservative $f_{gas} = 0.1$ (\citealt{Isbell2018}; higher $f_{gas}$ values would further increase their $\delta MS$). This gas fraction is consistent with the estimate we obtain considering the molecular gas mass inferred by ALMA data (\citealt{Pereira2021}; Lamperti et al., in prep) and the $M_*$ measurements  available for a few PUMA targets (see Table \ref{Tdisk}), $\bar f_{gas} = 0.11\pm 0.05$. For the dark matter fraction (within 2$R_e$) we assumed $f_{DM} = 0.26$, defined as $1- M_{bar}/M_{dyn}$, with $M_{bar} = M_* + M_{gas}= M_* /(1-f_{gas})$. The $f_{DM}$ estimate was derived for the PUMA and \citet{Johnson2016} SB galaxies for which stellar masses are available, and considering the dynamical masses within 2$R_e$ (see Sect. \ref{SMdyn}). Because of these assumptions, we considered a factor 3 uncertainties for the stellar mass measurements of (U)LIRGs. These uncertainties, however, play a minor role in the derived $\delta MS$: at low-$z$, the MS has a soft slope, and normal and massive MS galaxies have similar SFRs (e.g. at $z\sim 0.1$, galaxies of $10^{10}$ and $10^{11}$ M$_{\odot}$ have $SFR|_{MS} \sim 1$ and $\sim 3.5$ M$_\odot$ yr$^{-1}$,  respectively); on the other hand, local (U)LIRGs have much higher SFRs, from 10s to 100s M$_\odot$ yr$^{-1}$,  and therefore $\delta MS$ of the order of $10-100$. 
Finally, for the \citet{Alaghband2012} and \citet{Harrison2012} high-$z$ galaxies  we assumed that $M_* = 10^{11}$ M$_{\odot}$, following \citet{Harrison2012}.

Figure \ref{sigmaz}, bottom, shows a clear correlation between $\sigma_0/\sigma_{0, \ MS}$ and $\delta MS$ (Spearman rank correlation coefficient 0.6), suggesting that SB galaxies tend to have higher velocity dispersion than normal galaxies at given $z$ and stellar mass. 
A tentative evidence of such correlation was already reported by \citet{Wisnioski2015}, for the KMOS$^{3D}$ galaxies at $z \sim 2$, and by \citet{Varidel2020}, for nearby MS galaxies of the SAMI survey, but the lack of dynamical range in terms of starburstiness in these surveys maintained the correlation at low significance level (see also Figs. 15-16 in \citealt{Ubler2019}). 

By performing a linear regression fit, we derived   
\begin{equation}
    log (\sigma_0/\sigma_{0,\ MS}) = (0.27_{-0.02}^{+0.04}) \times log (\delta MS) +  (0.07_{-0.01}^{+0.02}).
\end{equation}
Given the correlation between $\sigma_0$ and SFR (e.g. \citealt{Varidel2020}) and the tight inter-relationship between $\delta MS$ and $f_{gas}$ (e.g. \citealt{Wisnioski2015,Tacconi2020}), it is  unsurprising that the starburstiness correlates with the excess in the velocity dispersion with respect to MS galaxies. The positive slope of $\sim 0.27$ is inconsistent with the one observed between $f_{gas}$ and $\delta MS$, $+0.5$ (\citealt{Tacconi2020}), suggesting that complex interactions between different physical drivers are responsible of the correlation observed in Fig. \ref{sigmaz} (bottom). As a final check, we studied the correlation between $\sigma_0$ and $v_{rot}$, as well as between $\sigma_0/\sigma_{0, \ MS}$ and $v_{rot}$, obtaining Spearman coefficients of $\sim - 0.2$ (with {\it p-value} $\sim 0.05$); this shows that not even $v_{rot}$ can have a significant role in determining the increase in the velocity dispersion of SB galaxies.

The strong correlation $\sigma_0/\sigma_{0, \ MS} - \delta MS$ suggests that the SFR of galaxies above the MS is taking place in an ISM significantly more unsettled than in normal (i.e. MS) galaxies. This is likely due to the presence of interactions and mergers which enhance SFR while simultaneously increase the velocity dispersion of the ISM. The absence of a strong correlation between SFR and elevated velocity dispersion in star-forming clumps both in local (U)LIRGs and high-z SFGs (\citealt{Arribas2014,Genzel2011}) further suggests that the extreme dispersion cannot be simply related to the strong SFR in these systems.

Our arguments are consistent with the parsec-resolution hydrodynamical simulations of major mergers presented by \citet{Renaud2014}: they found that the increase of ISM velocity dispersion precedes the star formation episodes. Therefore, this enhancement is not a consequence of stellar feedback but instead has a gravitational origin. In this scenario, $\delta MS$ can be interpreted as a tracer of the strength of gravitational torques: stronger gravitational torques during the interactions lead the gas to flow inwards, both increasing the velocity dispersion and the efficiency in converting gas into stars.

We finally mention that AGN outflows, which are ubiquitous in these systems (\citealt{Perna2021,Pereira2021}), can also contribute to increase the velocity dispersion of the disks, as suggested by high-resolution hydrodynamic simulations (e.g. \citealt{Wagner2013,Cielo2018}).

A more detailed investigation of the physical meaning of the correlations reported in Fig. \ref{sigmaz} goes beyond the purpose of this study; here we just stress that, by selecting a (relatively small) sample of MS and SB disk galaxies in the redshift range $0.03-2.6$, a more significant correlation is observed between $\sigma_0/\sigma_{0, \ MS}$ and $\delta MS$ rather than between $\sigma_0$ and SFR. 
We argue that this result might be even more evident considering the entire population of (U)LIRGs (i.e. without excluding targets with no evidence of rotating disk; see e.g. Table \ref{Tproperties}), and measuring the velocity dispersion without excluding possible contribution from outflows and streaming motions (see Figs. \ref{I00188linemaps}-\ref{I22491linemaps}).

\subsection{Dynamical masses}\label{SMdyn}

In this section we derive the dynamical masses of our PUMA sub-sample, and compare them with those of other (U)LIRGs from the literature.  
Assuming that the source of the gravitational potential is spherically distributed, we can estimate the dynamical mass within a radius R as:

\begin{equation}\label{EMdyn}
    M_{dyn} = \frac{v_{circ}^2}{G} R = 2.33 \times 10^5 v_{circ}^2 R, 
\end{equation}
where $G$ is the gravitational constant,  $v_{circ}$ is the circular velocity in km/s, and $R$ is given in kpc. 

We used the near-IR continuum 2$R_e$ as the radius to calculate $M_{dyn}$, which for an exponential profile contains $85\%$ of the total flux. The effective radius of I10190 W, I13120 and I17208 is computed with the Isophote package of Astropy (Sect. \ref{Sphot}); IZw1 $R_e$ is instead taken from \citet{Veilleux2006}, who performed multi-component two-dimensional image decomposition to separate the host galaxy from its bright active nucleus; for the remaining three targets, we adopted the ULIRGs average $R_e$ derived by \citet{Bellocchi2013}, as the presence of nearby nuclei (I07251 and I12072) and strong tidal features (I14348 NE) do not allow us to model the continuum with isophotal ellipses.

To infer the circular velocity we consider both the rotation and dispersion motions traced by the narrow H$\alpha$. In particular, we included the asymmetric drift term, which represents an extra component due to the dispersion of
the gas around the disk of the galaxy,
\begin{equation}
    v_{circ}^2 = v_{rot}^2 + \eta \sigma^2.
\end{equation}
The term $\eta$ is a constant and can vary between approximately 1.5 and 6, depending on the mass distribution and kinematics of the galaxy: higher values indicate higher turbulence in the ISM of a rotating disk (\citealt{Neeleman2021}). 
We assumed $\eta = 3$, following \citet{Dasyra2006}, which is very close to the value expected for an exponential, turbulent pressure-supported disk ($\eta = 3.4$), and considered an uncertainty of $1.5$ to take into account the large range of possible values. We note that, on average, our dynamical masses would be a factor 1.15 lower (1.32 higher) assuming $\eta = 1.5$ (6).

The assumption of a spherically distributed ISM is at odds with that we used in Sect. \ref{S3DB} to measure the rotational velocities in our systems. In order to account for this, 
following \citet{Neeleman2021}, we conservatively increased the dynamical mass uncertainty by 20\% toward lower masses. This corresponds to consider that the effective total mass distribution falls somewhere in between a thin disk and a sphere.  

The measured dynamical masses of our PUMA systems range from $\sim 2$ to $\sim 7\times 10^{10}$ M$_{\odot}$, consistent with the median value derived by \citet{Bellocchi2013} for ULIRG systems, $4.8\times 10^{10}$ M$_\odot$, confirming that ULIRGs are intermediate mass systems like previously suggested (i.e., \citealt{Colina2005,Rodriguez2010}). 

In this final part, we further discuss about the dark matter fraction reported in the previous section, and defined as $f_{DM} = 1 - M_{bar}/M_{dyn}$. Using a sub-sample of 27 SB disk galaxies with available $M_{*}$ measurements, and assuming $f_{gas} = 0.1$, we obtained a median value $f_{DM} = 0.26$. We note however that, for a few sources (6/27, mostly from the PUMA sample), $M_*$ > $M_{dyn}$ (Table \ref{Tdisk}; see also Table 6 in \citealt{Rodriguez2010}). This can either suggest that the systems are not relaxed (due to the interaction/mergers) and $M_{dyn}$ is unreliable, or that $M_*$ measurements have high uncertainties. As the $M_{dyn}$ estimates are in agreement with previous works, and because of the fact that for binary PUMA systems in Table \ref{Tdisk} the available $M_*$ measurements are obtained without separating the contribution of the merging galaxies, we favour the second interpretation, i.e. that $M_*$ measurements are highly uncertain though a combination or both may also be possible. These arguments led us to consider significant (factor of 3) uncertainties in the determination of the stellar masses for the entire sample of local (U)LIRGs, required to estimate their $\delta MS$. The conclusions reported in the previous section are however not affected by these uncertainties, because of extreme SFR in (U)LIRGs targets.

\section{Summary and conclusions}\label{Sconclusions}

The project called Physics of ULIRGs with MUSE and ALMA (PUMA) is a survey of 25 nearby ULIRGs observed with MUSE and ALMA. This is a representative sample that covers the entire ULIRG luminosity range, and it includes a combination of systems with AGN and SB nuclear activity in (advanced) interacting and merging stages. \citet{Perna2021} presents the first MUSE results on the spatially resolved stellar kinematics and the incidence of ionized outflows in nuclear spectra; \citet{Pereira2021} analyzes high-resolution (400 pc) $\sim 220$ GHz continuum and CO(2–1) ALMA observations to constrain the hidden energy sources of ULIRGs. In this paper, we investigated the presence of ionised gas rotational dynamics in PUMA targets, to understand if, as predicted by models, rotation disks can be preserved during the merging
process (or rapidly regrown after coalescence) and, if so, which are their main properties.  
Our results are summarised below.

(a) We presented the spatially resolved \ha flux and kinematic maps for the entire PUMA sample, obtained from multi-component Gaussian fit analysis (Fig. \ref{I00188linemaps}- \ref{I22491linemaps}). Irregular large scale ionized gas velocity fields associated with tidally-induced motions and outflows are found in almost all targets; \ha velocities ($v50$) up to $\sim \pm 300$ \kms \ are detected in the MUSE FOV, while \ha line-widths $W80$ range from $\sim 100$ to $\sim 1500$ \kms. 
[NII] (and [OIII]) line transitions are even more affected by perturbed motions, as tidal streams and outflows. 

(b) We studied the \ha kinematics to infer the presence of rotating disk signatures. A kinematic decomposition is performed by selecting in the $\Delta V_j - FWHM_j$ plane all best-fit Gaussian components with relatively small velocities, and  constructing new narrow \ha data cubes.  
In these newly generated data cubes the emission associated to gas components with extreme velocities (likely due to outflows and/or tidally driven flows) is minimized.  

(c) By studying the gas kinematics along the major axes of our galaxies in the innermost regions ($\sim 5-20$ kpc), we found that 27\%  (8/29) individual nuclei are associated with disk-like motions. This has to be considered as a lower limit, as  the presence of vigorous winds and gravitational torques, as well as observational limitations (in terms of spatial and spectral resolution, and S/N), limit our capabilities in isolating more regular, disk-like kinematics through a multi-component Gaussian fit decomposition. This is supported by the fact that 5 merger remnants in our sample present stellar disk motions but highly perturbed gas kinematics. Indeed, the incidence of ionised gas rotating disks is a factor $\lesssim 2$ smaller than that of stellar disk-like motions (\citealt{Perna2021}). This is possibly suggesting that i) we are actually missing a significant fraction of sources with gas rotation because of the above mentioned limitations, or ii) gas component is more affected by winds and gravitational interactions and the probability of preserving a gas disk is lower than that of a stellar disk.  In both instances, our results show that, as predicted by models, rotation disks can be preserved during the merging process and/or rapidly regrown after coalescence.

(d)  For the 8 galaxies with evidence of disk-like motions, we modelled the narrow \ha data cubes with 3D-Barolo, and derived rotational velocities $v_{rot}\in $ [$70-300$] \kms. By combining them with the measured velocity dispersion $\sigma_0$ ($\in [30-80]$ \kms), we derive  
$v_{rot}/\sigma_0$ values in the range 1-8, providing further indication of rotationally supported gas motions in these ULIRGs. 
We also derived their $M_{dyn}$, obtaining values in the range $(2-7) \times 10^{10}$ M$_\odot$, consistent with $M_{dyn}$ of other ULIRGs in the literature.

(e) We compared the narrow \ha velocity dispersion $\sigma_0$ of our 8 PUMA disk galaxies with those of other SB and normal MS disk galaxies at low and high-$z$. We found that all SB galaxies tend toward higher $\sigma_0$ values compared to MS galaxies at the same redshift. Interestingly, when we normalise $\sigma_0$ to the value expected for MS galaxies (at the same $z$), considering the \citet{Ubler2019} evolutionary trend $\sigma_{0, \ MS}$, we found a significant correlation between $\sigma_0/\sigma_{0, \ MS}$ and the starburstiness $\delta MS$. In particular, SB galaxies display up to a factor $\sim 4$ higher velocity dispersion than normal MS galaxies at same redshift. The relatively poor correlation between $\sigma_0$ and the SFR (Fig. \ref{sigmaz}, middle)   suggests that stellar activity cannot be the main responsible for the $\sigma_0$ enhancement observed in SB galaxies, and other mechanisms possibly related to interactions and mergers should be taken into account (see e.g.  \citealt{Renaud2014}).

We note however that most of the SB galaxies at $z \gtrsim 0.4$ collected from the literature are consistent with $\delta MS = 1$ once homogeneous recipes are used to derive the SFR, and measurement uncertainties are taken into account. As a result, the correlation reported in the figure is mostly driven by the comparison between $z \sim 0.03-0.4$ (U)LIRGs and KMOS$^{3D}$ MS galaxies at $z \sim 0.6-2.6$. This makes highly desirable a further investigation of gas dynamical conditions in SB galaxies at $z > 0.4$.
The JWST NIRSpec IFS, with its wide spectral range (from 0.6 to 5.3 $\mu$m) and sub-arcsec resolution,  will allow a comprehensive characterisation of the ionised gas dynamical conditions in such systems.

\begin{acknowledgements}
 
We thank the referee for an expert review of our paper.
The authors thanks Elena Valenti for her  support when preparing the observations, and Giustina Vietri for useful discussion on spectral analysis of type 1 AGN. 
MP is supported by the Programa Atracci\'on de Talento de la Comunidad de Madrid via grant 2018-T2/TIC-11715. MP, SA, CTC and LC acknowledge support from the Spanish Ministerio de Econom\'ia y Competitividad through the grant ESP2017-83197-P, and PID2019-106280GB-I00. 
MPS and IL acknowledge support from the Comunidad de Madrid through the Atracci\'on de Talento Investigador Grant 2018-T1/TIC-11035 and PID2019-105423GA-I00 (MCIU/AEI/FEDER,UE). H{\"U} gratefully acknowledges support by the Isaac Newton Trust and by the Kavli Foundation through a Newton-Kavli Junior Fellowship. LC acknowledges financial support from Comunidad de Madrid under Atracci\'on de Talento grant 2018-T2/TIC-11612 and the Spanish Ministerio de Ciencia, Innovaci\'on y Universidades through grant PGC2018- 093499-B-I00. 
EB acknowledges support from Comunidad de Madrid through the Attracci\'on de Talento grant 2017-T1/TIC-5213. 
SC acknowledge financial support from the State Agency for Research of the Spanish MCIU through the \lq\lq Center of Excellence Severo Ochoa\rq\rq \ award to the Instituto de Astrof{\'i}sica de Andaluc{\'i}a (SEV-2017-0709).
ACG acknowledges support from the Spanish Ministerio de Econom\'ia y Competitividad through the grant BES-2016-078214. 
RM acknowledges ERC Advanced Grant 695671 ``QUENCH''  and support by the Science and Technology Facilities Council (STFC).
JPL acknowledges financial support by the Spanish MICINN under grant AYA2017-85170-R.

\end{acknowledgements}

\begin{appendix}

\section{Multi-component Gaussian fit results}\label{AHa}

\begin{figure*}[t]
%
\centering
\includegraphics[width=16.cm,trim= 0 20 0 10,clip]{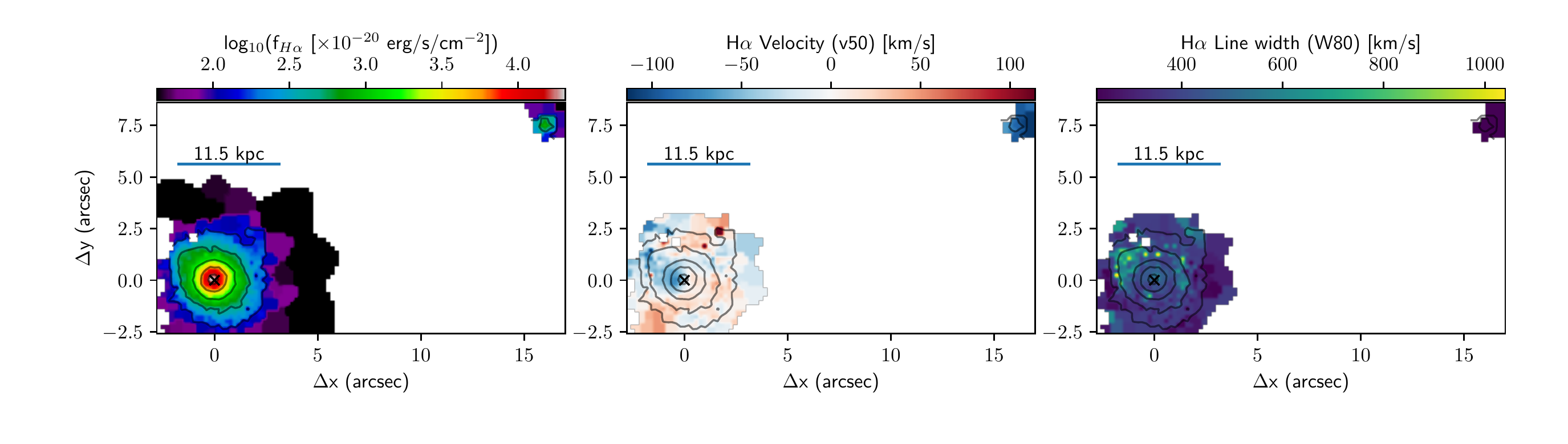}

\caption{\small I00188 maps: \ha integrated flux  (left), \ha centroid ($v50$, centre) and line-width ($W80$, right) obtained from the multi-component Gaussian fit.
The first solid contour is 3$\sigma$ and the jump is 0.5 dex. The cross marks the nucleus. North is up and West is right.
}
\label{I00188linemaps}
\end{figure*}

\begin{figure*}[t]
%
\centering
\includegraphics[width=16.cm,trim= 0 50 0 35,clip]{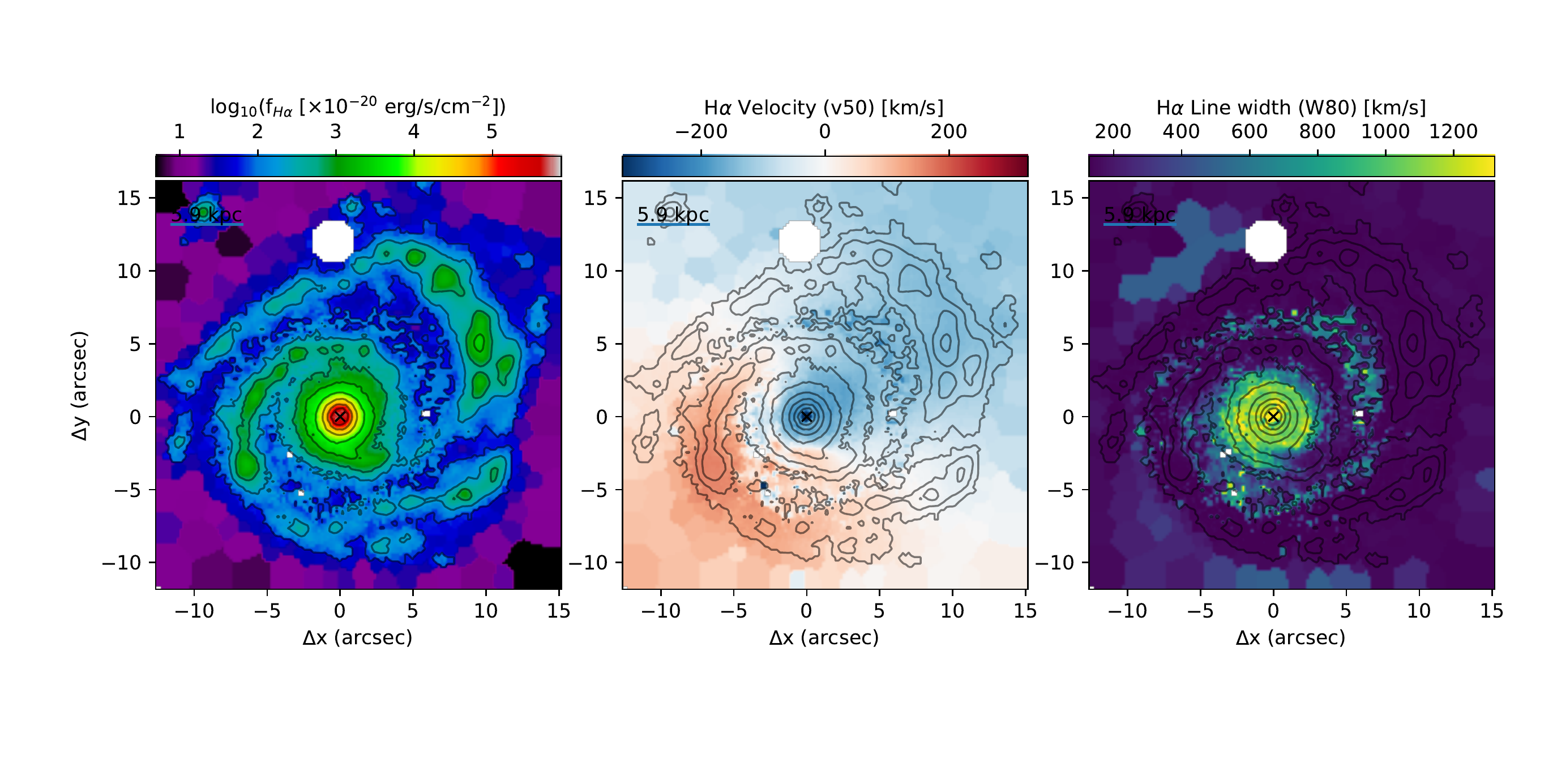}

\caption{\small IZw1 maps. See Fig. \ref{I00188linemaps} for details.
}
\label{izw1linemaps}
\end{figure*}

\begin{figure*}[t]
%
\centering
\includegraphics[width=16.cm,trim= 0 15 0 30,clip]{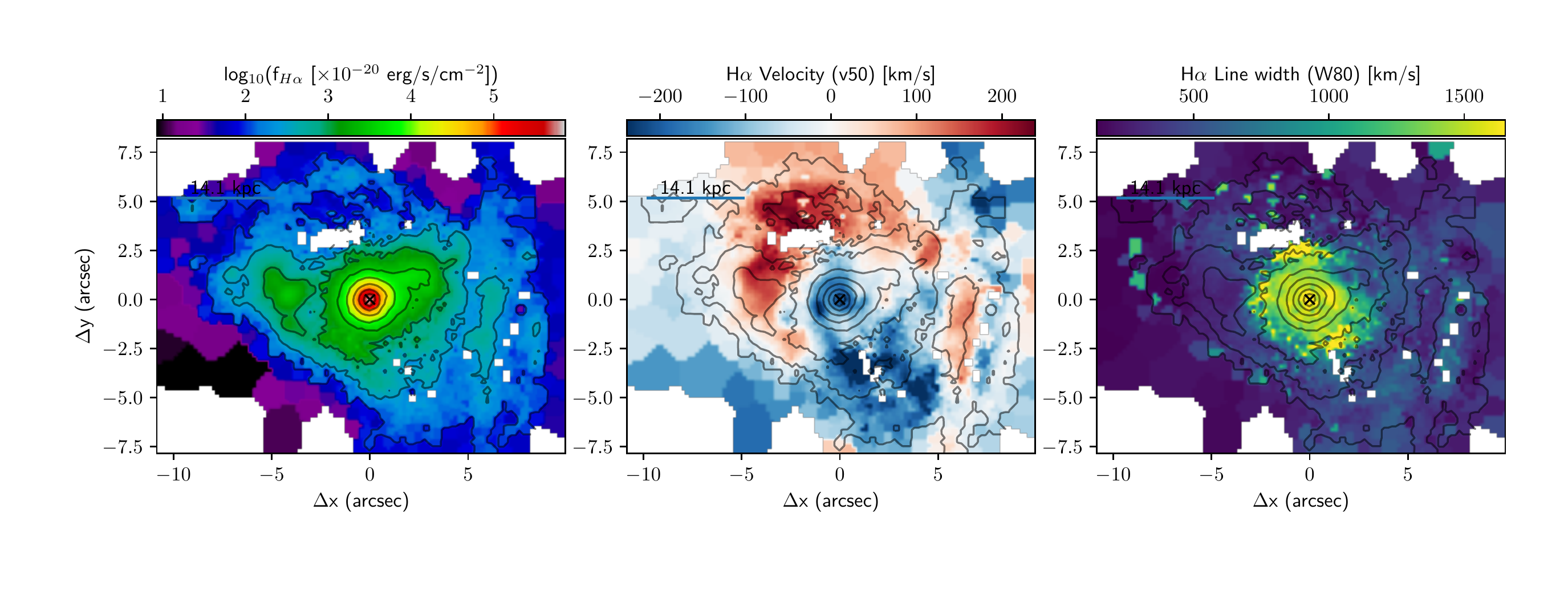}

\caption{\small I01572 maps. See Fig. \ref{I00188linemaps} for details.
}
\label{mrk1014linemaps}
\end{figure*}

\begin{figure*}[t]
%
\centering
\includegraphics[width=16.cm,trim= 0 80 0 65,clip]{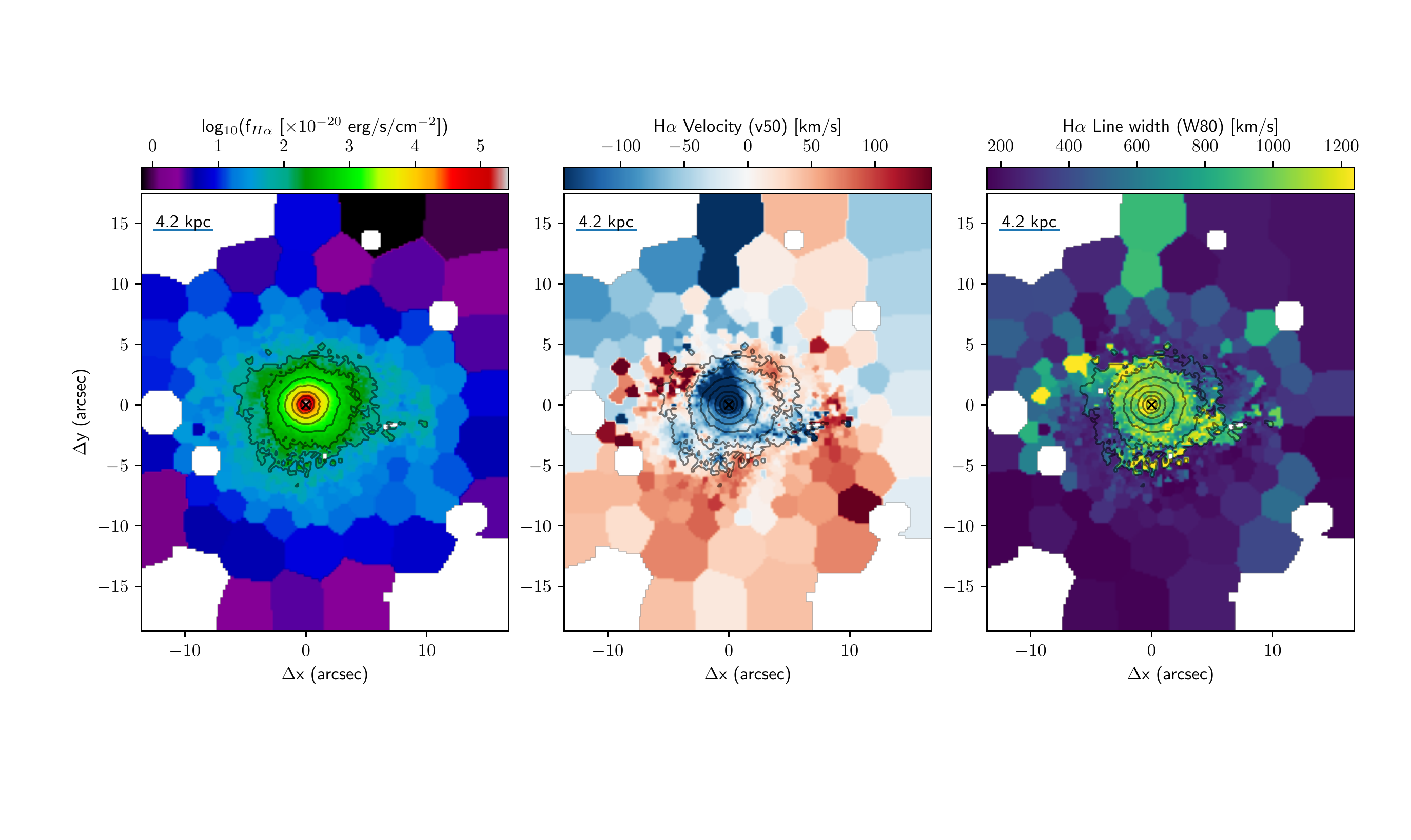}

\caption{\small I0589 maps. See Fig. \ref{I00188linemaps} for details.
}
\label{I0589linemaps}
\end{figure*}

\begin{figure*}[t]
%
\centering
\includegraphics[width=16.cm,trim= 0 20 0 35,clip]{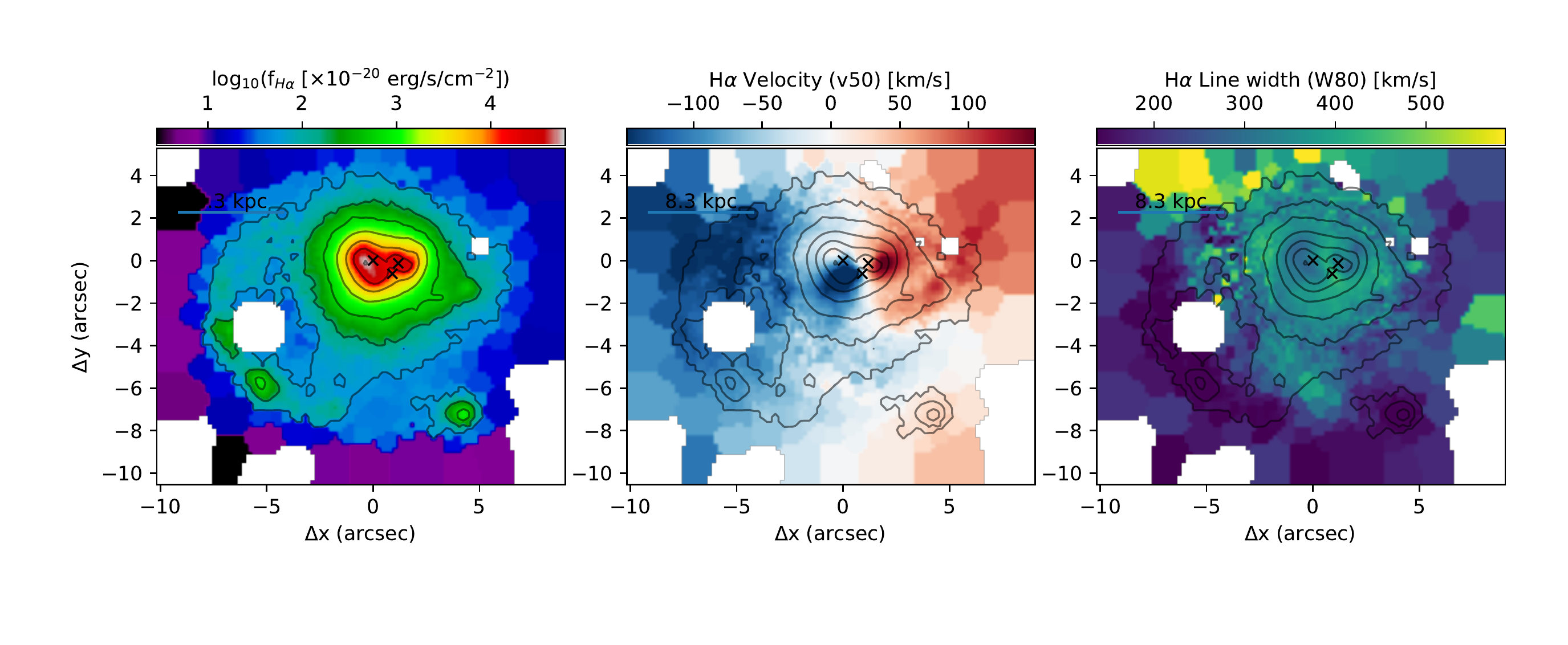}

\caption{\small I07251 maps. See Fig. \ref{I00188linemaps} for details.
}
\label{I07251linemaps}
\end{figure*}

\begin{figure*}[t]
%
\centering
\includegraphics[width=16.cm,trim= 0 50 0 25,clip]{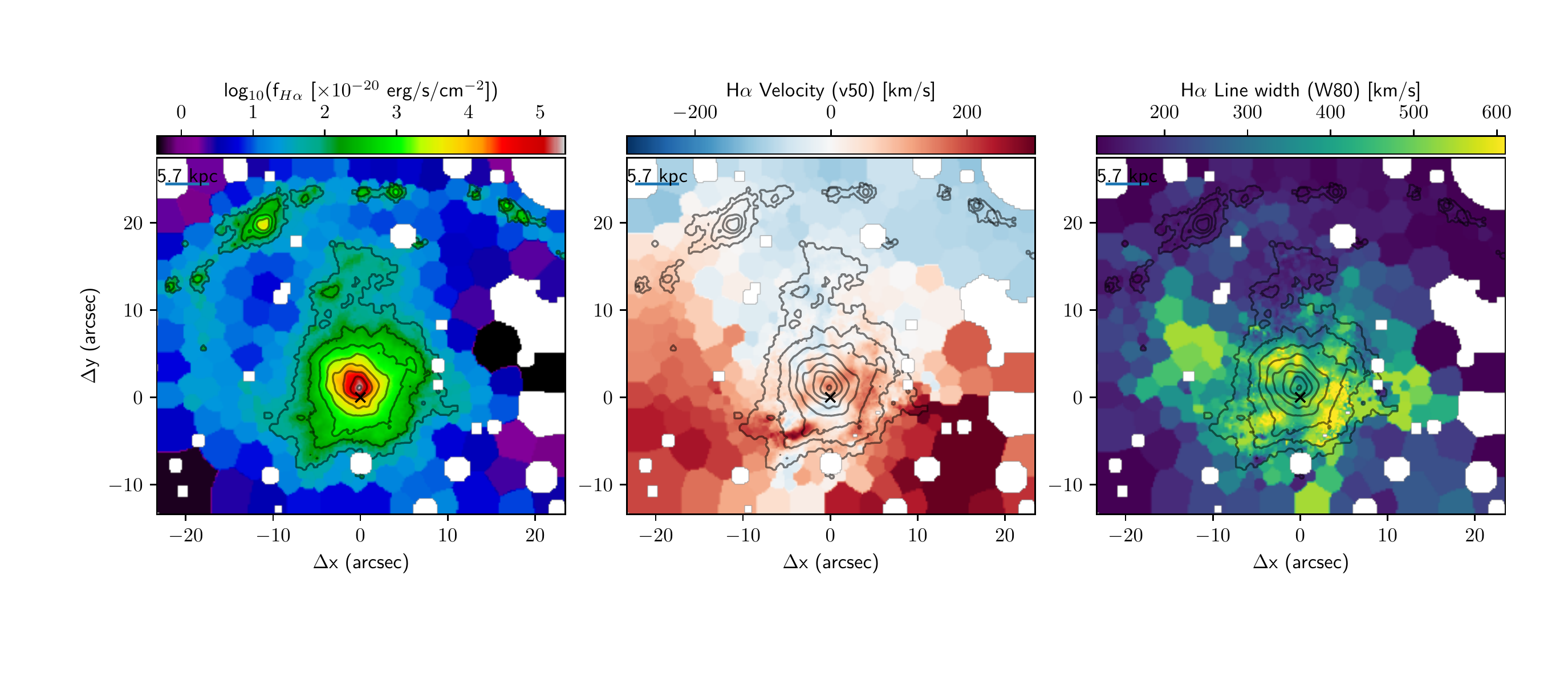}

\caption{\small I09022 maps. See Fig. \ref{I00188linemaps} for details.
}
\label{I090221linemaps}
\end{figure*}

\begin{figure*}[t]
%
\centering
\includegraphics[width=16.cm,trim= 0 45 0 28,clip]{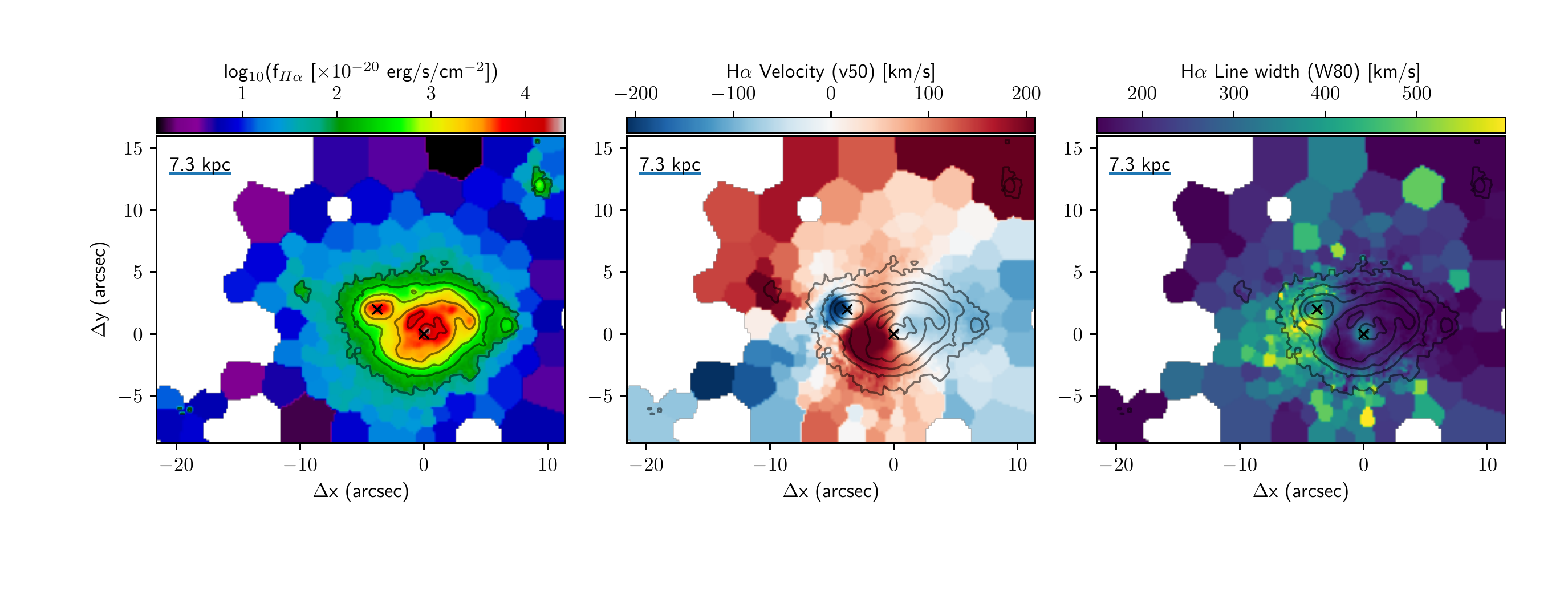}

\caption{\small I10190 maps. See Fig. \ref{I00188linemaps} for details.
}
\label{I10190linemaps}
\end{figure*}

\begin{figure*}[t]
%
\centering
\includegraphics[width=16.cm,trim= 0 45 0 35,clip]{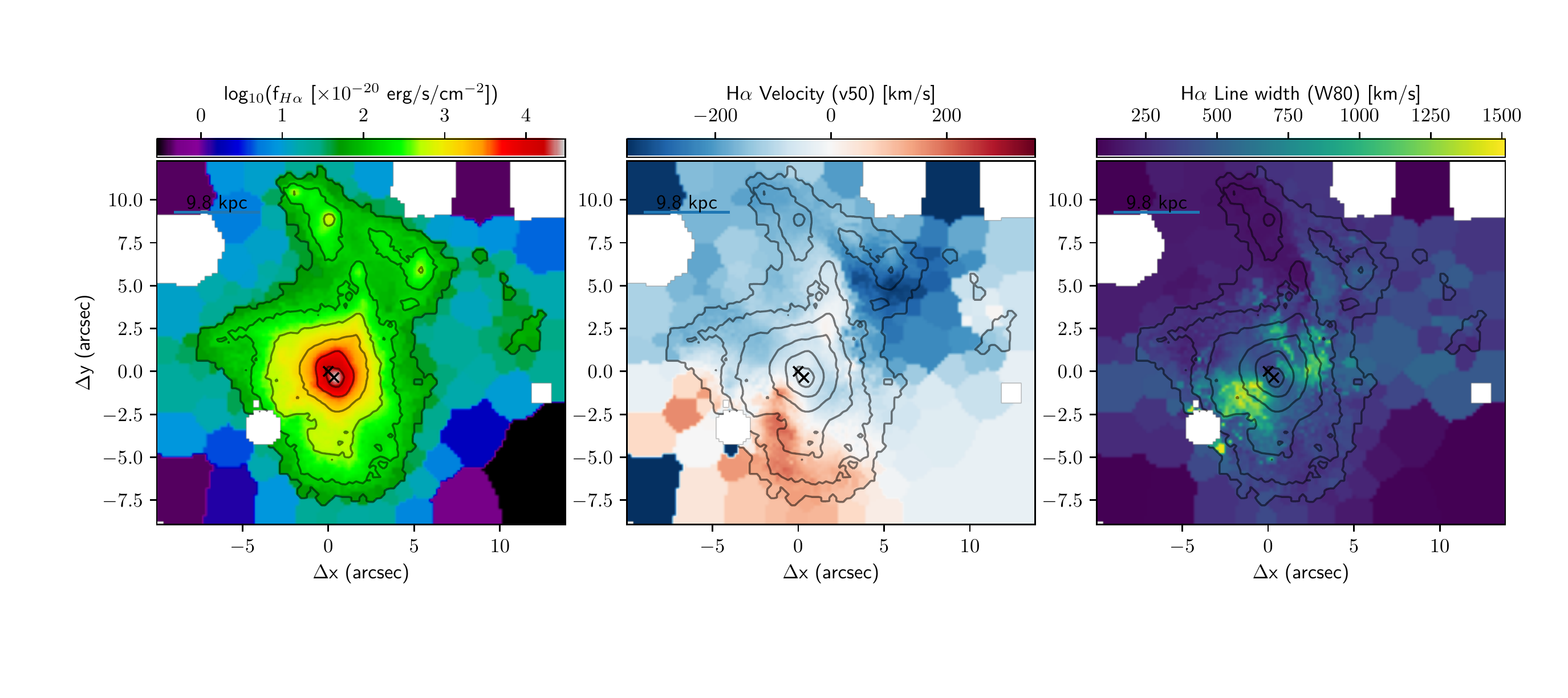}

\caption{\small I11095 maps. See Fig. \ref{I00188linemaps} for details.
}
\label{I11095linemaps}
\end{figure*}

\begin{figure*}[t]
%
\centering
\includegraphics[width=16.cm,trim= 0 70 0 65,clip]{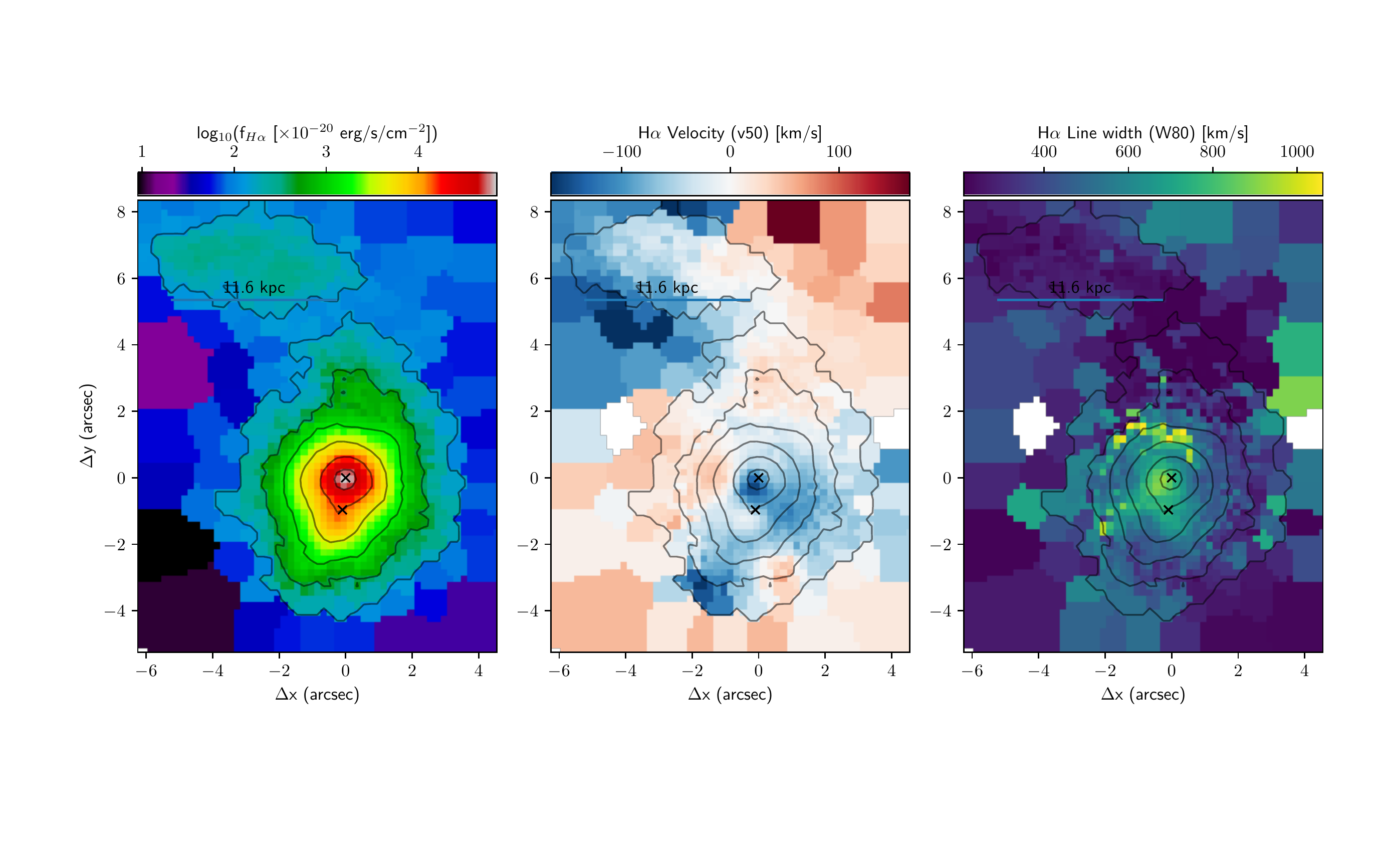}

\caption{\small I12072 maps. See Fig. \ref{I00188linemaps} for details.
}
\label{I12072linemaps}
\end{figure*}

\begin{figure*}[t]
%
\centering
\includegraphics[width=16.cm,trim= 0 30 0 25,clip]{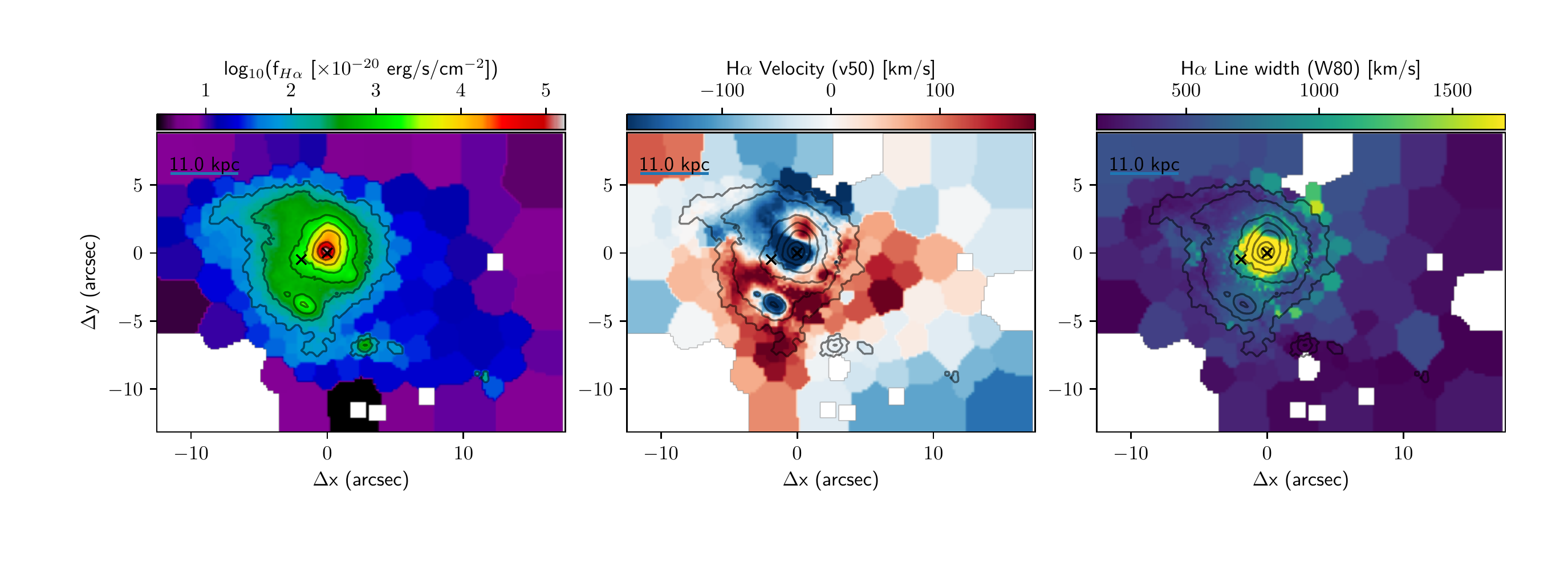}

\caption{\small I13451 maps. See Fig. \ref{I00188linemaps} for details.
}
\label{I13451linemaps}
\end{figure*}

\begin{figure*}[t]
%
\centering
\includegraphics[width=16.cm,trim= 0 56 0 55,clip]{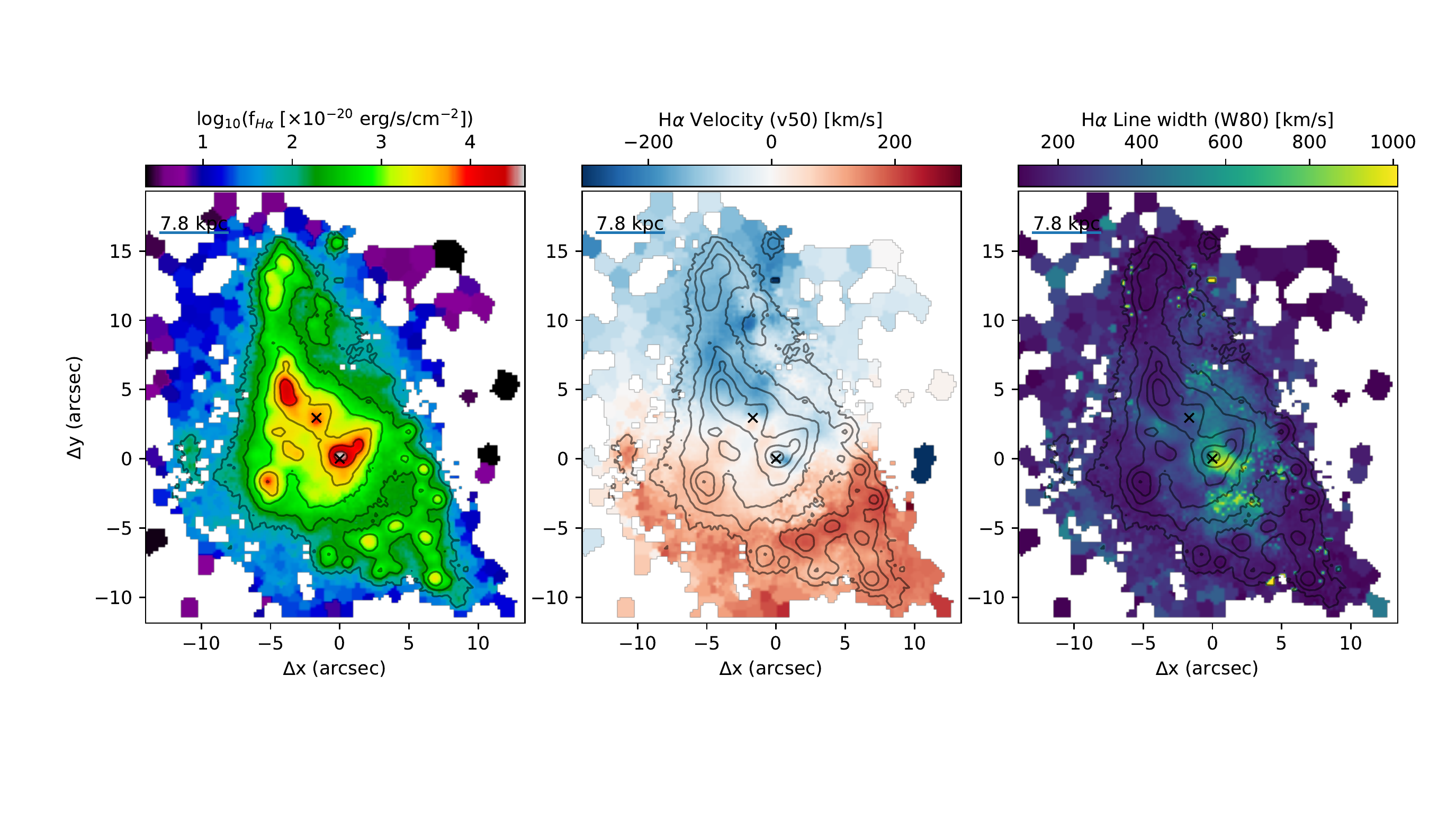}

\caption{\small I14348 maps. See Fig. \ref{I00188linemaps} for details.
}
\label{I14348linemaps}
\end{figure*}

\begin{figure*}[t]
%
\centering
\includegraphics[width=16.cm,trim= 0 26 0 25,clip]{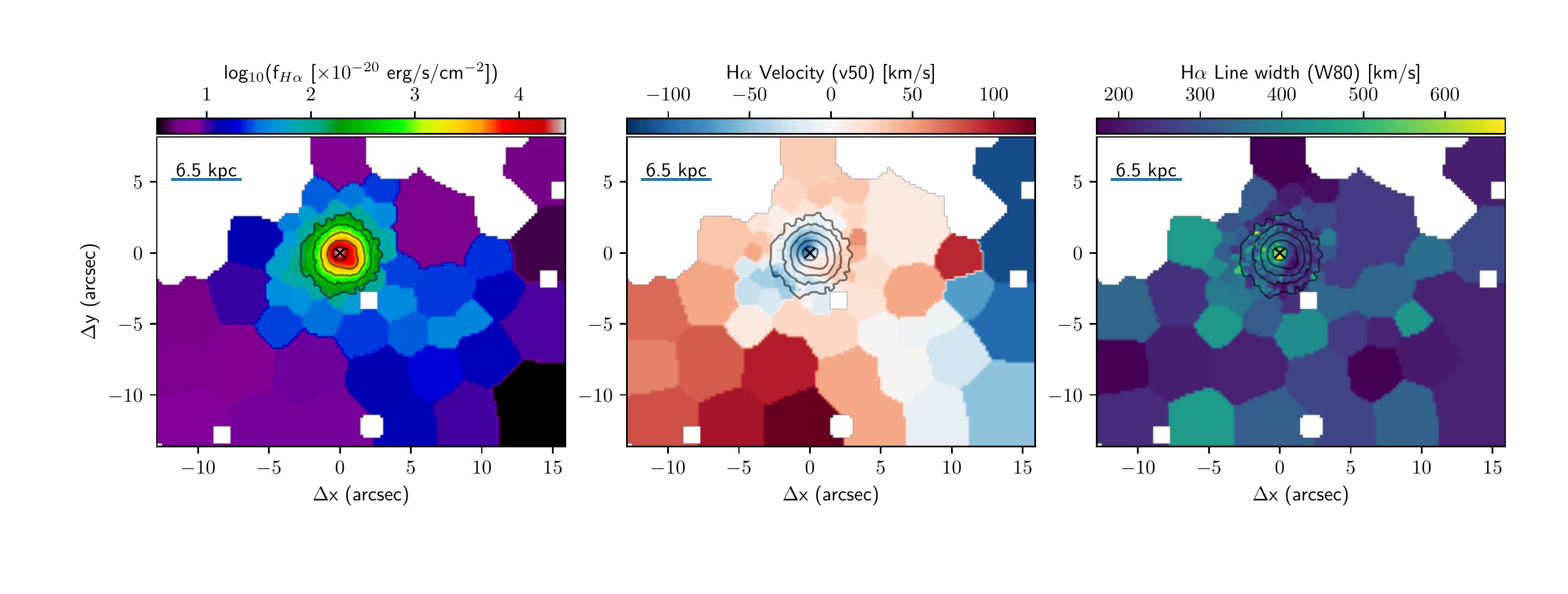}

\caption{\small I14378 maps. See Fig. \ref{I00188linemaps} for details.
}
\label{I14378linemaps}
\end{figure*}

\begin{figure*}[t]
%
\centering
\includegraphics[width=16.cm,trim= 0 60 0 45,clip]{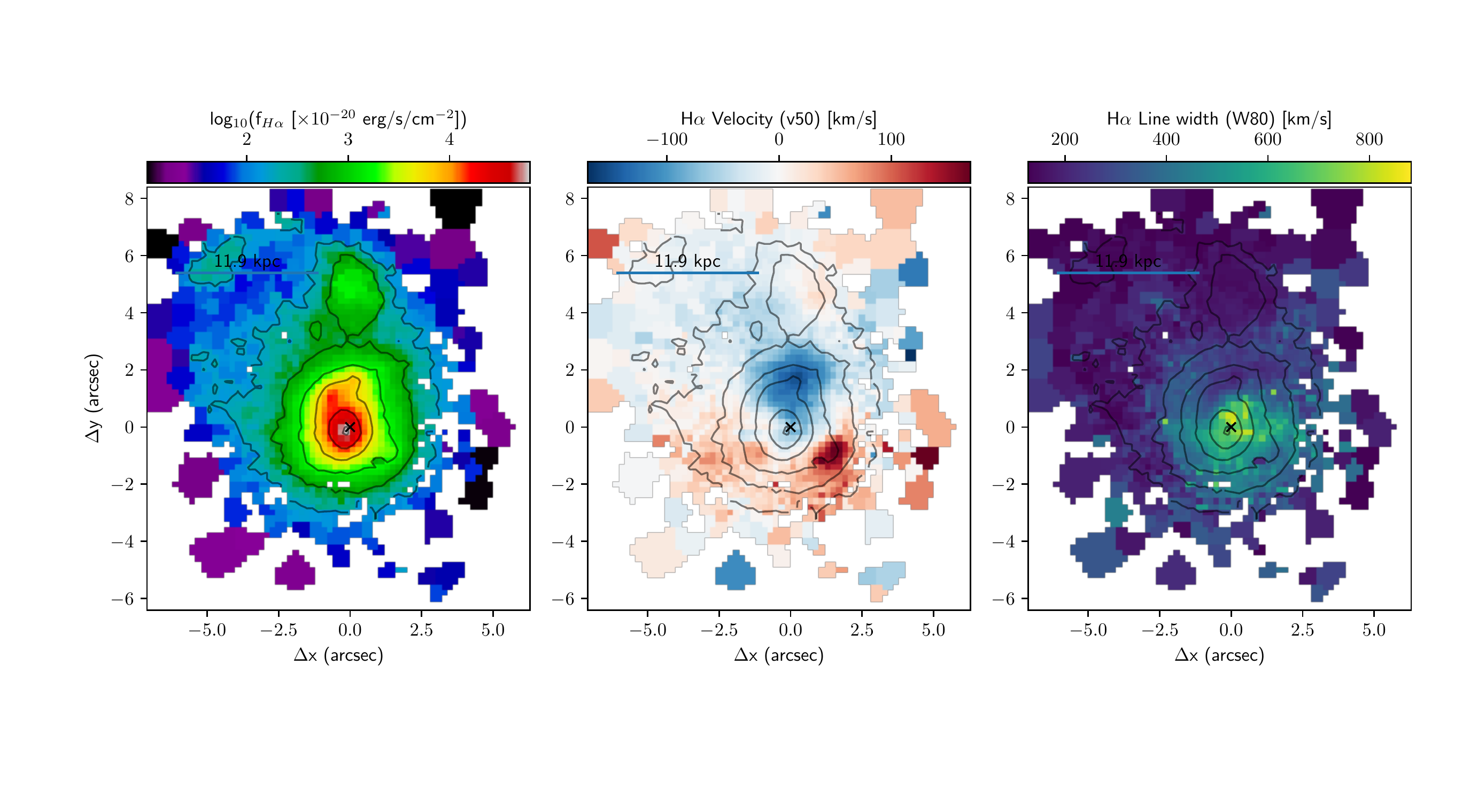}

\caption{\small I16090 maps. See Fig. \ref{I00188linemaps} for details.
}
\label{I16090linemaps}
\end{figure*}

\begin{figure*}[t]
%
\centering
\includegraphics[width=16.cm,trim= 0 60 0 45,clip]{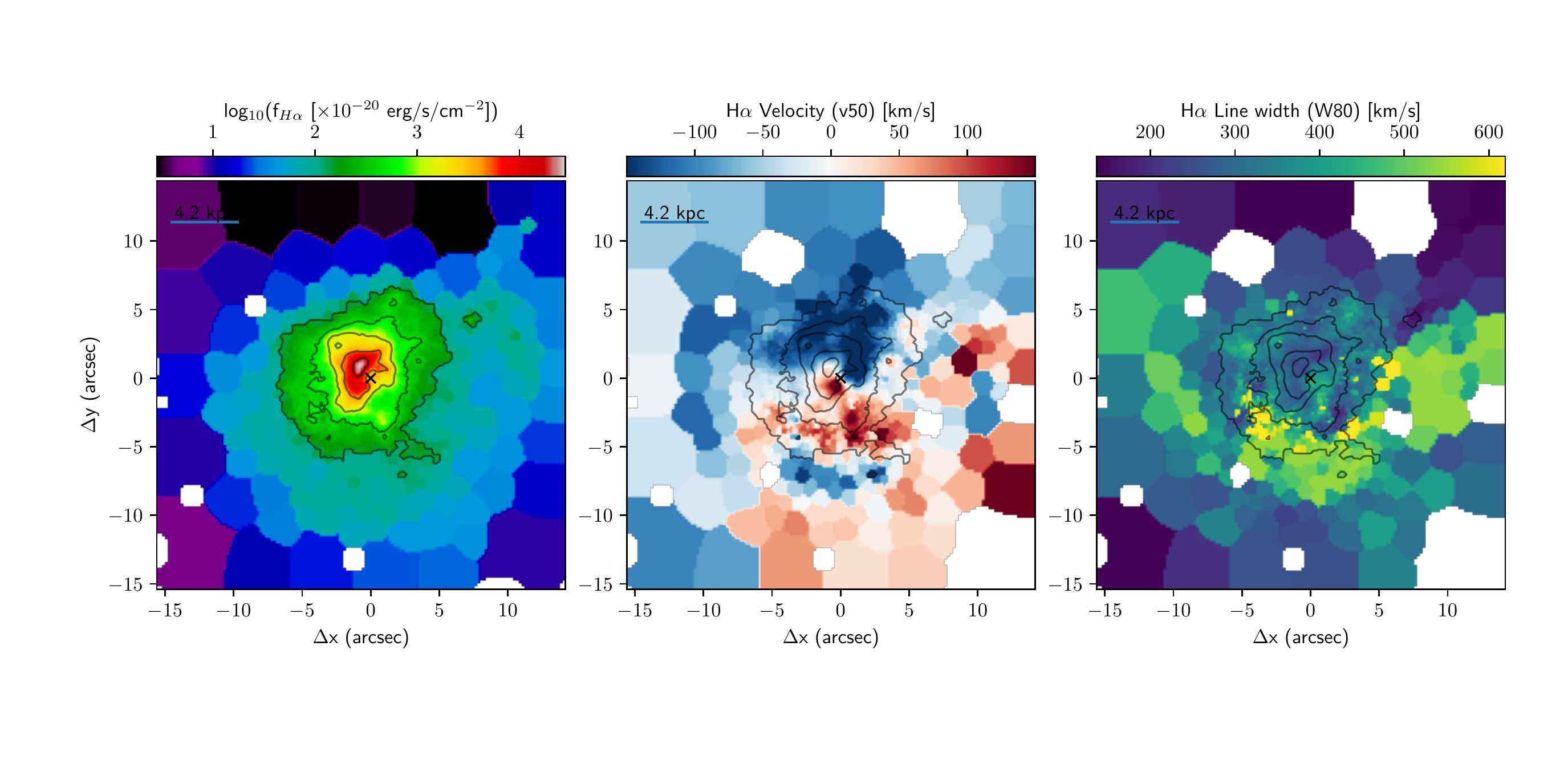}

\caption{\small I17208 maps. See Fig. \ref{I00188linemaps} for details.
}
\label{I17208linemaps}
\end{figure*}

\begin{figure*}[t]
%
\centering
\includegraphics[width=16.cm,trim= 0 50 0 45,clip]{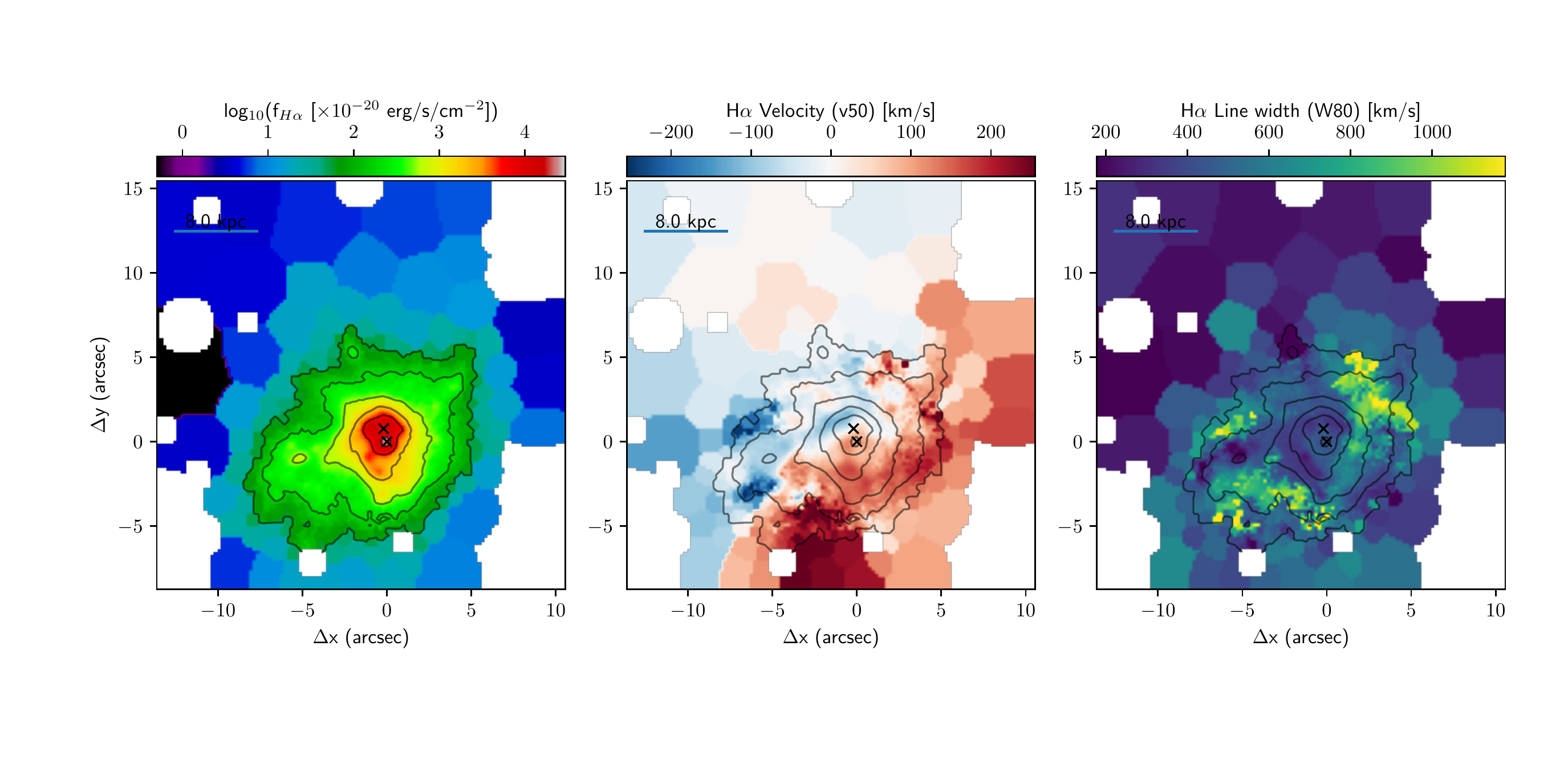}

\caption{\small I19297 maps. See Fig. \ref{I00188linemaps} for details.
}
\label{I19297linemaps}
\end{figure*}

\begin{figure*}[t]
%
\centering
\includegraphics[width=16.cm,trim= 0 70 0 55,clip]{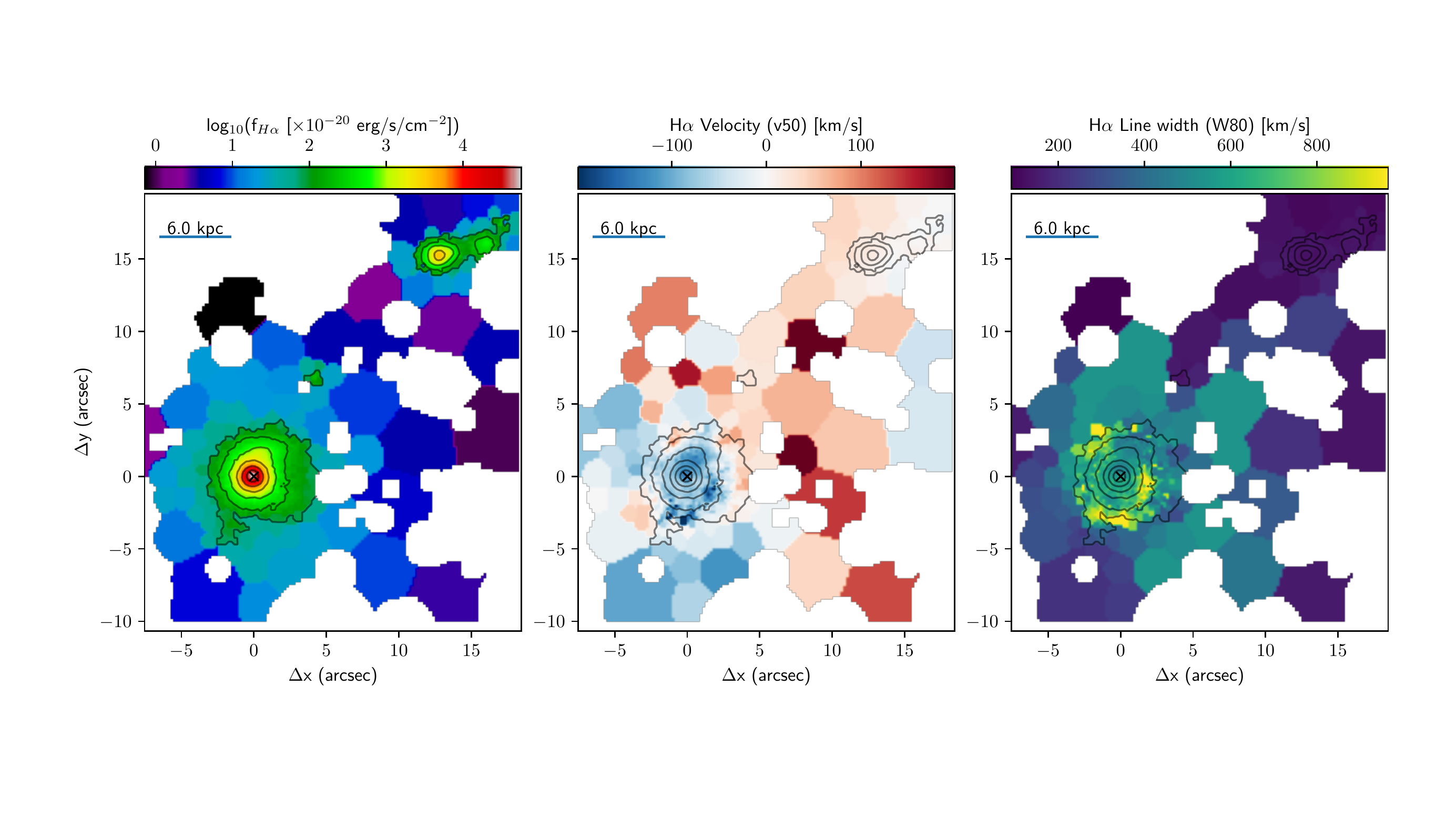}

\caption{\small I19542 maps. See Fig. \ref{I00188linemaps} for details.
}
\label{I19542linemaps}
\end{figure*}

\begin{figure*}[t]
%
\centering
\includegraphics[width=16.cm,trim= 0 30 0 35,clip]{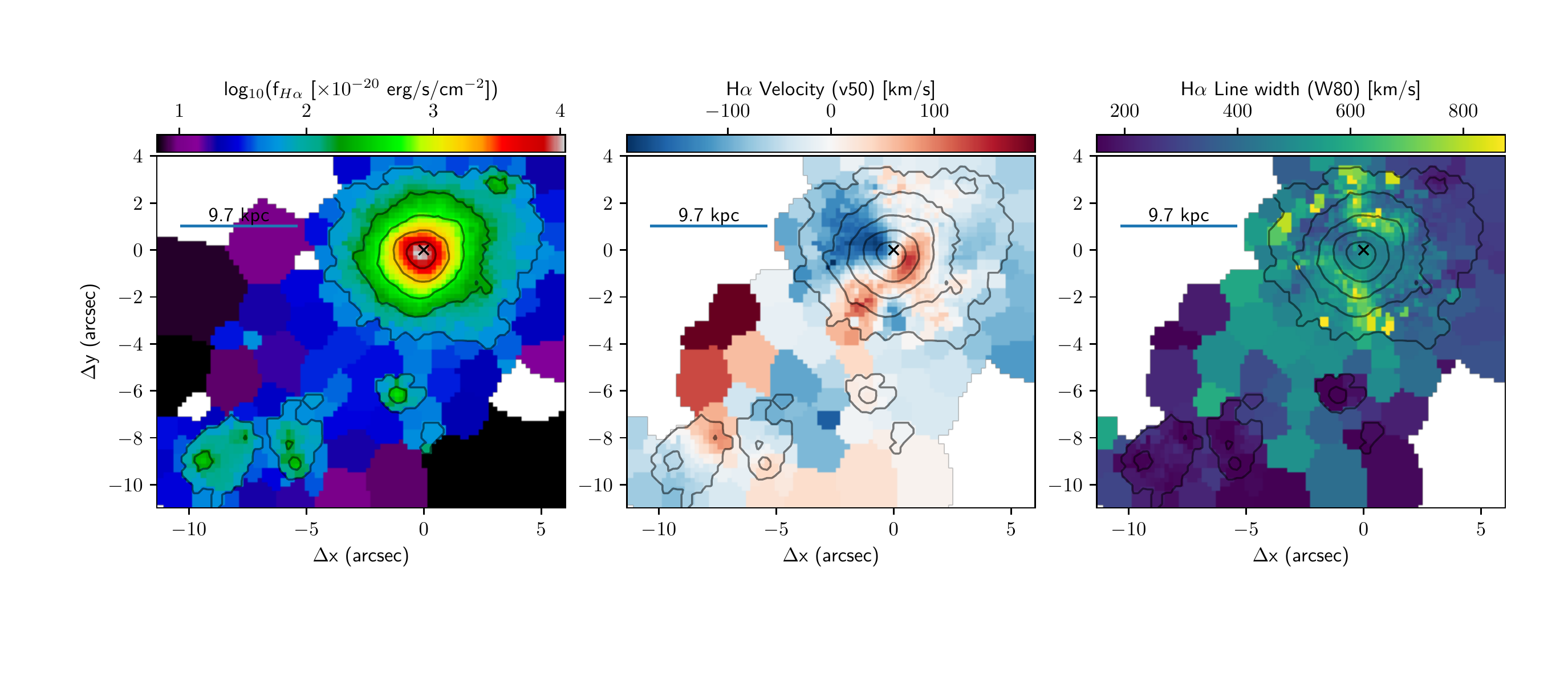}

\caption{\small I20087 maps. See Fig. \ref{I00188linemaps} for details.
}
\label{I20100linemaps}
\end{figure*}

\begin{figure*}[t]
%
\centering
\includegraphics[width=16.cm,trim= 0 30 0 25,clip]{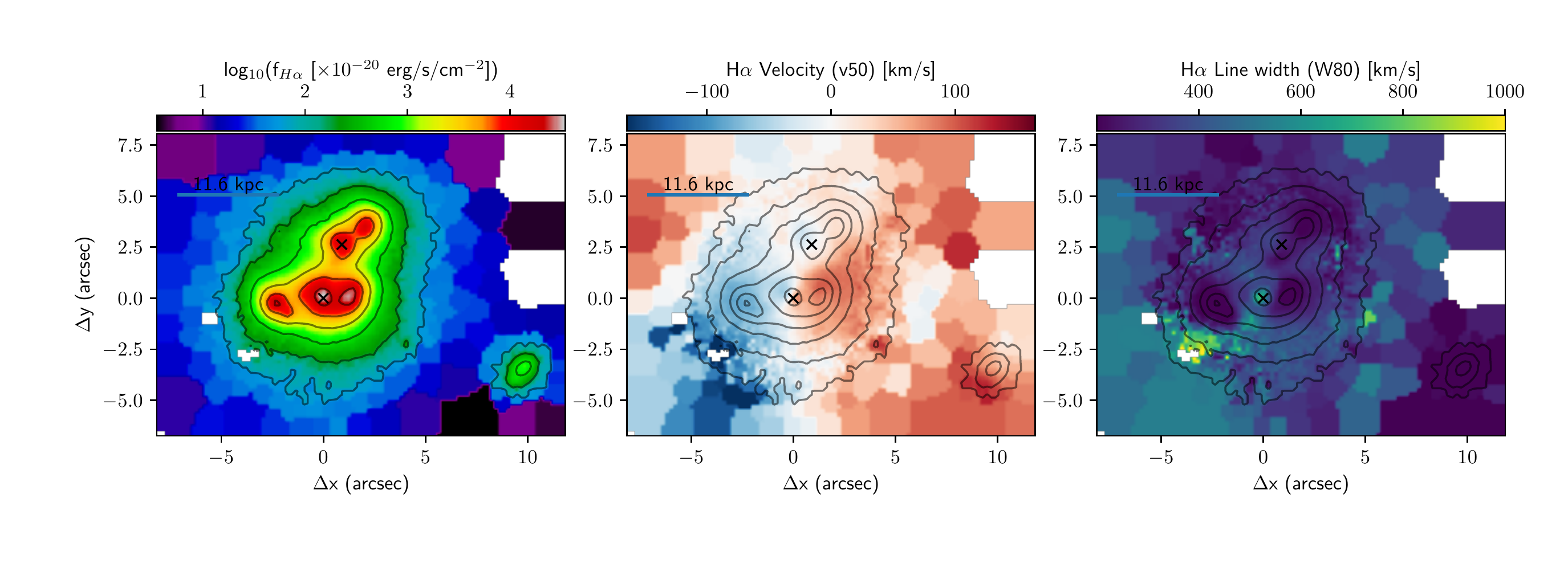}

\caption{\small I20100 maps. See Fig. \ref{I00188linemaps} for details.
}
\label{I20100linemaps}
\end{figure*}

\begin{figure*}[t]
%
\centering
\includegraphics[width=16.cm,trim= 0 50 0 45,clip]{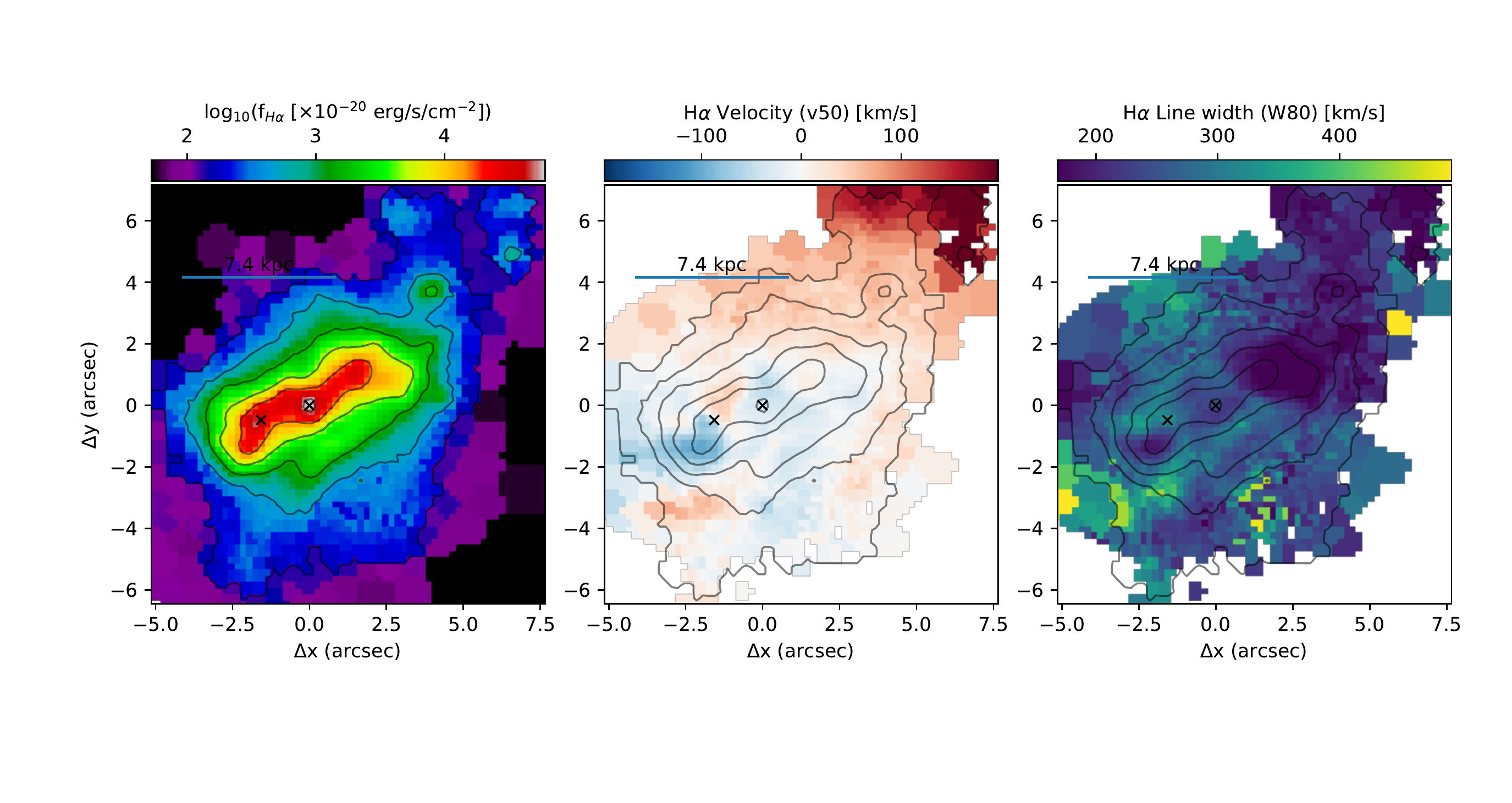}

\caption{\small I22491 maps. See Fig. \ref{I00188linemaps} for details.
}
\label{I22491linemaps}
\end{figure*}

\section{Position-Velocity diagrams}\label{APV}
In Fig. \ref{PV_all} we show the comparison between the (total and narrow) \ha and stellar velocities along their kinematic major axis, for all targets for which we can observe a clear velocity gradient (and measure a PA$^{kin}$)  together with a peak in the velocity dispersion diagram at the position of the nucleus for at least one component (i.e. gas or stars). The nuclear positions are inferred from registered HST/F160W images (\citealt{Perna2021}). The PV plots show only a small portion of the total extension of the ULIRG systems, in order to exclude the contribution from tidal tails, extended outflows or second nuclei, and better identify rotation-like signatures.

The major axis PA$^{kin}$ measurements have been obtained with the python PaFit package (\citealt{Krajnovic2006}), for both stellar and gas kinematics (columns 5 and 6 in Table \ref{Tproperties}). The difficulty in measuring reliable PA$^{kin}$ and obtaining regular velocity profiles, led to the exclusion of the following systems: I07251 E and W, I09022, 
I11095, I12072 S, I13451 E and W, 19297 S, 20100 NW, I22491 E and W.

For the sources for which the gas and stellar major axes agree within the errors, the PV diagrams have been extracted considering a PA$^{kin}$ equal to the weighted mean of PA$^{kin}_{gas}$ and PA$^{kin}_*$, with the weighting factors equal to the inverse of the quadratic uncertainty on the PA$^{kin}$ measurements; for those sources for which no PA$^{kin}$ measurement can be obtained for the gas, we considered the PA$^{kin}_*$ (I01572, I05189, I19297 N, I19542); vice versa, PA$^{kin}_{gas}$ has been chosen when the stars do not show a clear velocity gradient (I07251 E, I22491 E). Finally, for those sources with different stellar and gas kinematics (I10190 E, I14348 NE, I16090, I20087 and I20100 SE)  we extracted the velocity profiles  along different PAs.

Before briefly discussing the PV diagrams of individual targets shown in the figure, we note that a clear but shallow velocity gradient (i.e. with $\delta v \ll 100$ \kms) could translate in a flat $\sigma$ profile under the MUSE observing conditions (i.e. with an angular resolution $\approx 0.6^{\prime\prime}$). This would determine the exclusion of disk-like candidates, according to two selection criteria mentioned at the beginning of this section. However, most of PUMA individual systems show $\delta v \gtrsim 100$ \kms (see Table \ref{Tproperties}); the few systems with lower $\delta v$ (in particular, the gas component in I00188, I22491 E, I14348 SW and I14378) also present disturbed kinematics, excluding the presence of disks even in these conditions.

Notes on individual targets in Fig. \ref{PV_all}: 

{\bf I00188} shows a regular gradient in $V_{*}$ and a peak in $\sigma_{*}$ at the position of the nucleus, as expected for rotation-dominated kinematics; on the other hand, the narrow \ha velocity profiles is irregular, while its velocity dispersion is close to $\sim 110$ km/s along the entire extension of the major axis, indicating highly perturbed gas kinematics. 

{\bf IZw1} The \ha in the  central pixel is saturated; this translates in a gap in the PV slices shown in the figure. The presence of strong Sy1 emission also prevents us to correctly infer proper $V_*$ and $\sigma_*$ measurements in the vicinity of the nucleus; in particular, the $\sigma_*$ values might be strongly overestimated. 

{\bf I01572} As for IZw1, the presence of a strong Sy1 prevents us to properly infer stellar velocity and velocity dispersion in the vicinity of the nucleus. Moreover, the gas kinematics are strongly disturbed by the nuclear outflow and tidal motions. 

{\bf I05189} shows a clear gradient in $V_{*}$ and a peak in $\sigma_{*}$ at the position of the nucleus, as expected for rotation-dominated kinematics; on the other hand, the narrow \ha velocity profiles is irregular,  while its velocity dispersion is close to $\sim 110$ km/s along the entire extension of the major axis, indicating highly perturbed gas kinematics. 

{\bf I07251} presents disk-like motions, but with a kinematic centre not coincident with the position of either the two nuclei of this interacting system (see also Fig. \ref{BBaroloI07251}, top panels). The PV diagrams presented in Fig. \ref{APV} have been therefore extracted considering the kinematic centre obtained from the 3D-Barolo analysis. 


{\bf I10190}. This system shows two rotating disks associated with the two nuclei, separated by $\sim 7.2$ kpc.  However, both the stellar and gas kinematics in the vicinity of the E nucleus are strongly affected by the presence of the W system motions. 


{\bf I12072} This  source  presents  two  nuclei,  at  a  projected distance  of  2.3  kpc.  Disk-like  motions  are  presumably  associated  with  the  N nucleus. Broadly regular PV diagrams are observed for both stellar and gas components. 

{\bf I13120} This target is extensively presented in the main text of this work. 


{\bf I14348} is an interacting ULIRG with two nuclei, at a projected distance of $\sim 5$ kpc. The NE nucleus presents broadly regular PV diagrams for both gas and stellar components; the SW nucleus presents broadly regular stellar kinematics, and more irregular gas motions due to the presence of a strong outflow. 

{\bf I14378} shows a clear gradient in $V_{*}$ and a peak in $\sigma_{*}$ at the position of the nucleus, as expected for rotation-dominated kinematics; on the other hand, the narrow \ha velocity profiles is irregular, and  $\sigma_{gas} \approx 90$ km/s along the entire extension of the major axis indicates highly perturbed gas kinematics. 

{\bf Arp220} shows a broadly regular velocity gradient in both gas and stellar components (see also \citealt{Scoville1997}). However, the presence of the two nuclei with distinct rotation features on scales of $\sim 100$ pc (at a distance of $\sim 370$ pc),  and a kpc-scale,  wide-angle outflow prevents a detailed characterisation of the gas kinematics in this system (see e.g. Fig. 19 in \citealt{Perna2020}). For completeness, in Fig. \ref{PV_all} we report the PV plots extracted along a PA$^{kin} \sim 40^{\circ}$, with respect to the position of the two nuclei. 

{\bf I16090} shows broadly regular gradient in $V_{*}$ and $V_{gas}$, and a peak in $\sigma_{*}$ at the position of the nucleus, as expected for rotation-dominated kinematics; on the other hand, $\sigma_{gas} \approx 130$ km/s along the entire extension of the major axis indicates highly perturbed gas kinematics. 

{\bf I17208} shows broadly regular velocity profiles in both stellar and gas components; the velocity dispersion profiles are instead more irregular, with higher values toward the south-east direction, probably due to streaming motions (see also Fig. \ref{BBaroloI17208}).  

{\bf I19297} No evidence of disk-like motions are present in the vicinity of the two nuclei of this system. The very high gas velocity dispersion ($> 100$ km/s) suggests the presence of strongly disturbed kinematics. 

{\bf I19542} shows a clear gradient in $V_{*}$ and a peak in $\sigma_{*}$ at the position of the nucleus, as expected for rotation-dominated kinematics. On the other hand, the narrow \ha velocity profiles is irregular, mostly associated with blueshifted emission, and the $\sigma_{H\alpha} \approx 110$ km/s along the entire extension of the major axis indicates highly perturbed gas kinematics. 

{\bf I20087} shows a clear gradient in $V_{*}$ and a peak in $\sigma_{*}$ at the position of the nucleus, as expected for rotation-dominated kinematics. The gas velocities broadly resemble the $V_{*}$ profile, but reaching maximum velocities at $\sim 2$ kpc from the nuclear position a factor of $\sim 2.6$ higher than $V_{*}$. This behaviour might be due to the presence of a bi-conical outflow, and will be better investigated in a forthcoming paper.  

{\bf 20100 SE} shows broadly regular velocity profiles in both stellar and gas components. We however observe a significant misalignment between the gas and stellar major axis PAs. 


\begin{figure*}[t]
%
\centering
\includegraphics[width=9.1cm,trim= 10 0 10 0,clip]{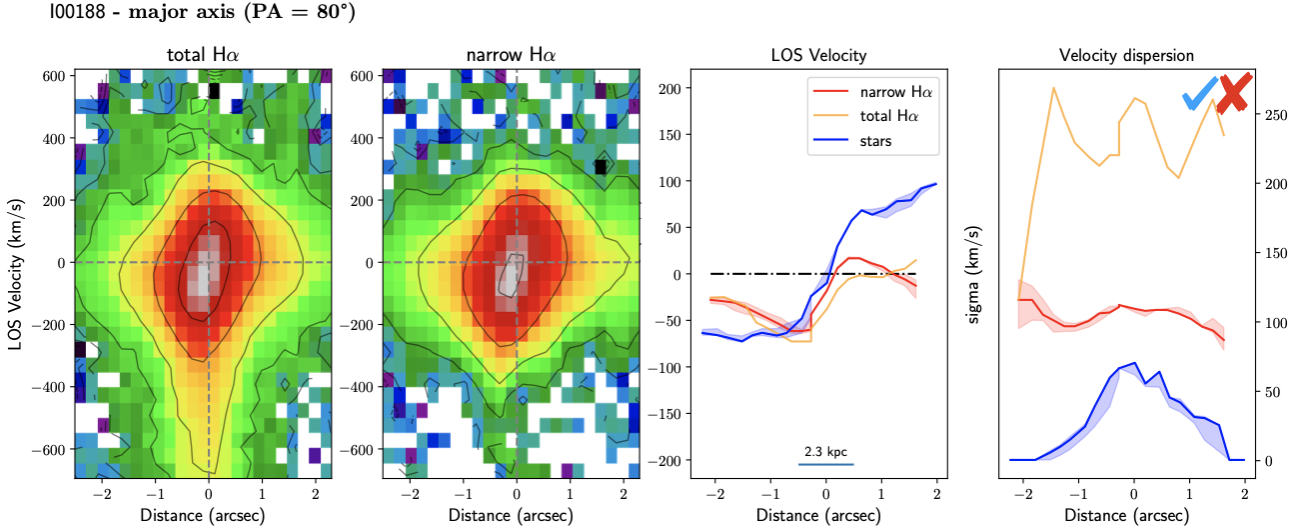}
\includegraphics[width=9.1cm,trim= 10 0 10 0,clip]{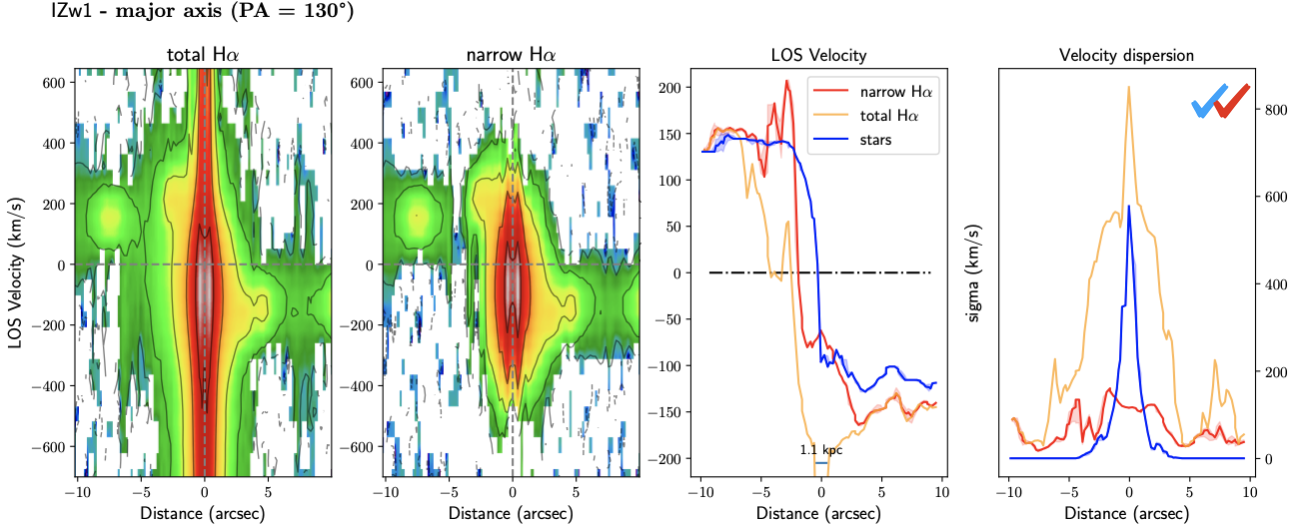}
\includegraphics[width=9.1cm,trim= 10 0 10 0,clip]{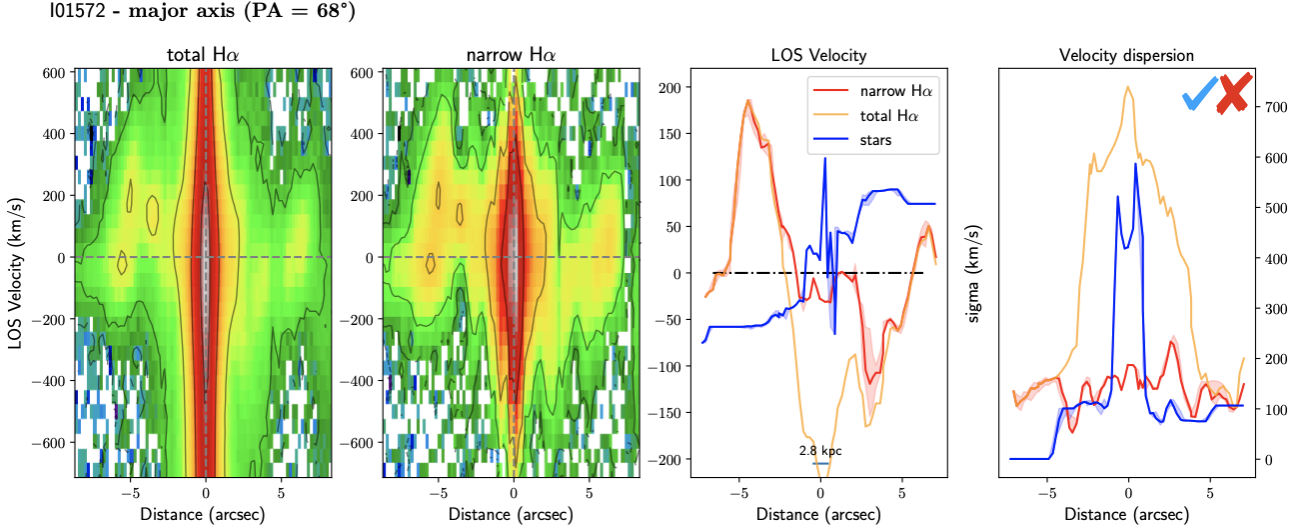}
\includegraphics[width=9.1cm,trim= 10 0 10 0,clip]{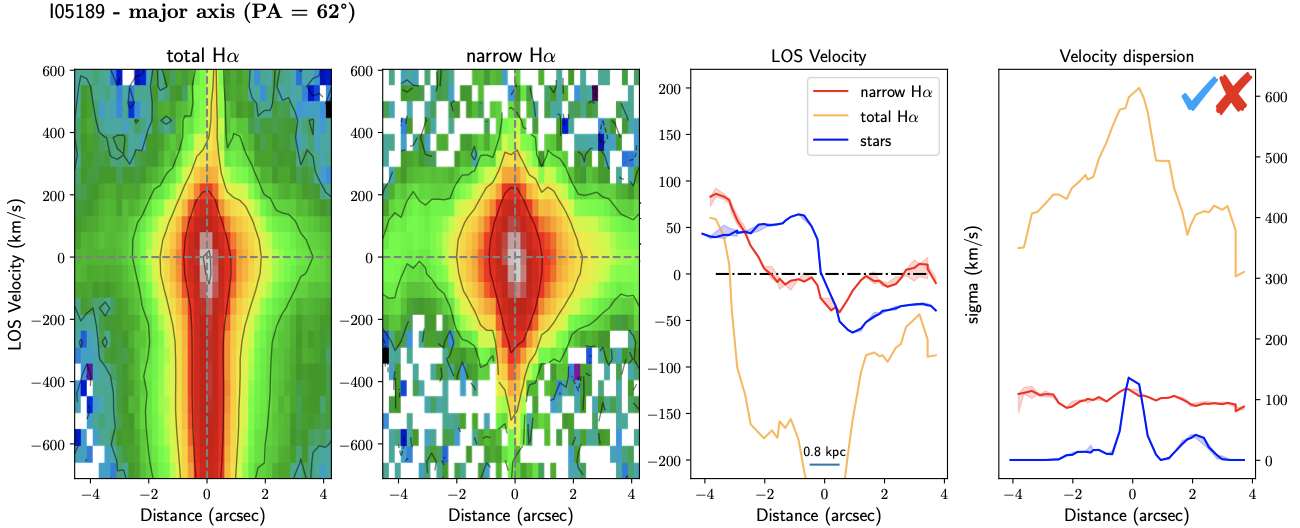}
\includegraphics[width=9.1cm,trim= 10 0 10 0,clip]{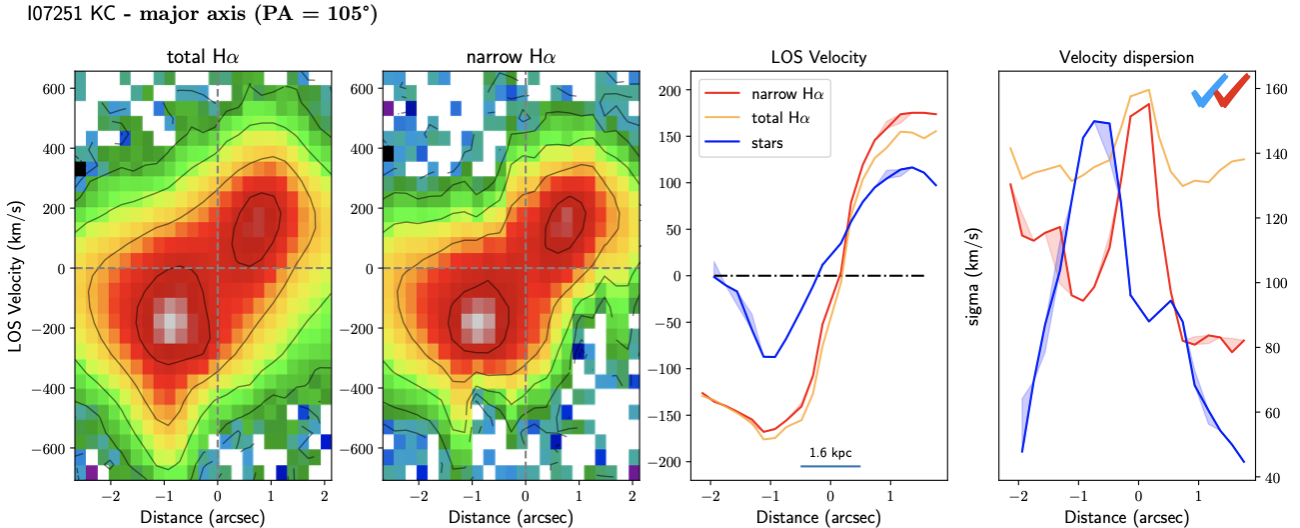}
\includegraphics[width=9.1cm,trim= 10 0 10 0,clip]{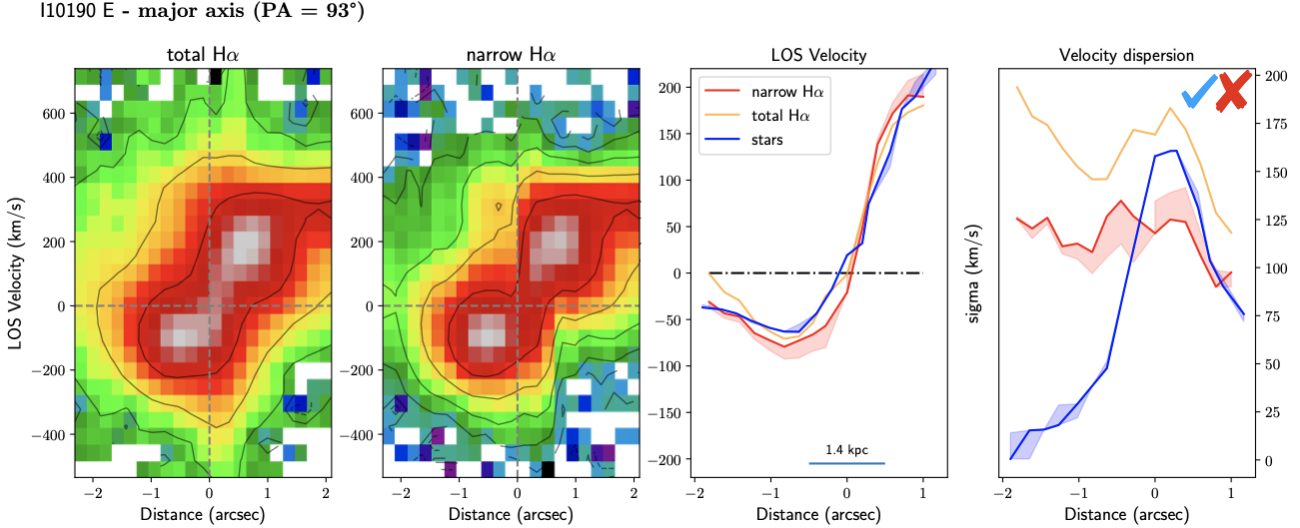}
\includegraphics[width=9.1cm,trim= 10 0 10 0,clip]{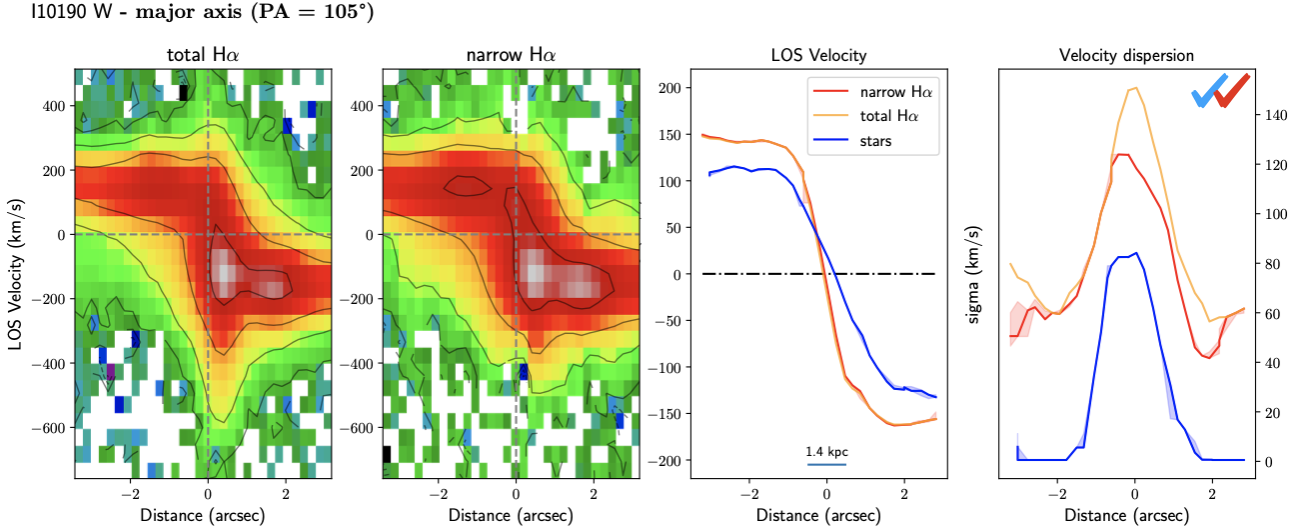}
\includegraphics[width=9.1cm,trim= 10 0 10 0,clip]{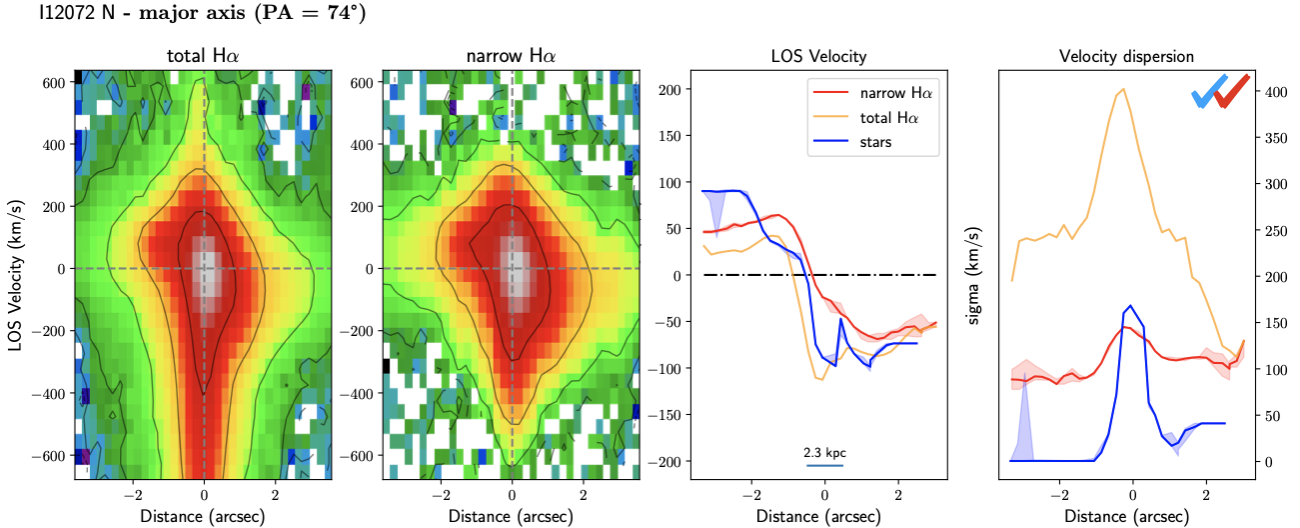}
\includegraphics[width=9.1cm,trim= 10 0 10 0,clip]{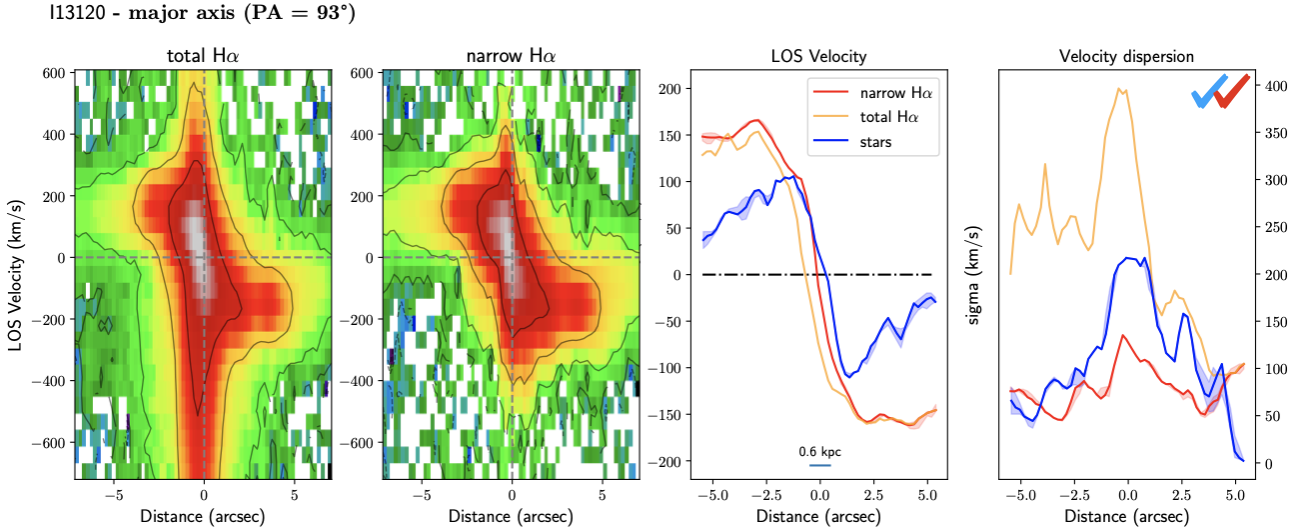}
\includegraphics[width=9.1cm,trim= 10 0 10 0,clip]{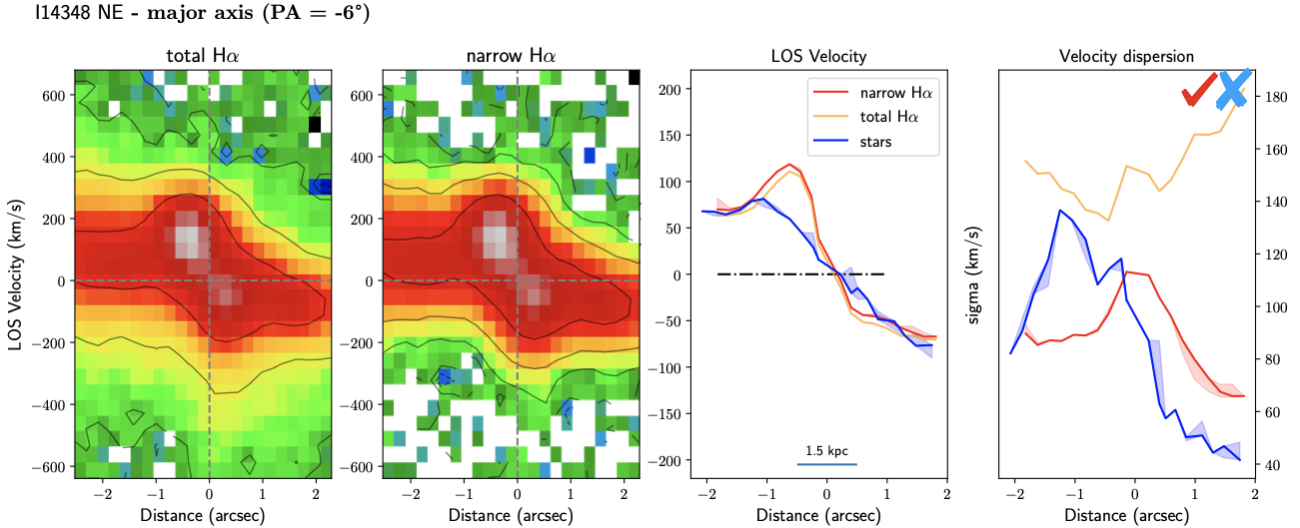}

\caption{\small 
Position-velocity diagrams along the galaxy major axis for all PUMA systems showing i) a well defined velocity gradient along the major axis, and ii) a peak in the velocity dispersion diagram close to the position of the nucleus. These two conditions provide initial evidence for  rotation-dominated kinematics. All systems which gas (stellar) kinematics satisfy the two conditions are marked with a blue (red) check-mark in the top-right corner of the velocity dispersion panel; on the contrary, the systems not satisfying at least one of the two conditions are marked with a cross symbol; more uncertain kinematics are marked with a question mark. 
The system for which the gas {\it and} the stellar components are not satisfying the two conditions are not reported in figure. Line velocity centroids and line widths of narrow \ha, total \ha and stars are reported for each target, as labeled. See Fig. \ref{PV_I13120} for further details about the individual panels. 
}
\label{PV_all}
\end{figure*}

\begin{figure*}[h]
\ContinuedFloat
\centering
\includegraphics[width=9.1cm,trim= 0 0 0 0,clip]{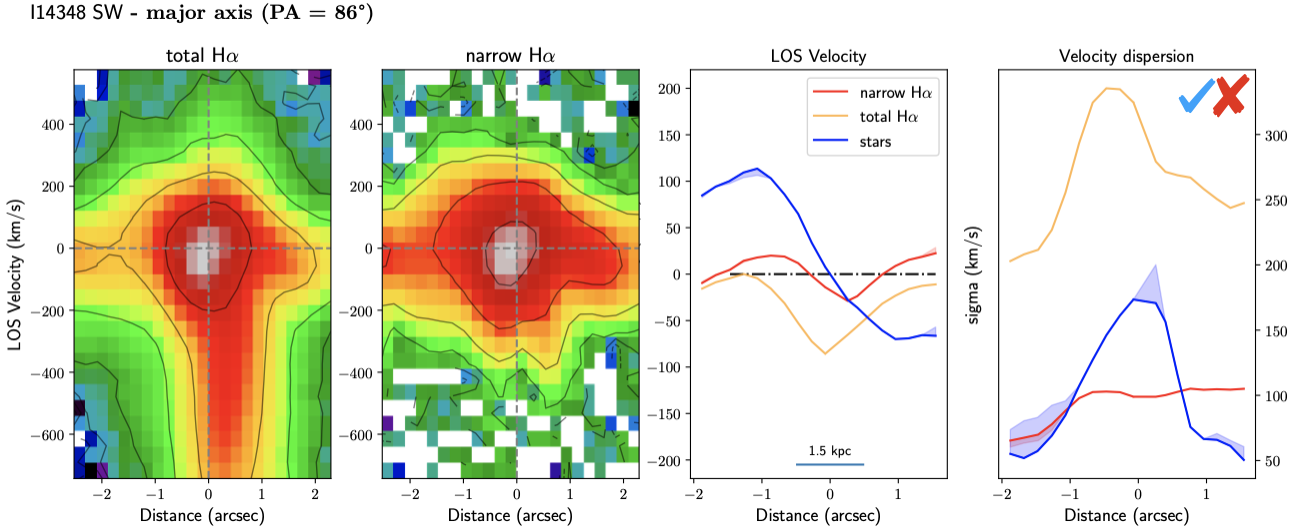}
\includegraphics[width=9.1cm,trim= 0 0 0 0,clip]{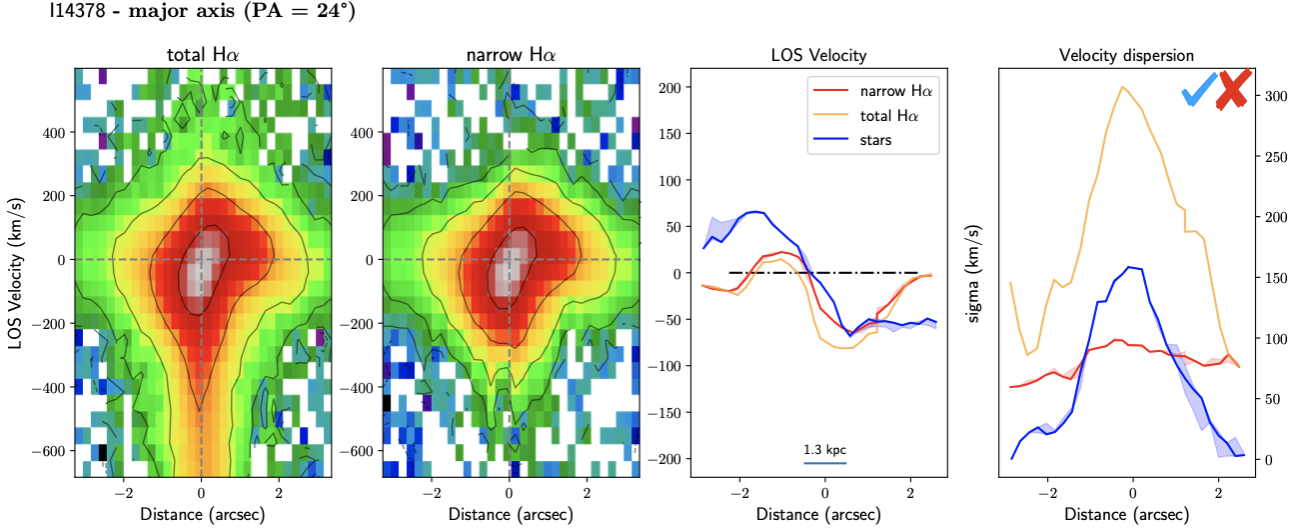}
\includegraphics[width=9.1cm,trim= 0 0 0 0,clip]{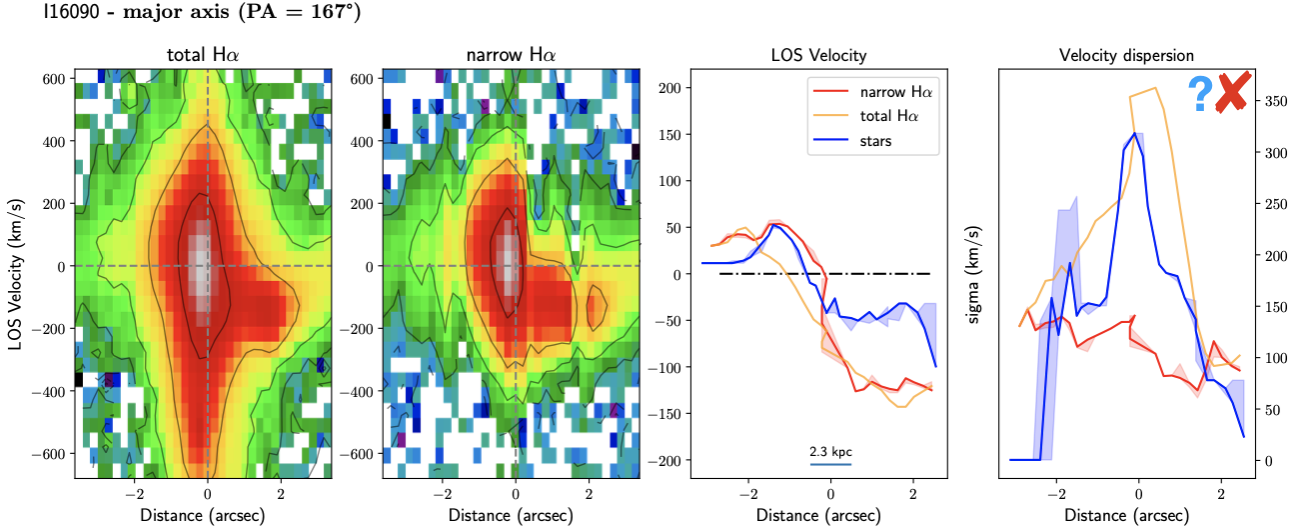}
\includegraphics[width=9.1cm,trim= 0 0 0 0,clip]{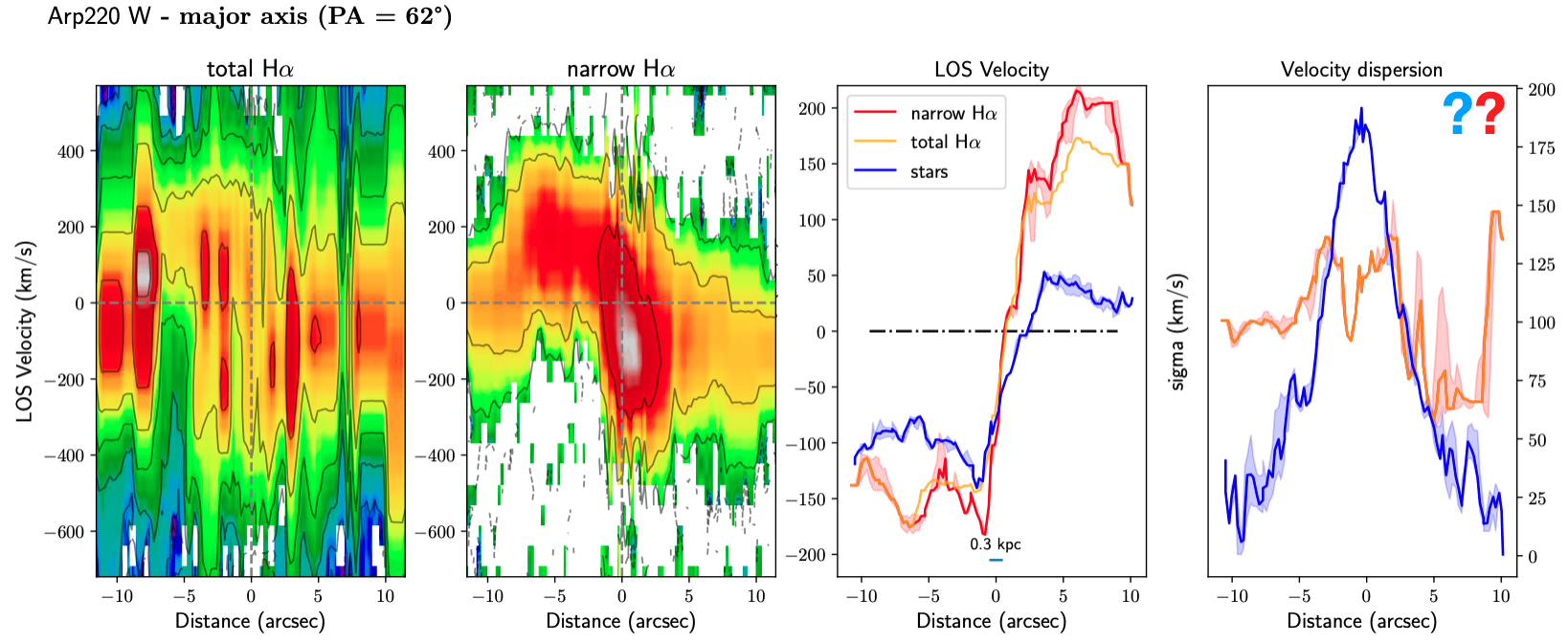}
\includegraphics[width=9.1cm,trim= 0 0 0 0,clip]{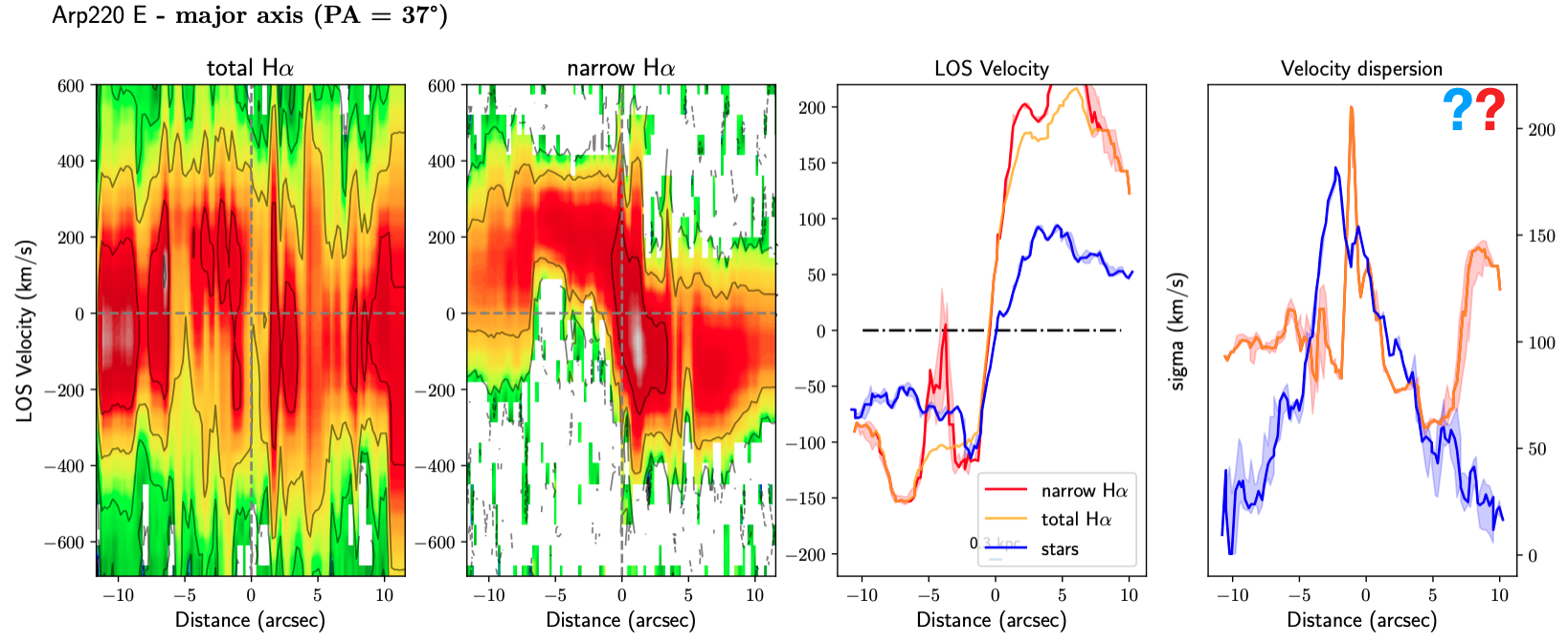}
\includegraphics[width=9.1cm,trim= 0 0 0 0,clip]{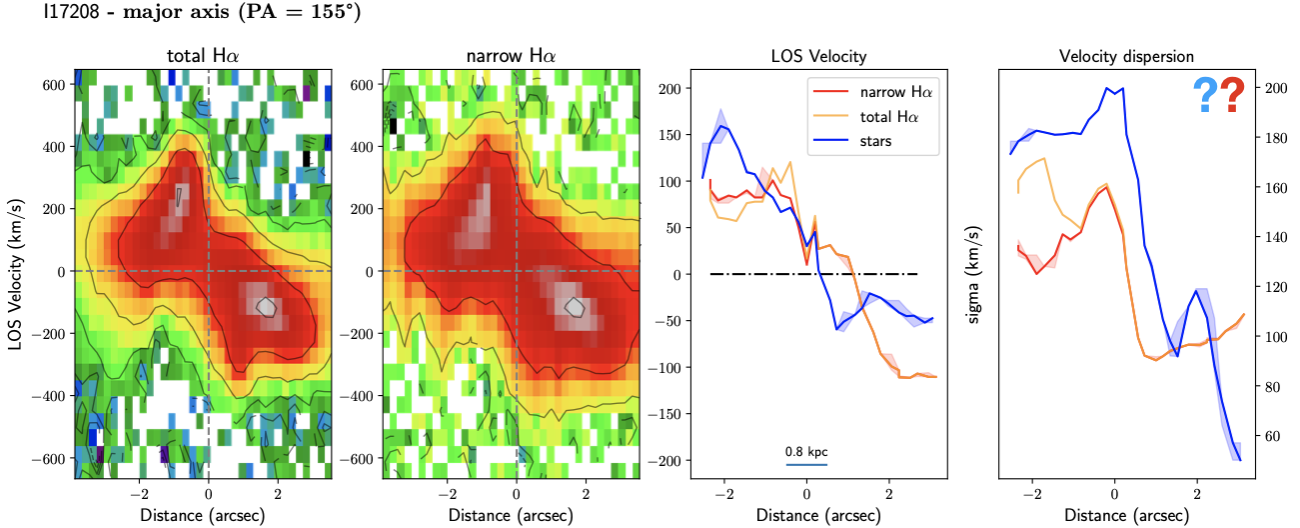}
\includegraphics[width=9.1cm,trim= 0 0 0 0,clip]{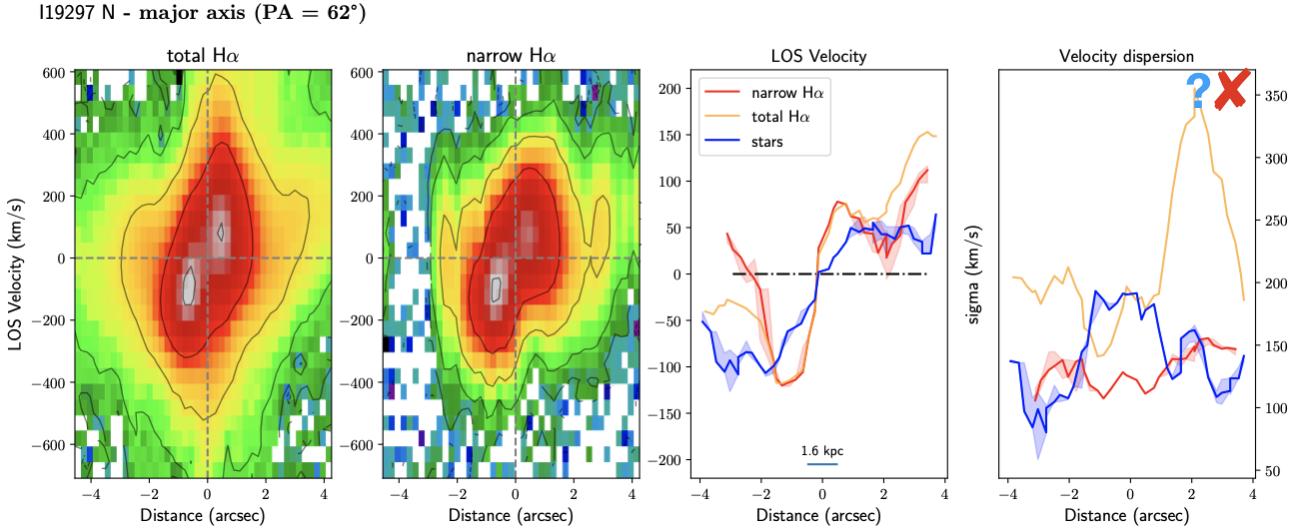}
\includegraphics[width=9.1cm,trim= 0 0 0 0,clip]{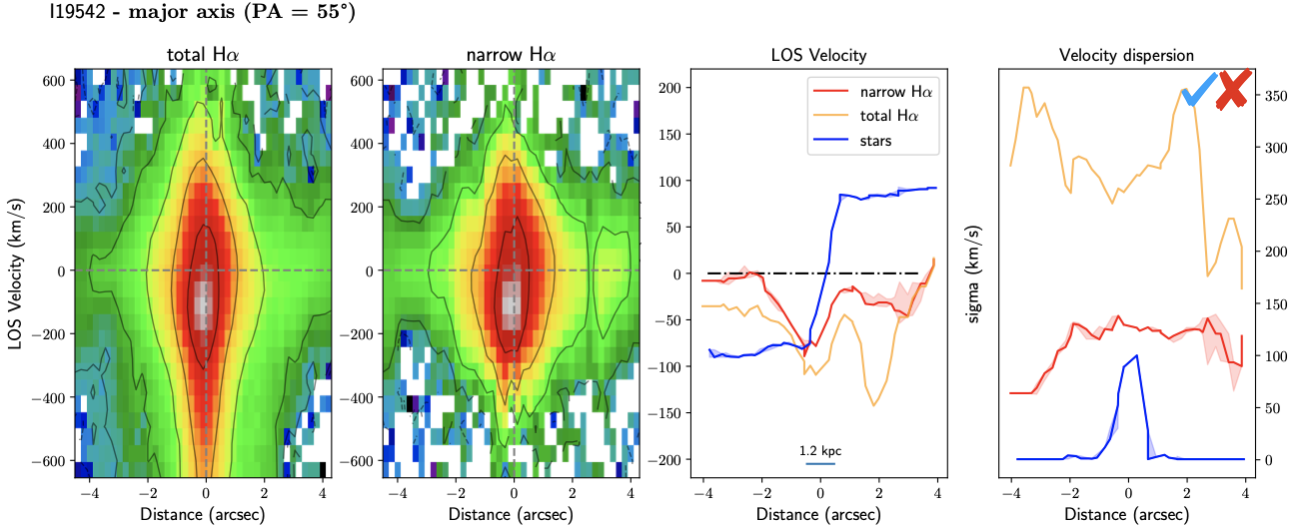}
\includegraphics[width=9.1cm,trim= 0 0 0 0,clip]{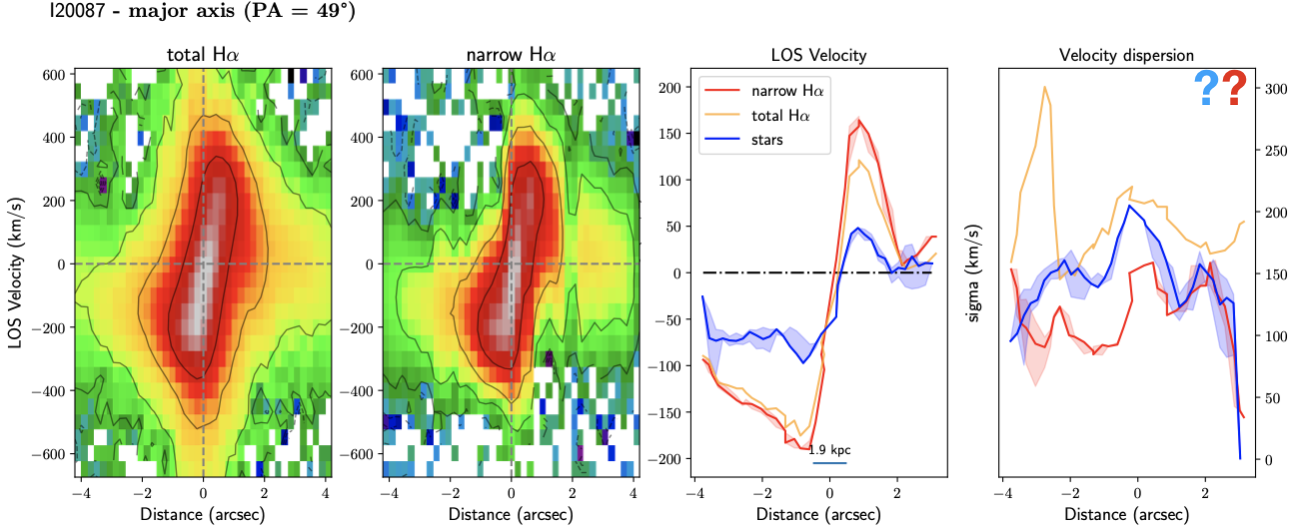}
\includegraphics[width=9.1cm,trim= 0 0 0 0,clip]{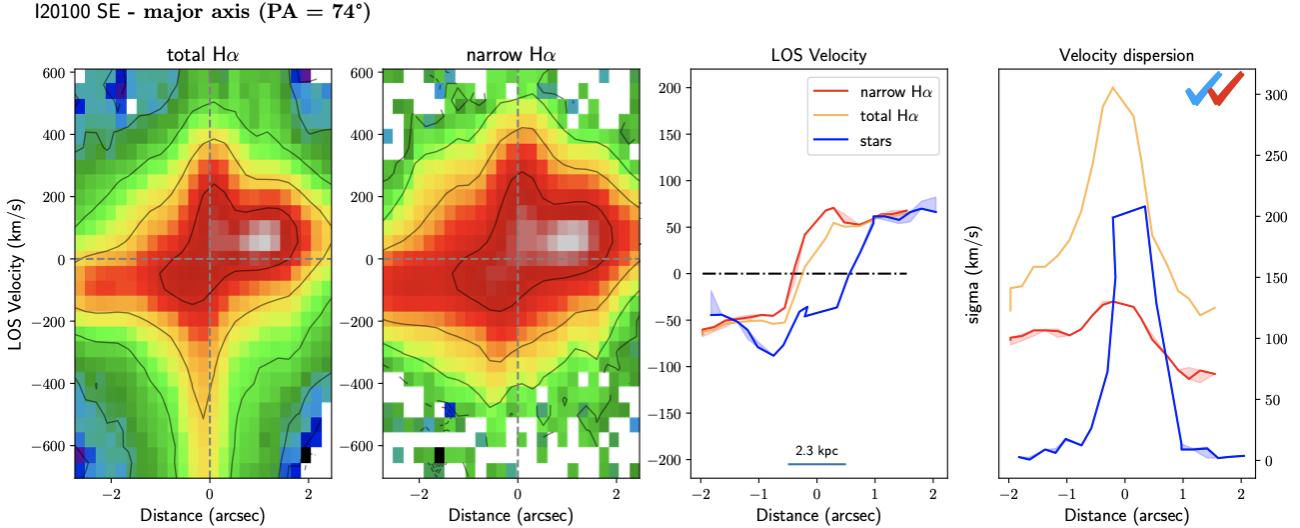}

\caption{\small 
Continued.
}
\end{figure*}

\section{3D-Barolo analysis}\label{A3DBarolo}

As already reported in Sect. \ref{S3DB}, we followed two different methods to derive 3D-Barolo best-fit results. With the first method, we first tried to constrain the disk inclination using different azimuthal models spanning the almost entire range of  inclinations ($i \in 5-85^{\circ}$), selecting the $i$ value with minimal residual. Finally, we ran 3D-Barolo with local models, fitting the rotation velocity $v_{rot}$, the velocity dispersion $\sigma$, and the major axis PA $\phi$, using the disk inclination derived in the first step as initial guess. With the second method, we directly fit all disk kinematic parameters with a local model, by initialising the inclination to the value derived from the \Isophote \ modelling of HST data (Sect. \ref{Sphot}).  
The general fitting procedure is here explained in more details for each target in our sub-sample of PUMA systems with evidence of rotation. 

\begin{figure}[t]
%
\centering
\includegraphics[width=8.cm,trim= 50 975 485 0,clip]{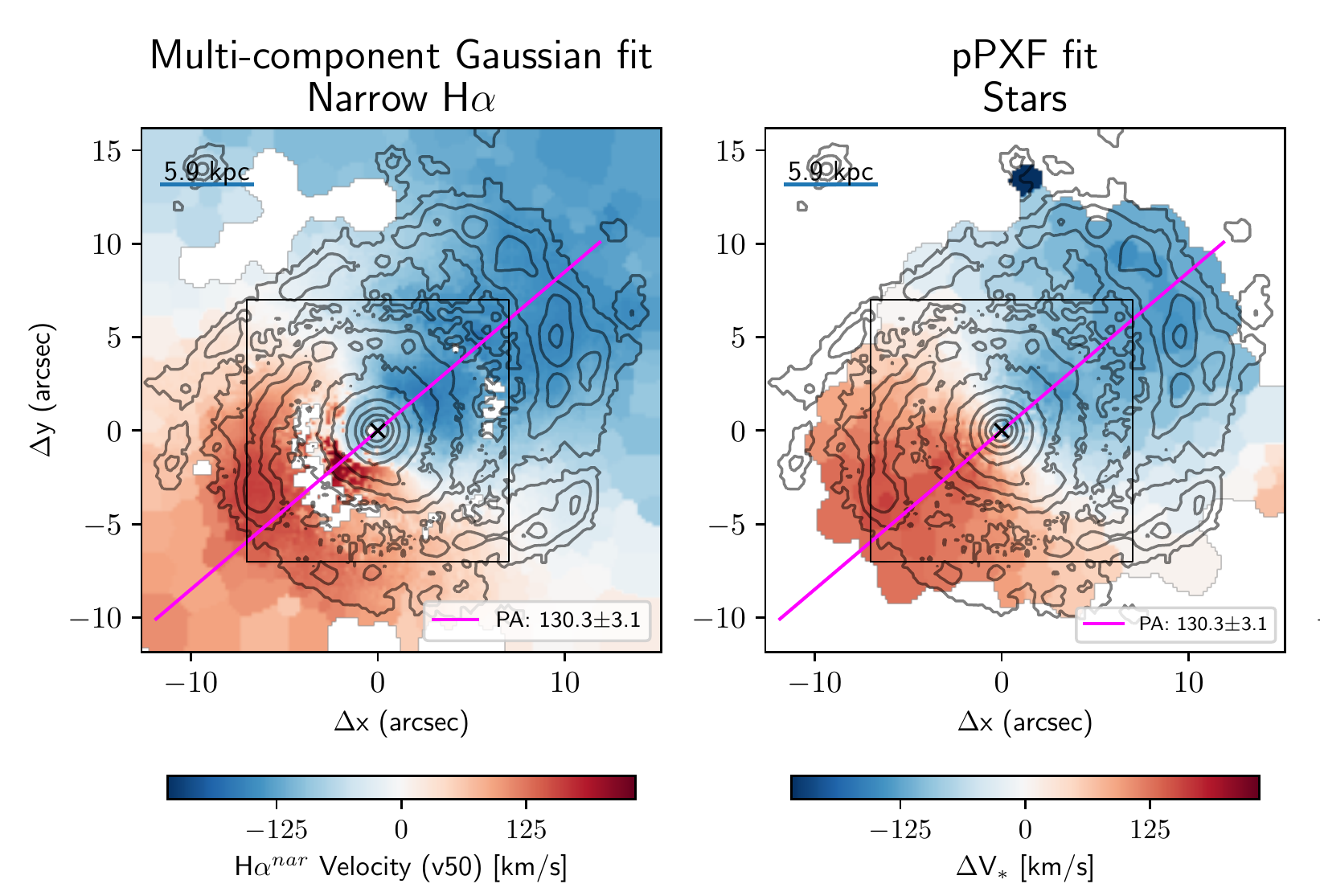}
\includegraphics[width=9.cm,trim= 0 0 0 0,clip]{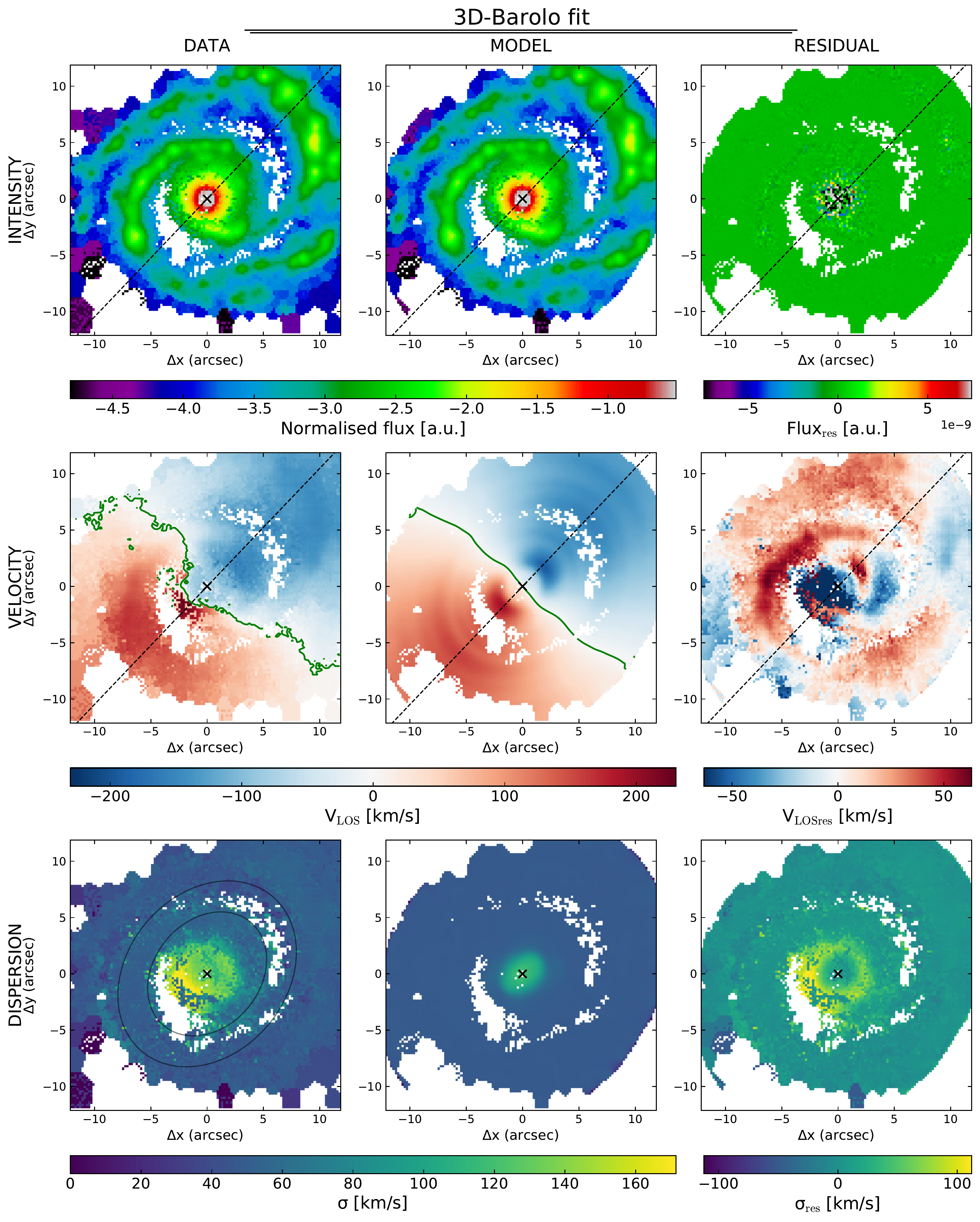}
\includegraphics[width=9.cm,trim= 0 0 0 27,clip]{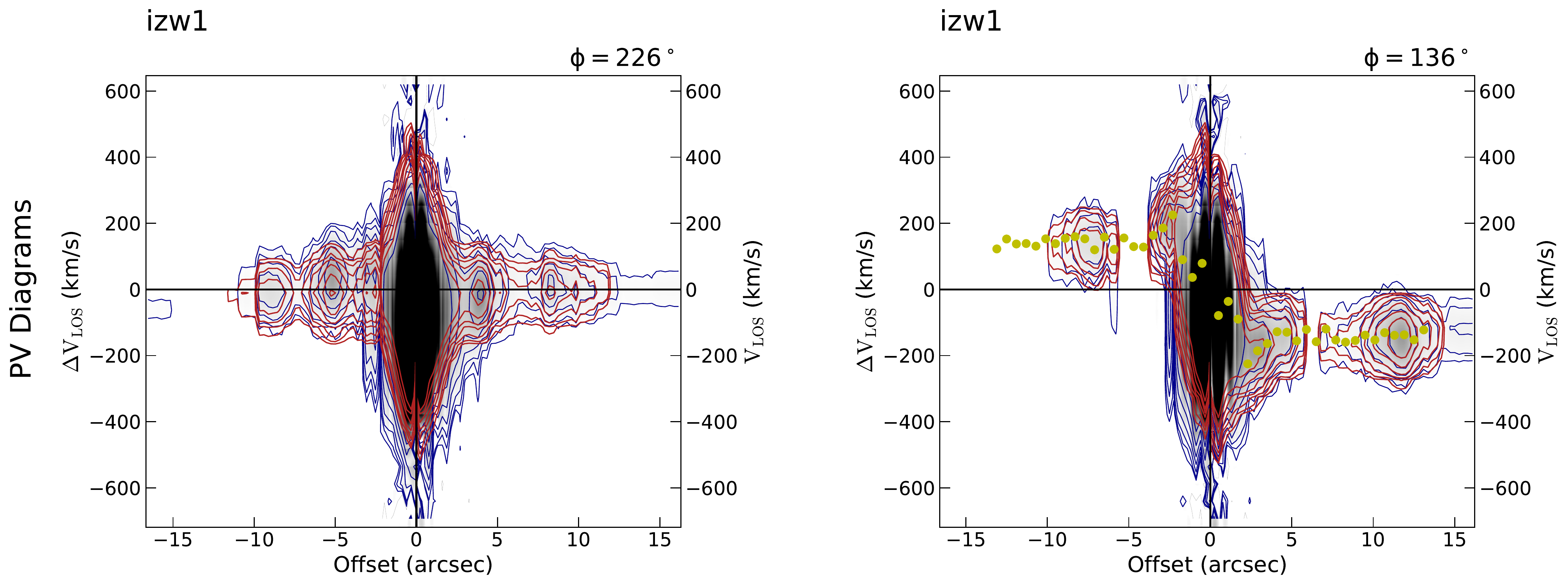}

\caption{\small 
IZw1 velocity maps and 3D-Barolo disk kinematic best-fit.
{\it Top panels}:    narrow \ha (left) and pPXF stellar (right) velocity maps. The magenta line identifies the major axis PA measurement, computed within the black box region; the black cross identifies the nucleus. {\it Second-to-fourth row panels}: comparison of 3D-barolo data and model moment maps, as labeled in the figure. In the intensity and velocity maps, we reported the major axis PA (dashed line), and the position of the kinematic centre (black cross); the green curve in the velocity maps represent the zero-velocity axis. In the dispersion map on the left, the black curves show the region from which the median $\sigma_0$ value has been derived. {\it Bottom panels}: PV diagram along the major axis (right) and minor axis (left) for both data (grey map and blue contours) and best-fit model (red contours, and yellow dots associated to individual concentric rings used to model the data).
}
\label{BBaroloizw1}
\end{figure}

{\bf IZw1}: 
In this target, a correct ionised gas kinematic decomposition between disk and outflow components in the innermost nuclear regions is challenging. The strong outflow, the BLR and iron emission close to the \ha are responsible of a strong degeneracy in the multi-component Gaussian fit results. As a consequence, the narrow \ha map shown in the top-left panel of Fig. \ref{BBaroloizw1} display a  blueshifted kinematic component in the innermost nuclear regions not associated with disk kinematics. Similarly, the more extreme redshifted velocities in the south-east direction at $\sim 3^{\prime\prime}$ are reasonably due to the same fit degeneracy (see also Fig. \ref{izw1linemaps}). On the contrary, the stellar kinematics (top-right panel in Fig. \ref{BBaroloizw1}) are more regular and display a clear rotation pattern.

The results obtained with  I and II methods are broadly consistent (see Table \ref{Tdisk}), and point to an intermediate inclination of $\sim 40^{\circ}$ and a PA$^{kin} \sim 140^{\circ}$, both consistent with the results presented in \citet{Tan2019} and derived from ALMA observations of the CO(1-0) molecular gas emission. 3D-Barolo fit results are reported in Fig. \ref{BBaroloizw1}. Significant residuals are observed in the innermost nuclear regions, due to the AGN driven outflow, and close to the spiral arms, probably due to the fact that 3D-Barolo uses a concentric rings structure instead of spiral models to reproduce the observed kinematics.

{\bf I07251}: This target shows two interacting nuclei at a projected distance of $\sim 1.8$ kpc, and very extended tails and streaming gas with velocities from $\sim -150$ \kms (north-east) to $\sim +150$ \kms (north-west). In the innermost nuclear regions, it presents disk-like motions, but with a kinematic centre not coincident with none of its two nuclei (Fig. \ref{I07251linemaps}). I07251 flux distribution is very irregular, and cannot be modelled with elliptical isophotes. 

\begin{figure}[t]
\centering

\includegraphics[width=8.cm,trim= 50 975 485 30,clip]{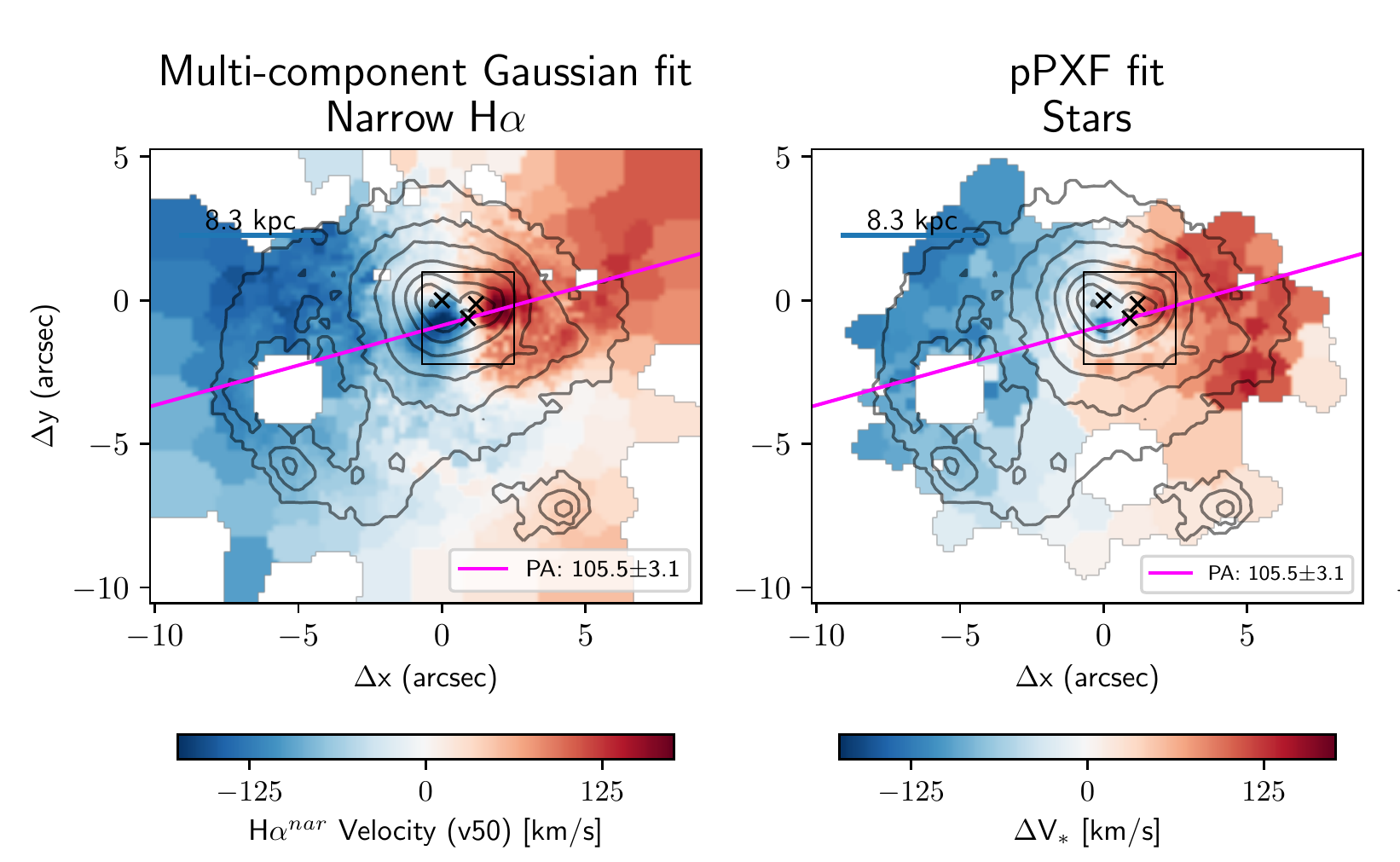}
\includegraphics[width=9.cm,trim= 0 0 0 0,clip]{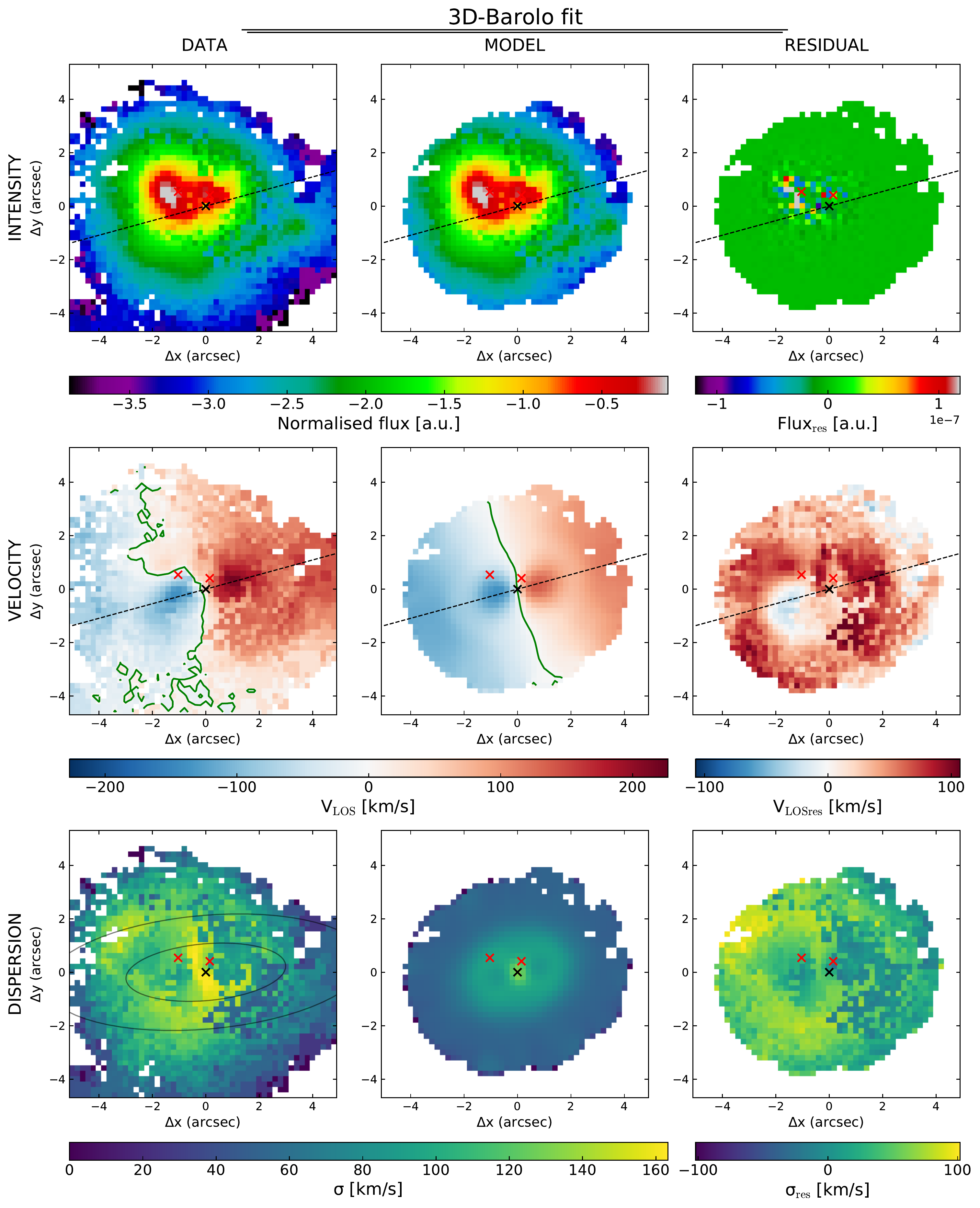}
\includegraphics[width=9.cm,trim= 0 0 0 27,clip]{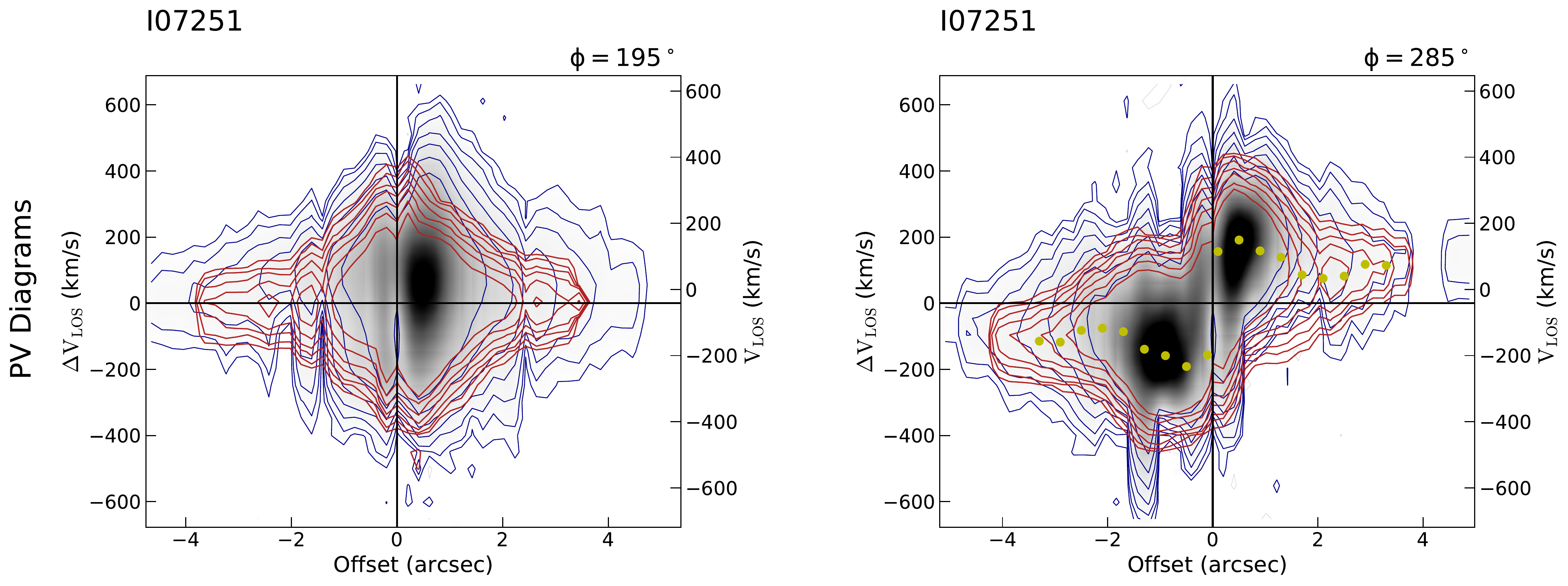}

\caption{\small 
I07251 velocity maps and 3D-Barolo disk kinematic best-fit. In this target, the kinematic centre is shown with a black cross, while the two ULIRG nuclei with red crosses. See Fig. \ref{BBaroloizw1} for further details. 
}
\label{BBaroloI07251}
\end{figure}

We fitted the narrow \ha data cube with 3D-Barolo, following the I method strategy, fitting $v_{rot}$, $\sigma$, $\phi$, and the kinematic centre position. Unfortunately, the data quality does not allow us to constrain the disk inclination in this target; we therefore performed the second step of the 3D-Barolo fit assuming a mean inclination of $52^{\circ}$ (\citealt{Bellocchi2013}) as initial guess. The 3D-Barolo best-fit results are shown in Fig. \ref{BBaroloI07251}. Also for this target, we observe significant residuals in the velocity and velocity dispersion maps, due to the complex nature of this interacting system. 

{\bf I10190 W}: I10190 shows two interacting nuclei at a projected distance of 7.2 kpc, and very extended tails and streaming gas with velocities from $\sim -200$ \kms (south-east) to $\sim +300$ \kms (north-west). Each nucleus shows disk-like kinematics on kpc scales (Fig. \ref{I10190linemaps}); however, both the stellar and gas kinematics in the vicinity of the E nucleus are strongly affected by the presence of the W system motions (see e.g. Fig. \ref{APV}). This limits the possibility to study the E nucleus kinematics.  

\begin{figure}[t]
\centering
\includegraphics[width=8.cm,trim= 50 975 485 30,clip]{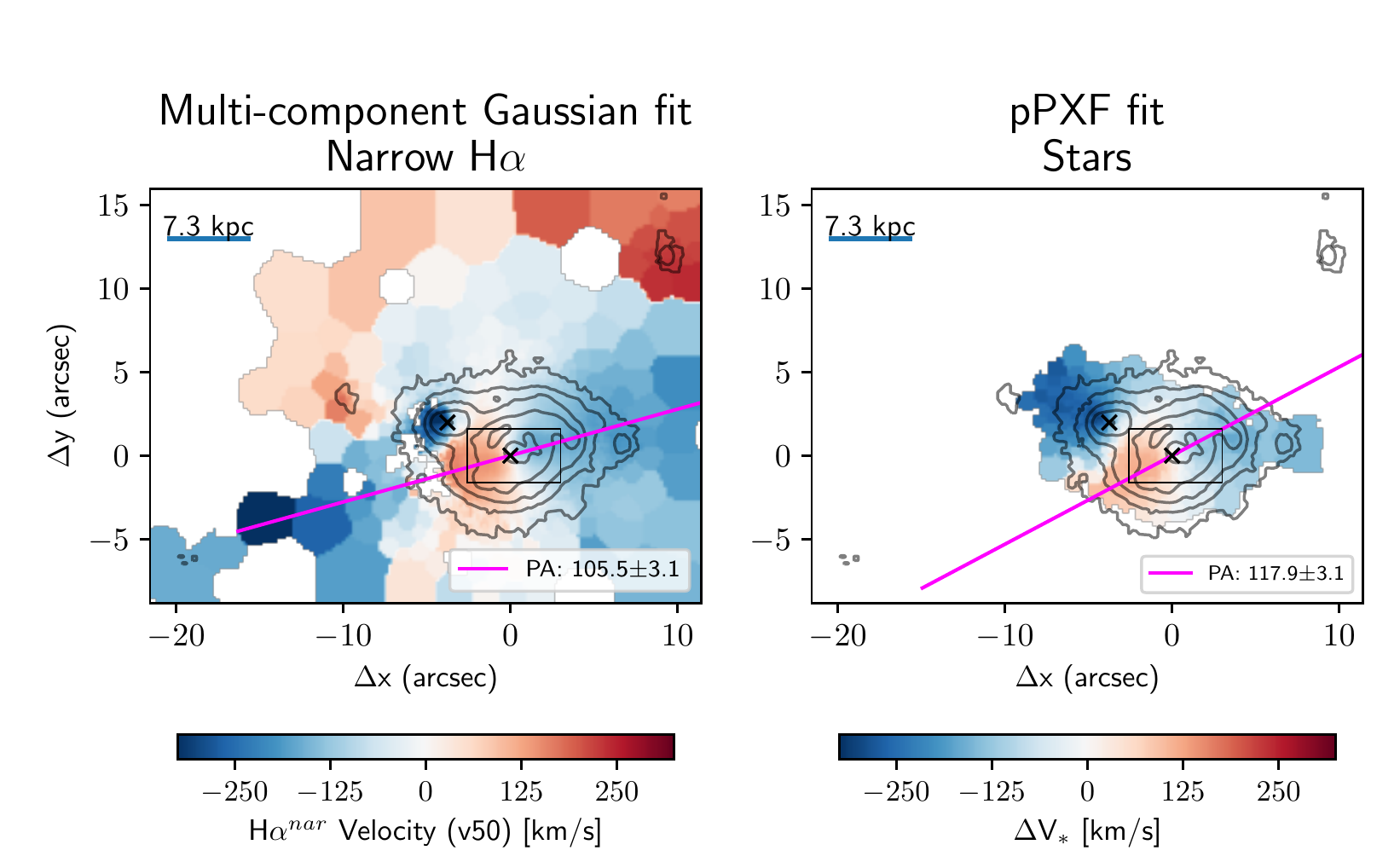}
\includegraphics[width=9.cm,trim= 0 0 0 0,clip]{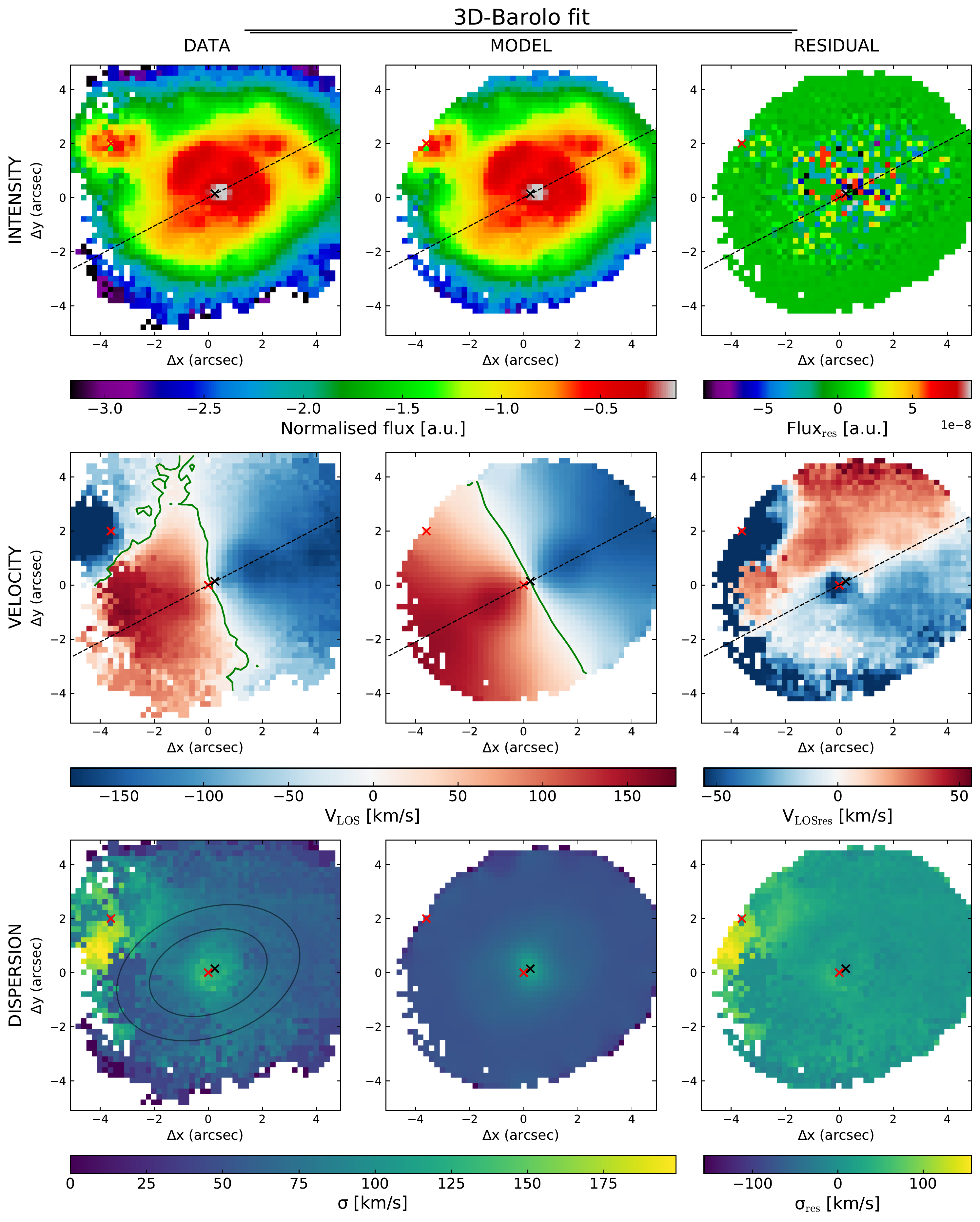}
\includegraphics[width=9.cm,trim= 0 0 0 27,clip]{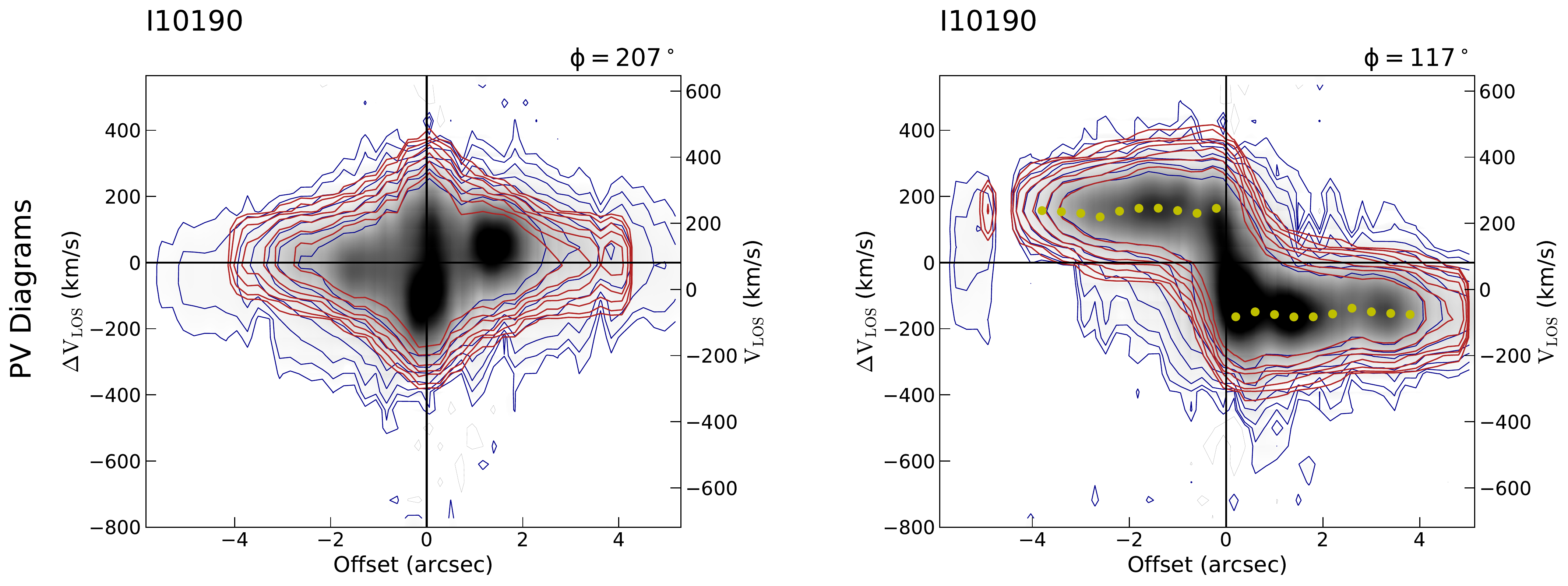}

\caption{\small 
I10190 W velocity maps and 3D-Barolo disk kinematic best-fit. See Fig. \ref{BBaroloizw1} for further details. 
}
\label{BBaroloI10190}
\end{figure}

We fitted the I10190 W gas kinematics with both I and II methods, obtaining totally consistent results (Fig. \ref{BBaroloI10190}). Also for this target, we found a slightly offset between the near-IR nucleus and the fitted kinematic centre ($\sim 0.3^{\prime\prime}$); significant residuals are also observed in the 3D-Barolo maps.  

\begin{figure}[t]
\centering
\includegraphics[width=8.cm,trim= 50 1195 485 10,clip]{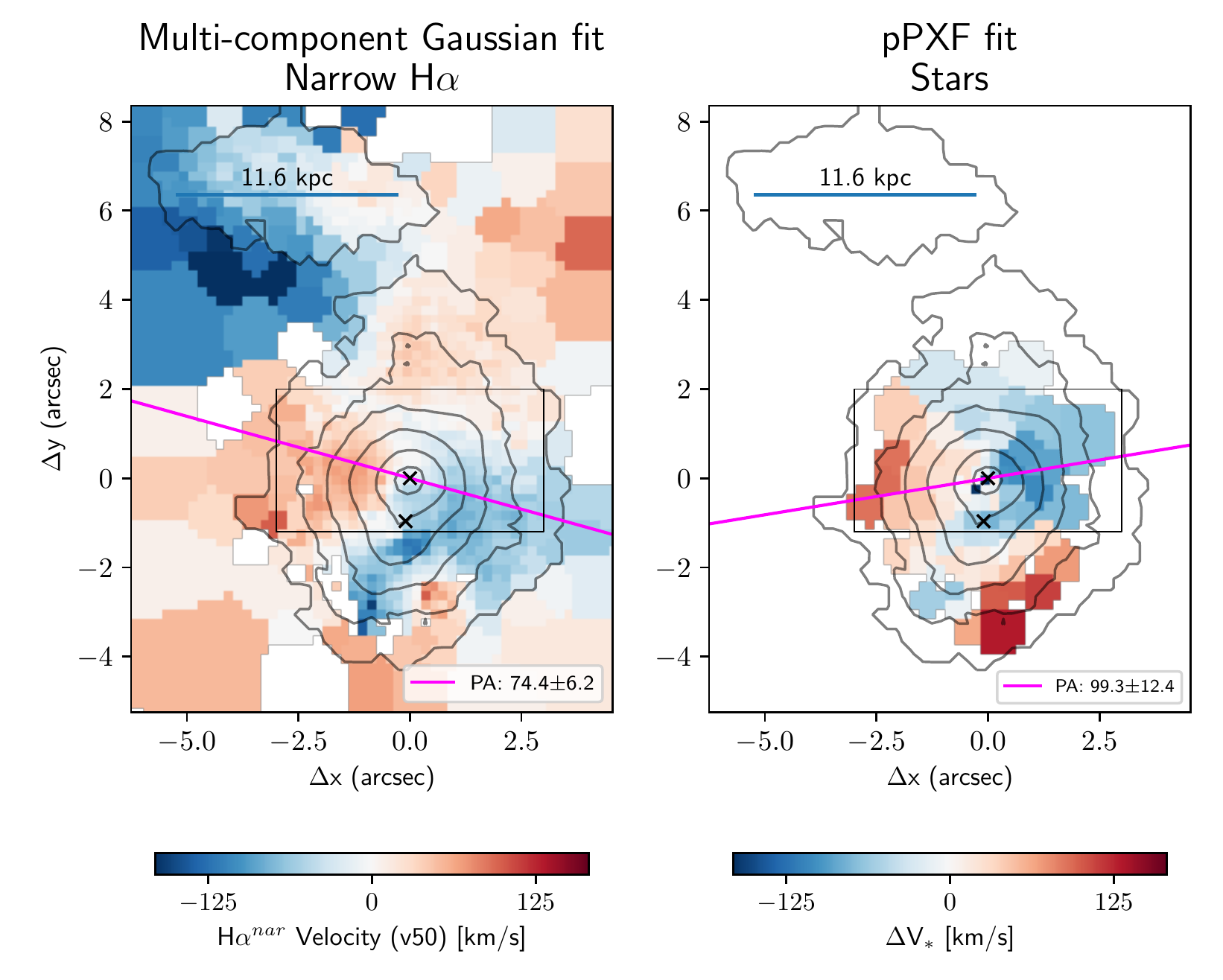}
\includegraphics[width=9.cm,trim= 0 0 0 0,clip]{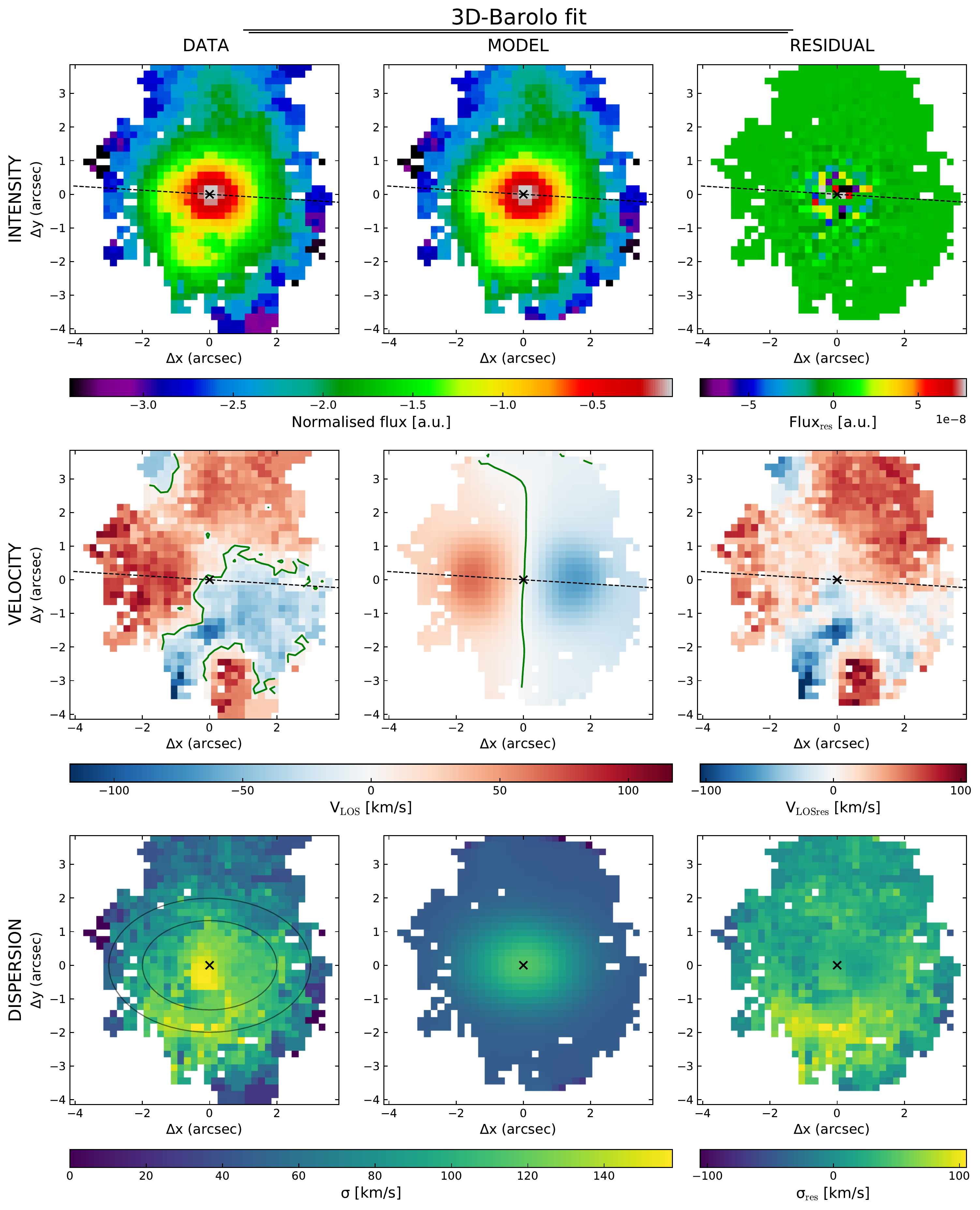}
\includegraphics[width=9.cm,trim= 0 0 0 27,clip]{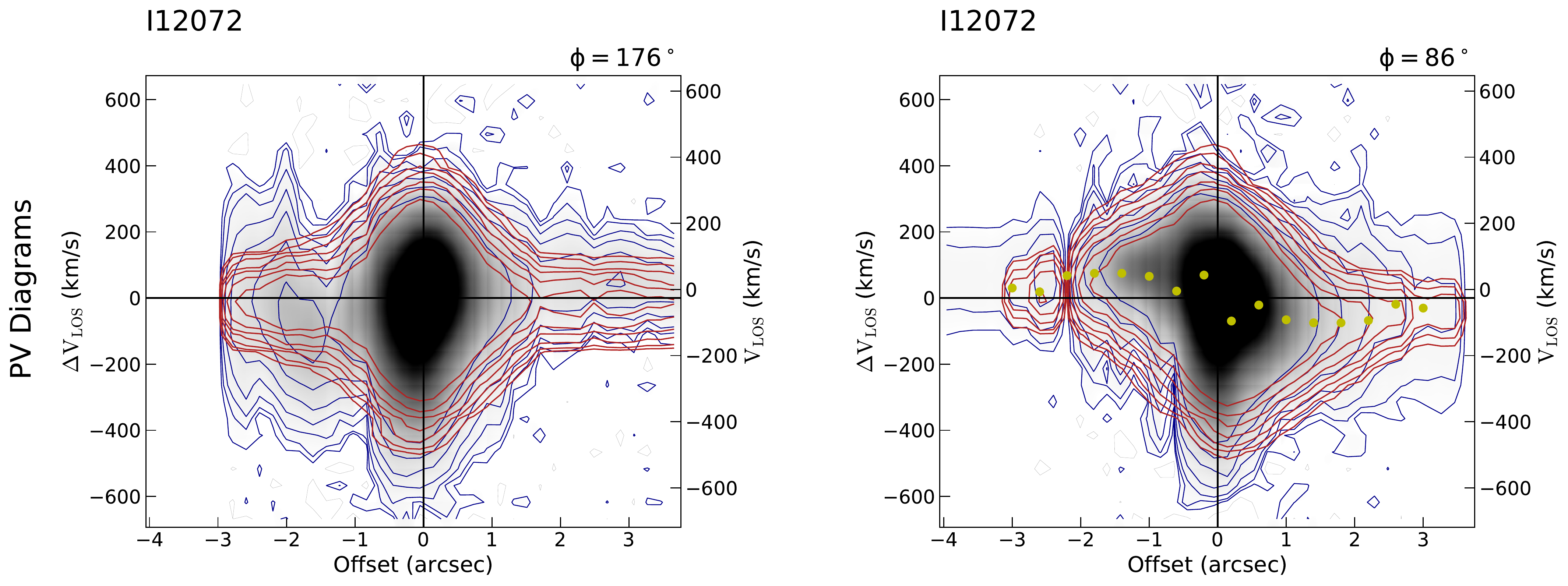}

\caption{\small 
I12072 velocity maps and 3D-Barolo disk kinematic best-fit. See Fig. \ref{BBaroloizw1} for further details. 
}
\label{BBaroloI12072}
\end{figure}

{\bf I12072 N} This source presents two nuclei, at a projected distance of 2.3 kpc, and a very extended plume in the north-east (Fig. \ref{I12072linemaps}). Disk-like motions are presumably associated with the N nucleus. The flux distribution is complex, and does not allow us to perform a robust  \Isophote \ modelling. We therefore fitted the narrow \ha data cube with the 3D-Barolo, applying the I method. The fit results are reported in Fig. \ref{BBaroloI12072}. Also in this case, velocity and velocity dispersion maps present significant residuals, due to the complex nature of this ULIRG.

{\bf I13120} The 3D-Barolo analysis is extensively presented in Sect. \ref{S3DB}.

\begin{figure}[t]
\centering
\includegraphics[width=8.cm,trim= 50 995 485 0,clip]{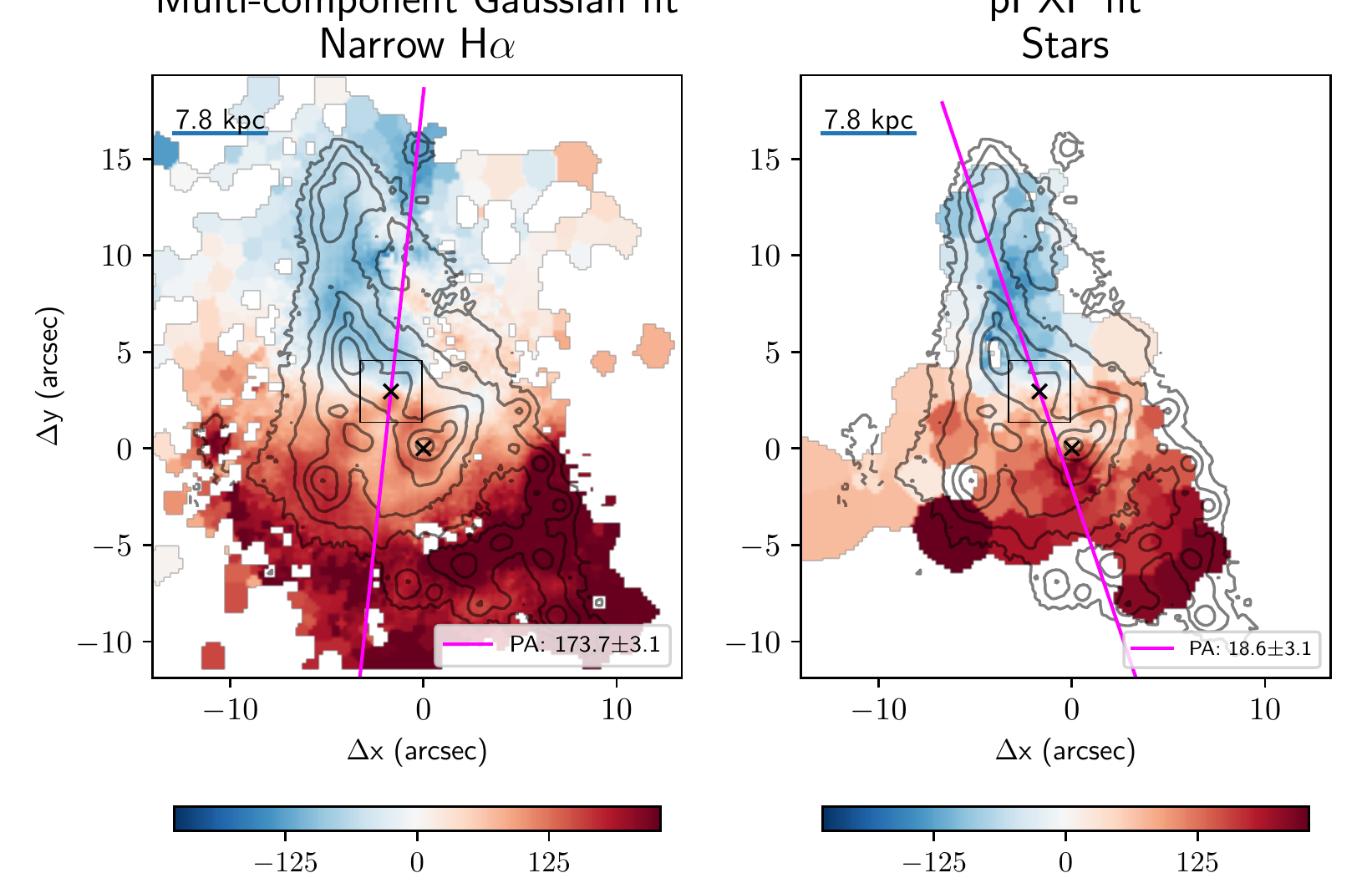}
\includegraphics[width=9.cm,trim= 0 0 0 0,clip]{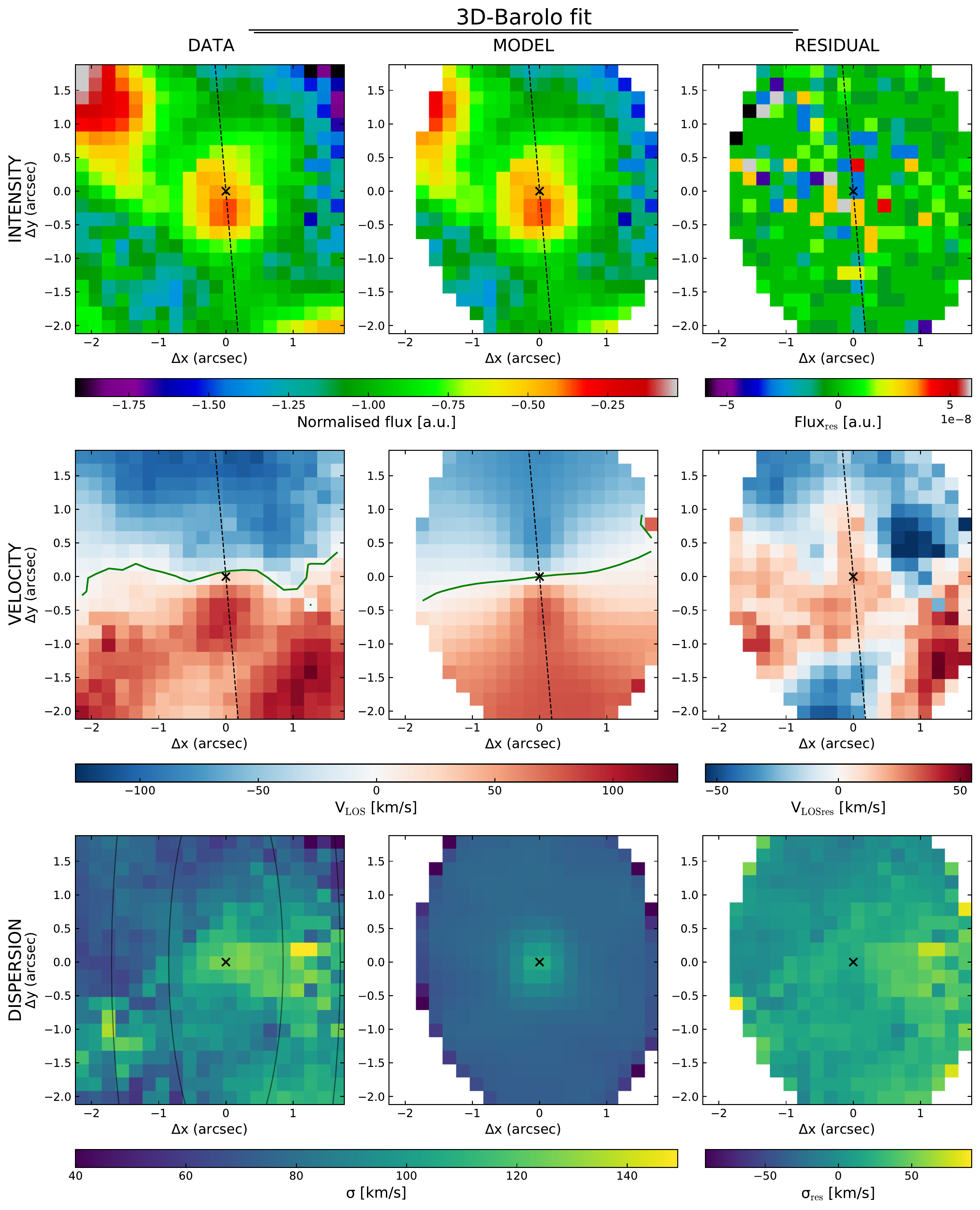}
\includegraphics[width=9.cm,trim= 0 0 0 27,clip]{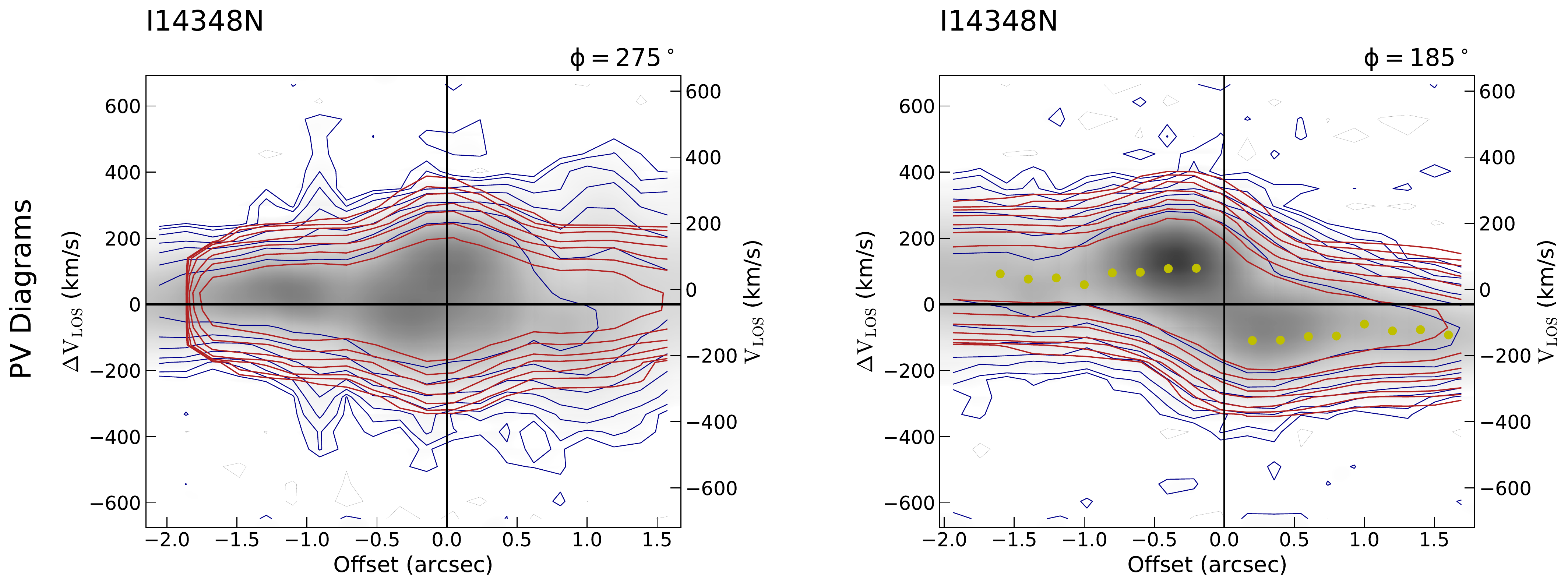}

\caption{\small 
I14348 NE velocity maps and 3D-Barolo disk kinematic best-fit. See Fig. \ref{BBaroloizw1} for further details. 
}
\label{BBaroloI14348NE}
\end{figure}

{\bf I14348 NE} I14348 is an interacting ULIRG with two nuclei at a projected distance of 5.3 kpc. The NE nucleus presents disk-like motions, while the SW kinematics are dominated by a strong outflow pointing to south-west; streaming motions along the north-east south-west direction are also present on scales of 10s kpc (Fig. \ref{I14348linemaps}). 
We fitted with 3D-Barolo the kinematics in the NE nuclear region, applying the I method to infer, as a first step, the inclination of the disk. 
In fact, the complex nature of this interacting system prevents robust \Isophote \ analysis and no morphological information is available. We decided to limit the 3D-Barolo fit analysis to the innermost nuclear regions ($\sim 3^{\prime\prime}\times3^{\prime\prime}$), in order to exclude the contribution from the large scale streaming motions. As a consequence, the disk inclination we determined with the I method, $52^\circ \pm 3^\circ$, has to be taken with caution.  
The fit results are reported in Fig. \ref{BBaroloI14348NE}. Significant residuals are observed in the velocity and velocity dispersion maps. The kinematic centre has been fixed at the position of the nucleus during the 3D-Barolo analysis, as no significant variations in fit results are observed considering the kinematic centre position as free parameter. 

\begin{figure}[t]
\centering
\includegraphics[width=8.cm,trim= 50 995 485 10,clip]{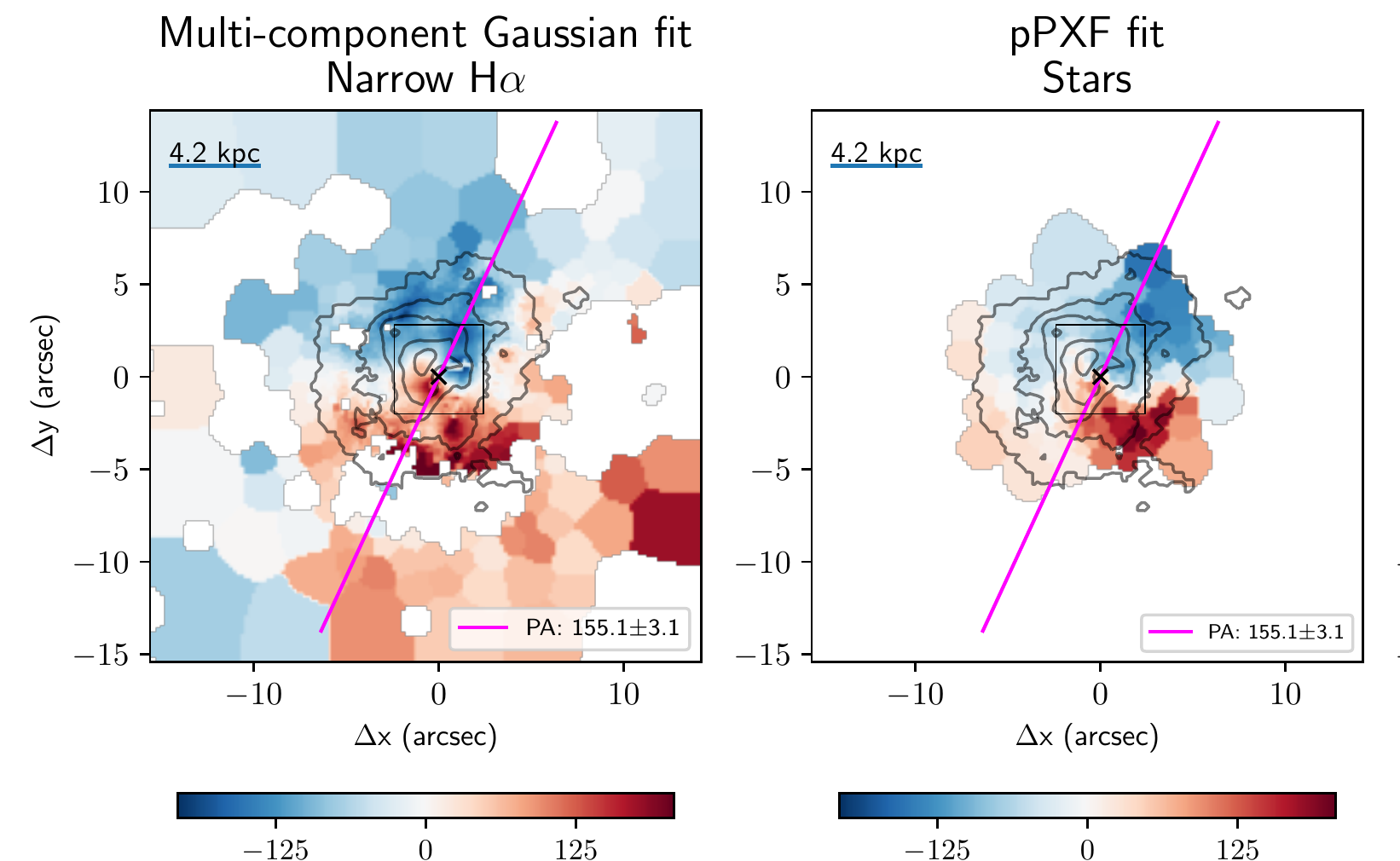}
\includegraphics[width=9.cm,trim= 0 0 0 0,clip]{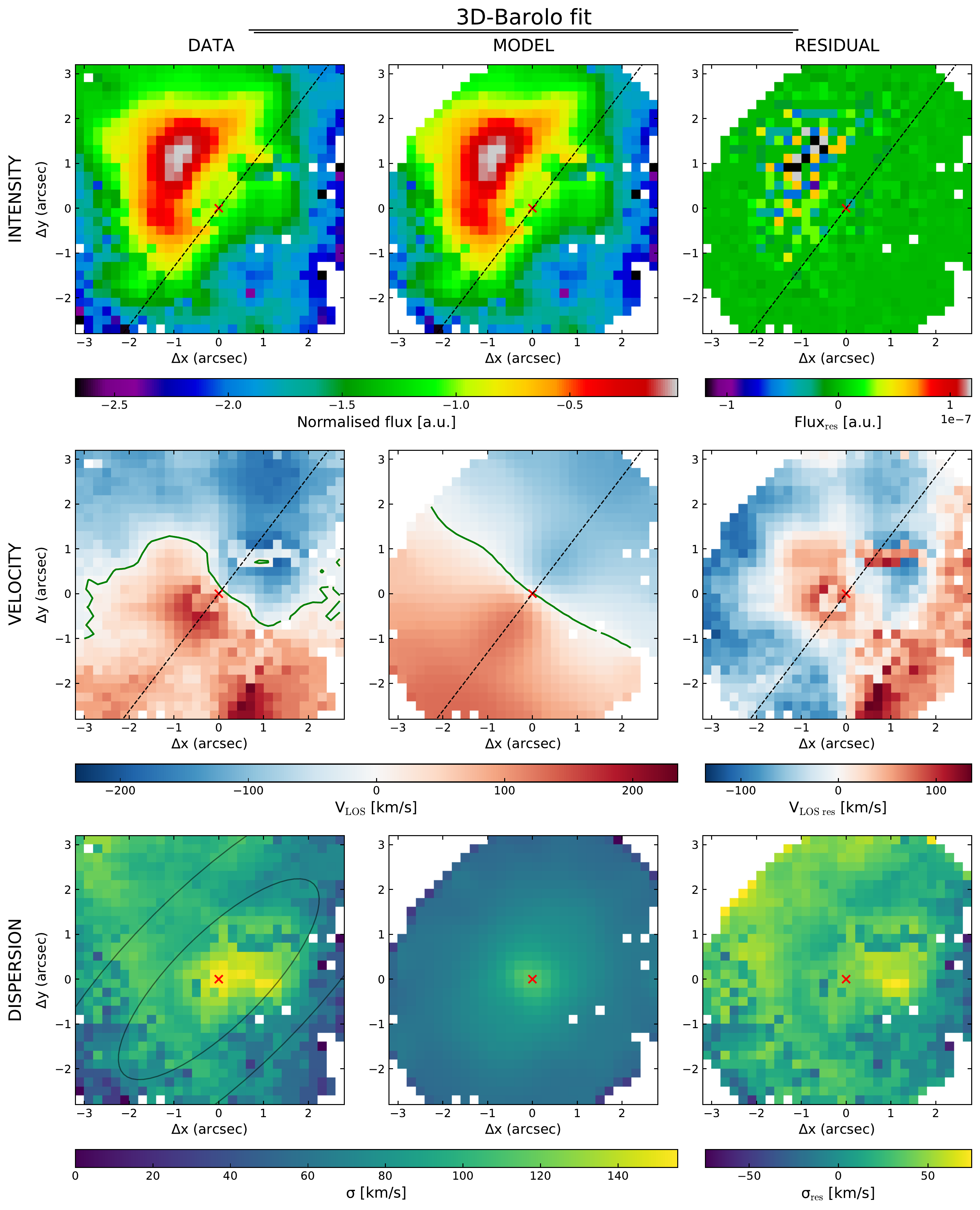}
\includegraphics[width=9.cm,trim= 0 0 0 27,clip]{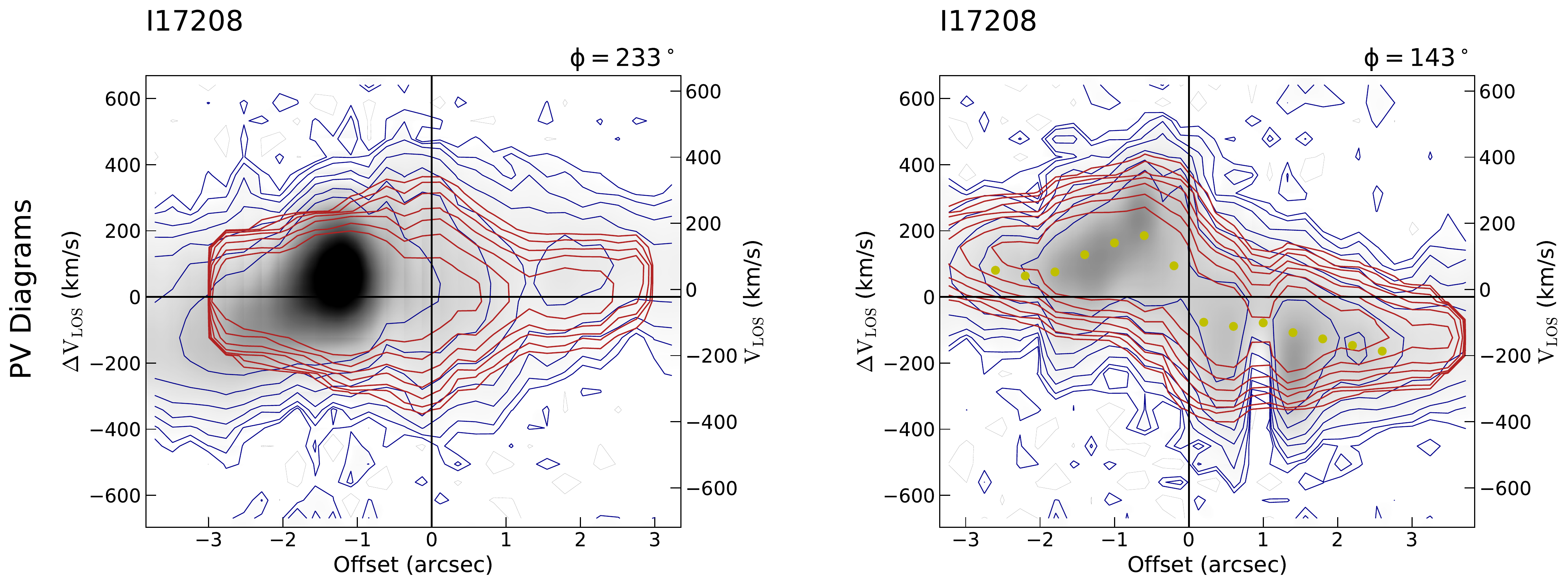}

\caption{\small 
I17208 velocity maps and 3D-Barolo disk kinematic best-fit. See Fig. \ref{BBaroloizw1} for details.  
}
\label{BBaroloI17208}
\end{figure}

{\bf I17208} This source presents disk-like kinematics and extended tidal tails with velocities from $\sim -150$ \kms (north) to $\sim +150$ \kms (south-west; see Fig. \ref{I17208linemaps}). For this target, we decided to exclude from the 3D-Barolo analysis the more external regions, associated with streaming motions; as a result, I method does not provide robust constraints for the disk inclination. We therefore applied the II method, assuming a disk inclination of $40^\circ$ (from \Isophote \ analysis) as initial guess. The 
3D-Barolo fit results are reported in Fig. \ref{BBaroloI17208}. The kinematic centre has been fixed at the position of the nucleus during the 3D-Barolo analysis; no significant variations in fit results are obtained adding the kinematic centre position as free parameter.

{\bf I20100 SE} The I20100 ULIRG is an interacting system with two nuclei at a projected distance of 6.5 kpc. The SE nucleus presents disk-like motions, while the NW kinematics are more complex and probably affected by the streaming motions (Fig. \ref{I20100linemaps}). We therefore decided to analyse with 3D-Barolo the kinematics in the vicinity of the SE nucleus only.  

\begin{figure}[t]
\centering
\includegraphics[width=8.cm,trim= 50 975 485 60,clip]{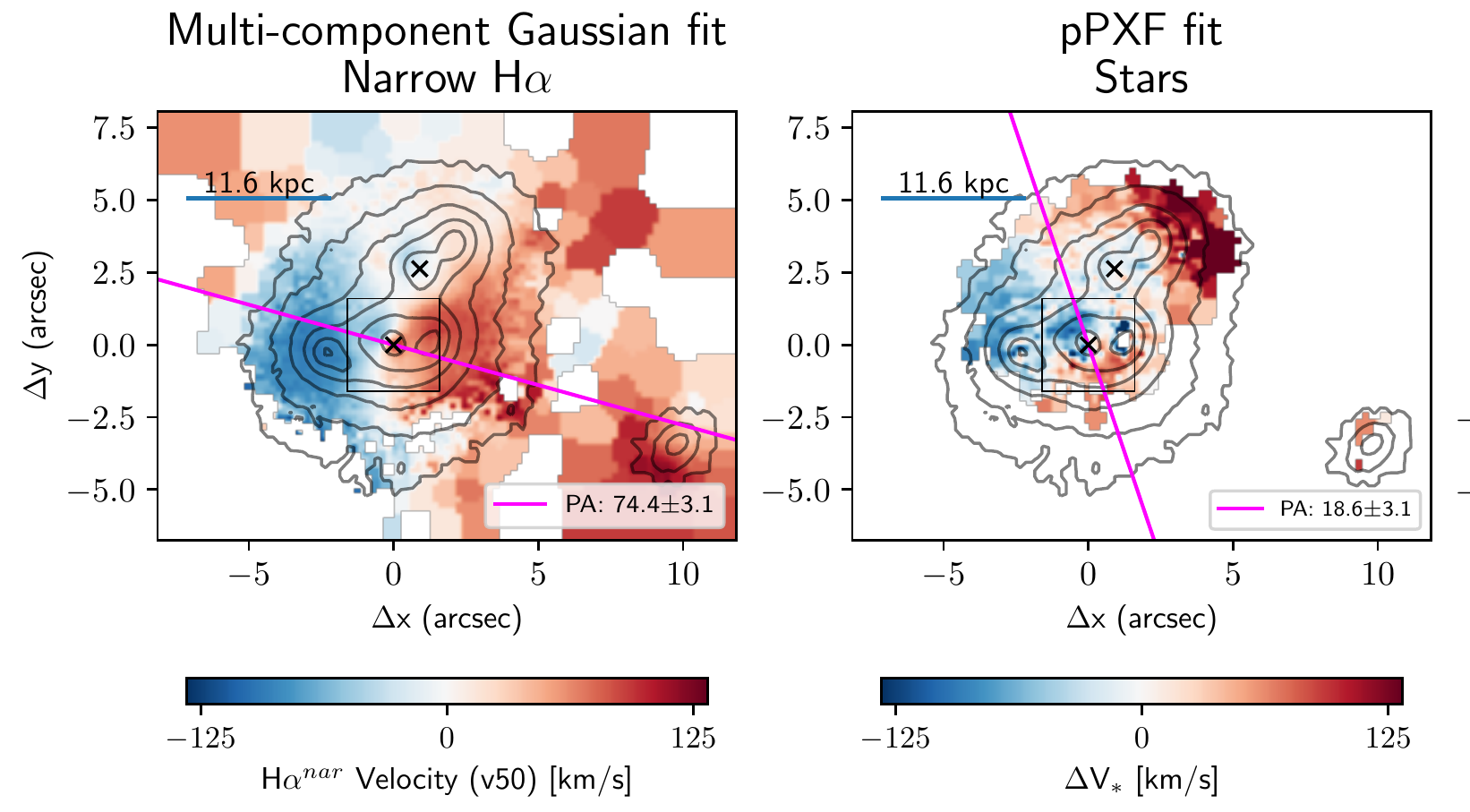}
\includegraphics[width=9.cm,trim= 0 0 0 0,clip]{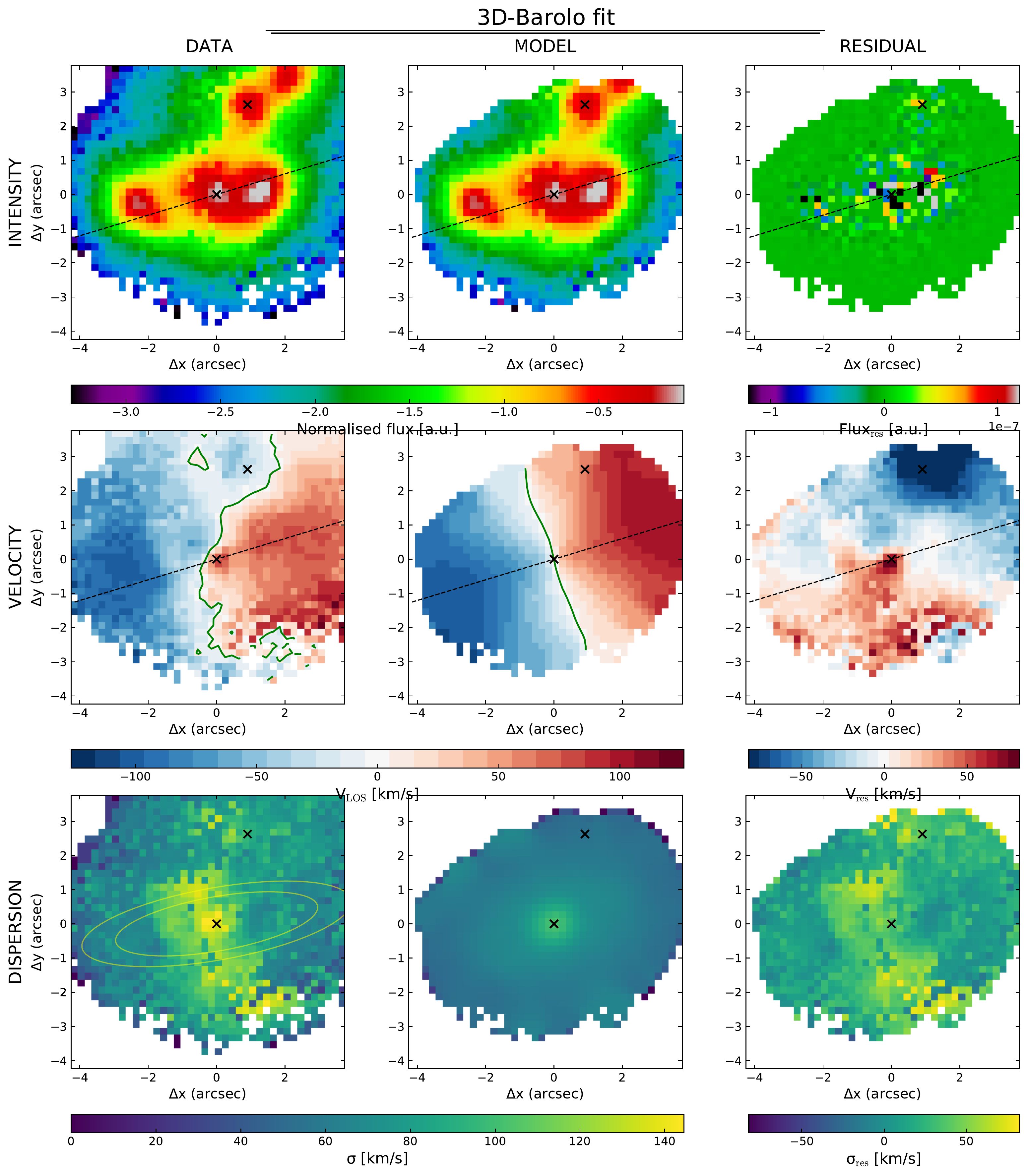}
\includegraphics[width=9.cm,trim= 0 0 0 27,clip]{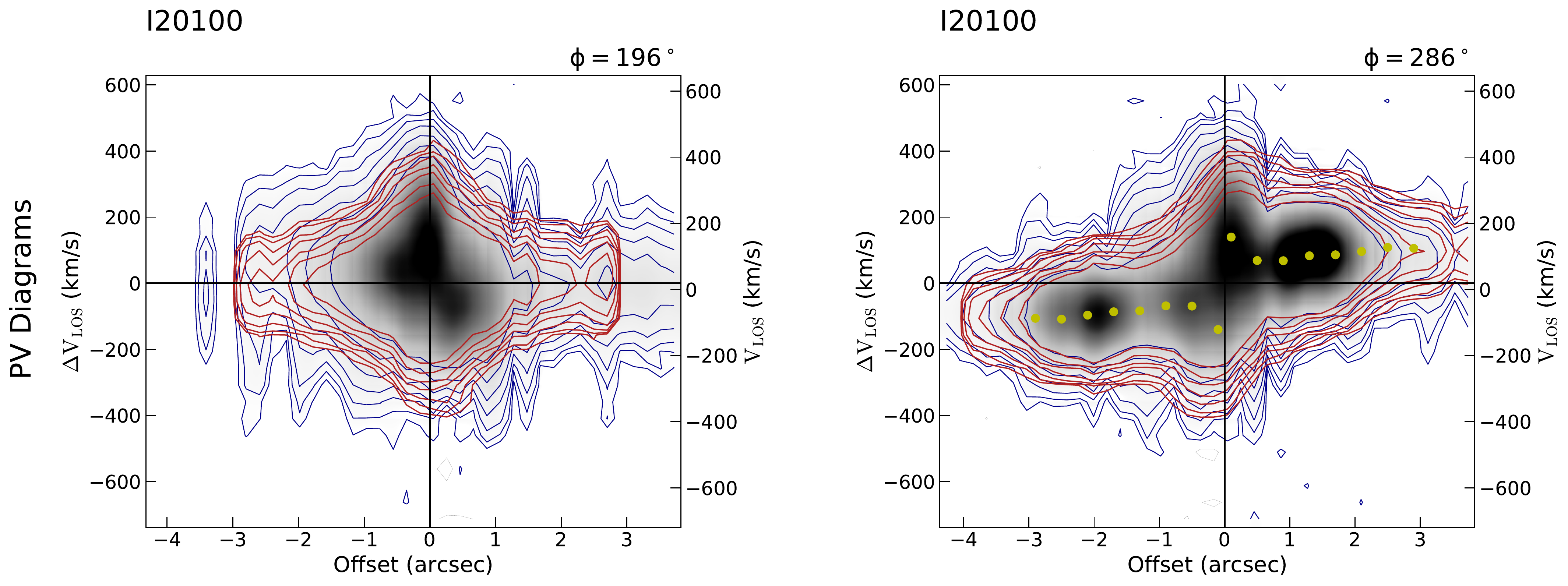}

\caption{\small 
I20100 SE velocity maps and 3D-Barolo disk kinematic best-fit. See Fig. \ref{BBaroloizw1} for details.  
}
\label{BBaroloI20100SE}
\end{figure}

The results obtained with methods I and II are broadly consistent (see Table \ref{Tdisk}), and point to an inclination of $\sim 58^{\circ}$ and a PA$^{kin} \sim 287^{\circ}$. The significant difference between gas and stellar PA$^{kin}$ might suggest more complex kinematics in the innermost nuclear regions of this source; indeed, we observe significant residuals in the velocity and velocity dispersion maps (Fig. \ref{BBaroloI20100SE}) along the gas kinematic minor axis. 

\section{Extended sample of SB disk galaxies}\label{Aextendedsample}

In this section, we introduce the additional individual ionised gas measurements of SB disk galaxies reported in Fig. \ref{sigmaz}. In particular, we select: 

\begin{itemize}
    \item the \ha (narrow component)  velocity dispersion of 33 LIRGs and one ULIRG with disk kinematics from \citet[][B13 in the figure]{Bellocchi2013}, observed with VLT/VIMOS, and the Br$\gamma$ velocity dispersion of 7 LIRGs from \citet[][C21 in the figure]{Crespo2021}, observed with VLT/SINFONI. All these systems are characterised by mean $\sigma$ of the order of several 10s \kms;  
    \item the \ha  velocity dispersion of 22 rotationally supported dusty galaxies of the cluster Cl0024+17 at $z \sim 0.4$ (\citealt[][J16 in the figure]{Johnson2016}), observed with VLT/FLAMES. These systems, with their median $v_{rot}/\sigma = 5\pm 2$ and their SFRs likely enhanced by the effects of ram pressure, also tend toward higher values compared to MS galaxies at the same redshift; 
    \item the \ha velocity dispersion of eight (U)LIRGs at $z \sim 0.2-0.4$ from \citet[][PS19 in the figure]{Pereira2019}, observed with the optical integral field spectrograph SWIFT. Most of them are interacting systems (6/8), and have relatively small $v_{rot}/\sigma $ ratios (from 0.4 to 3.2); 
    
    \item the Pa$\alpha$ velocity dispersion of three SB disk galaxies at $z \sim 0.15$ presented in \citet[][M20 in the figure]{Molina2020}, and observed with SINFONI;
    
    \item the \ha velocity dispersion of seven disk galaxies at $z\sim 2$ from the SINS/zC-SINF AO Survey, presented in \citet[][FS18 in the figure]{Schreiber2018}: ZC400528, ZC406690, ZC407302, ZC410123,  ZC411737, ZC413507, and ZC415876. These galaxies display an offsets by a factor of $\gtrsim 4$ in SFR from the MS in their Fig. 6, and $\sigma_0 \sim 30-60$ \kms;
    
    \item the \ha velocity dispersion of a starburst disk galaxy at $z = 2.028$, SMM J0217-0503b, merging with an AGN source (at a projected distance of $\sim 15$ kpc), from \citet[][A12 in the figure]{Alaghband2012}, and observed with VLT/SINFONI;
    
    \item the \ha velocity dispersion of a SB disk galaxy at $z = 2.24$, SHiZELS-14, from \citet[][Co21 in the figure]{Cochrane2021}. This source has been observed with the adaptive optics assisted VLT/SINFONI spectrograph;
    
    \item the [OIII] (narrow component) velocity dispersion of a submm bright disky galaxy hosting a broad line quasar, SMM J1237+6203 at $z = 2.075$, from \citet{Harrison2012}, and observed with Gemini-North NIFS.
\end{itemize}

All these measurements have been obtained from IFS data. For the B13, C21, J16, A12 and H12 sources, $\sigma_0$ was computed as a mean velocity dispersion across the galaxy extension in the FOV; PS19 $\sigma_0$ were inferred from GalPak$^{3D}$ modelization (correcting for the beam-smearing and line-spread-function  broadening), which assumes a spatially constant velocity dispersion in the disk; FS18 $\sigma_0$ were measured along the kinematic major axis at the largest radii possible, away from the central peak caused by the steep inner disk velocity gradient; M20 and Co21 $\sigma_0$ were instead derived as mean velocity dispersion in the external regions of the sources, excluding the innermost regions affected by beam smearing.

\end{appendix}

\end{document}